\documentclass[12pt,a4paper]{article}
\usepackage{cite}
\usepackage{url}
\usepackage{amsfonts}
\usepackage{amsmath}
\usepackage{amssymb}
\usepackage{graphicx}
\usepackage{epsfig}
\usepackage{lineno}
\usepackage[usenames]{color}
\usepackage{hyperref}
\usepackage{subfigure}
\usepackage[normalem]{ulem} % for strikethroughs (\sout{})
\usepackage{color}
\usepackage{units}
\usepackage{xcolor}
\usepackage[T1]{fontenc}
\usepackage{titling}
\graphicspath{{./pics/}}
\newif\iflatexdiff
\newif\ifcomment
\newif\ifplotpold
\plotpoldtrue
\newif\ifarxiv
\arxivtrue
\newcommand{\pp}           {pp}

\newcommand{\pA}           {pA}

\newcommand{\pPb}          {pPb}

\newcommand{\PbPb}         {PbPb}

\newcommand{\Pythia}       {PYTHIA}
\newcommand{\MPI}          {MPI}
\newcommand{\MPIs}         {MPIs}    
\newcommand{\dd}           {\rm d}
\newcommand{\pT}           {\ensuremath{p_{\rm T}}}
\newcommand{\pt}           {\pT}
\newcommand{\dNdeta}       {\ensuremath{{\dd}}N/{\dd}\eta}

\newcommand{\Ncoll}        {\ensuremath{N_{\rm coll}}}

\newcommand{\Nmpi}         {\ensuremath{N_{\rm mpi}}}

\newcommand{\s}            {\ensuremath{\sqrt{s}}}
\newcommand{\avg}[1]       {\ensuremath{\left<#1\right>}}  

\newcommand{\hrefurl}[1]   {\href{#1}{\url{#1}}}
\newcommand{\Refe}[1]      {Ref.~\cite{#1}}

\newcommand{\Fig}[1]       {Fig.~\ref{#1}}

\newcommand{\co}[1]        {}

\newcommand{\pytstr}       {\Pythia~8~(Monash)}
\newcommand{\pytstrb}      {\Pythia~6~(Perugia 2011)}
\ifplotpold
\renewcommand{\pytstr}     {\Pythia~6~(Perugia 2011)}
\renewcommand{\pytstrb}    {\Pythia~8~(Monash)}
\fi
\begin{document}
%\title{Apparent strangeness enhancement from multiplicity selection in pp collisions}
\title{Apparent strangeness enhancement from multiplicity selection in high energy proton-proton collisions}
%\title{Enhanced strange-particle yield from multiplicity selection in high-energy proton-proton collisions}
%\title{Suppressed strange-particle yield from multiplicity selection in high-energy proton-proton collisions}
\author{
 Constantin Loizides \\
 \small{\textit{ORNL, Oak Ridge, TN, 37830, USA}} \\
 Andreas Morsch \\ 
 \small{\textit{CERN, 1211 Geneva 23, Switzerland}} \\
}
\date{\small \today}
\maketitle
%%%%%%%%%%%%%%%%%%%%%%%%%%%%%%%%%%%%%%%%%%%%%%%%%%%%%%%%%%%%%%%%%%%%%%%%%%%%%%%%%%%%%%%%%%%
\enlargethispage{1.5cm}
\vspace{-1.cm}
\thispagestyle{empty}
\begin{abstract}
The increase of strange-particle yields relative to pions as a function of the event charged-particle multiplicity in proton--proton~(\pp) collisions at the LHC is argued to indicate the relevance of the final system produced in the collision.
The rise with multiplicity~(referred to as strangeness enhancement) is usually described by microscopic or hydrodynamical models as a result of the increasing density of produced partons or strings and their interactions.
Instead, in this paper, we consider the multiple partonic interaction~(\MPI) picture originally developed in the context of the \Pythia\ event generator to describe the rich structure of the underlying event in \pp\ collisions.
We find that strangeness enhancement  in \Pythia\ is hidden by a large excess of low-$\pt$ multi-strange baryons, which mainly results from the hadronization of $u$-quark, $d$-quark and gluon~($udg$) strings.
Strange baryons produced in strings formed from parton showers initiated by strange quarks ($s$-fragmentation), however, describe well the spectral shapes of $\Xi$ and $\Omega$ baryons and their multiplicity dependence. 
Since the total particle yield contains contributions from soft and hard particle production, which cannot be experimentally separated, we argue that the correct description of the  $\pt$-spectra is a minimum requirement for meaningful comparisons of multiplicity dependent yield measurements to \MPI\ based\co{ Monte Carlo} calculations.
We demonstrate that the $s$-fragmentation component describes the increase of average $\pt$ and yields with multiplicity seen in the data, including the approximate multiplicity scaling for different collision energies. 
When restricted to processes that reproduce the measured $\pt$-spectra, the \MPI\ framework exhibits a smooth evolution from strictly proportional multiplicity scaling~($K_{\rm S}^0$, $\Lambda$, where the $udg$-hadronization component dominates) to linearity ($s$-fragmentation) and on to increasingly non-linear behavior~($c$-, $b$-quark and high-$\pt$ jet fragmentation), hence providing a unified approach for particle production across all particle species in \pp\ collisions.
%In \Pythia, the multiplicity threshold for yields of $s$-fragmentation is found to originate from a bias towards more central collisions, compared to yields which also receive contributions from soft processes, and in particular for low-multiplicity peripheral collisions also from particle production correlated with the multi-strange baryon.
\end{abstract}
%%%%%%%%%%%%%%%%%%%%%%%%%%%%%%%%%%%%%%%%%%%%%%%%%%%%%%%%%%%%%%%%%%%%%%%%%%%%%%%%%%%%%%%%%%%
\ifcomment /*Abstract on arXiv can only have 1920 characters*/
The increase of strange-particle yields relative to pions versus charged-particle multiplicity in proton-proton (pp) collisions at the LHC is usually described by microscopic or hydrodynamical models to result from the increasing density of produced partons or strings and their interactions.
Instead, we consider the multiple partonic interaction (MPI) picture originally developed in the context of the PYTHIA event generator.
We find that strangeness enhancement in PYTHIA is hidden by a large excess of low-$p_{\rm T}$ multi-strange baryons, which mainly results from the hadronization of $u$-quark, $d$-quark and gluon~($udg$) strings.
Strange baryons produced in strings formed from parton showers initiated by strange quarks ($s$-fragmentation), however, describe well the spectral shapes of $\Xi$ and $\Omega$ baryons and their multiplicity dependence.
Since the total particle yield contains contributions from soft and hard particle production, which cannot be experimentally separated, we argue that the correct description of the  $p_{\rm T}$-spectra is a minimum requirement for meaningful comparisons of multiplicity dependent yield measurements to MPI based calculations.
We demonstrate that the $s$-fragmentation component describes the increase of average $p_{\rm T}$ and yields with multiplicity seen in the data, including the approximate multiplicity scaling for different collision energies.
When restricted to processes that reproduce the measured $p_{\rm T}$-spectra, the MPI framework exhibits a smooth evolution from strictly proportional multiplicity scaling~($K_{\rm S}^0$, $\Lambda$, where the $udg$-hadronization component dominates) to linearity ($s$-fragmentation) and on to increasingly non-linear behavior ($c$-, $b$-quark and high-$p_{\rm T}$ parton fragmentation), hence providing a unified approach for particle production in pp collisions. 
\fi
\newpage
%%%%%%%%%%%%%%%%%%%%%%%%%%%%%%%%%%%%%%%%%%%%%%%%%%%%%%%%%%%%%%%%%%%%%%%%%%%%%%%%%%%%%%%%%%%
One of the pillars of the high-energy ultra-relativistic heavy-ion physics program is to study the evolution and onset of potential quark-gluon plasma (QGP) phenomena in smaller collision systems like \pp\ and \pA\ collisions at RHIC and LHC collision energies~\cite{Nagle:2018nvi}.
In high-energy \pp\ collisions at the LHC, the integrated yields of strange and multi-strange particles relative to pions were reported to increase significantly with the event\co{ charged-particle} multiplicity, reaching values similar to those in \pPb\ and \PbPb\ collisions~\cite{ALICE:2017jyt}. 
Strangeness enhancement was originally proposed as a signature of QGP formation in nuclear collisions~\cite{Koch:1986ud}.
%The yields in central \PbPb\ collisions can be described by thermal models.
In the picture of statistical hadronization, the increase with multiplicity in smaller systems, which is rather independent of collision species and energy~\cite{Acharya:2018orn,Acharya:2019kyh}, is a result of decreasing canonical suppression~\cite{Vislavicius:2016rwi} towards the grand-canonical ensemble realized in more central \PbPb\ collisions~\cite{ABELEV:2013zaa,Sharma:2018jqf}.
The strangeness data together with other measurements in high-multiplicity \pp\ collisions, in particular the ridge~\cite{Khachatryan:2010gv,Khachatryan:2016txc}, resemble collective properties as found in heavy-ion collisions and provoke speculation about the formation of a QGP in these collisions~\cite{Loizides:2016tew,Nagle:2018nvi}. 
Indeed, models incorporating final-state interactions originally developed for heavy-ion collisions, are also able to describe the \pp\ data~\cite{Habich:2015rtj,Weller:2017tsr,Srivastava:2018dye,Shao:2020sqr}.
Refined models based on the ``core/corona'' approach, where the core of the collisions is assumed to ``hydrodynamize'', while the corona is treated as a superposition of independent nucleon--nucleon collision, achieve a good description of data across all systems with a rather universal approach~\cite{Pierog:2013ria,Kanakubo:2019ogh}.

\ifcomment
An increasing number of details had to be implemented and tuned in modern Monte Carlo~(MC) generators to provide an accurate description of the wealth of minimum bias and underlying event data in \pp\ collisions~\cite{Sjostrand:2017cdm}.
%Historically, measurements of the properties of high-energy \pp\ collisions, like the charged particle multiplicity distributions, forward-backward multiplicity correlations, or the number of particles produced outside the jet-cone in events containing a jet~(jet pedestal) have influenced the development of the MPI picture. %~\cite{Sjostrand:1987su}.
%Multiple parton interactions also provide a natural interpretation of the fact that at high \s\ the leading order cross-section for $2 \rightarrow 2$ parton scatterings with a minimum transverse momentum scale $p_{\rm T,min}$ exceeds the total non-diffractive~(ND) \pp\ cross section in a \pT\ range where perturbative QCD is applicable.
%In the simplest independent interaction approach the mean number of scatterings per event $n_{\rm hard}$ is equal to the ratio of the hard and the total non-diffractive cross-section, $n_{\rm hard} = \sigma_{\rm hard}(\pT > p_{\rm T,min}) /   \sigma_{\rm ND}$ and event-by-event variations follow a Poissonian distribution. 
In particular, the accurate description of the dispersion of multiplicity distributions and the jet pedestal effect need an impact-parameter dependent rate of parton--parton scatterings~\cite{Sjostrand:1986ep}.
In recent years, effects like the fore-mentioned rise of strange-particle yield with multiplicity, as well as the azimuthal correlations over long range in pseudo-rapidity~(ridge) as well as the mass-dependent hardening of \pT\  distributions reinforced interest in \MPI\ models, either as a baseline or as a possibility to understand these effects as a consequence of coherence effects between multiple scatterings~(e.g.\ color reconnection~\cite{Corke:2010yf} and string fusion~\cite{Bierlich:2014xba}).
In these extensions of the \MPI\ model, strangeness enhancement is interpreted as a multi-string effect i.e.\ due to overlapping strings with increasing multiplicity driving the increase of strangeness production with multiplicity that however quickly saturates at higher multiplicity~\cite{Bierlich:2014xba,Bierlich:2015rha,Fischer:2016zzs,Pirner:2018ccp}.

In case a hot and dense medium is created in \pp\ collisions this could also lead to the modification of the production of so called hard probes (processes involving a large transverse momentum transfer in partonic scatterings) in high multiplicity pp collisions.
The measurement of such effects, which should be quantified like in \PbPb\ collisions by a centrality dependent modification factor is hampered by the fact that suitable normalisation factor, as the number of scatterings, can at present not be extracted from data. 
Moreover, multiplicity, especially when measured in the same region as the hard probe, can be modified by the presence of the hard probe itself (``auto-correlations'') or can itself be modified by density effects~(for example ``color reconnection'').
Nevertheless several measurements of so called self-normalized yields of hard probes (yields normalised by their mean value) as a function of multiplicity have been performed and compared to models.
\fi

In this paper, we focus on an alternative description of the \pp\ data~\cite{ALICE:2017jyt,Acharya:2019kyh} based on the \MPI\ model, as implemented in the \Pythia\ event generator~\cite{Sjostrand:2006za}.
Unlike other works~\cite{Fischer:2016zzs,Pirner:2018ccp,Nayak:2018xip}, we rely on the original string fragmentation, but include coherence effects between multiple scatterings~(e.g.\ color reconnection~\cite{Corke:2010yf,Bierlich:2015rha}).
We discuss the qualitative expectations of the model for multiplicity dependent measurements as well as the possible effects that can explain strangeness enhancement in pp collisions. 
An important ingredient for the \MPI\ model is the \pp\ impact parameter~($b$) dependence of the number of \MPI~($\Nmpi$), which was originally introduced for an accurate description of the dispersion of multiplicity distributions and the pedestal effect in underlying event measurements~\cite{Sjostrand:1986ep}. More recently it has been shown that the pp impact parameter dependence is also important for understanding the centrality dependence of hard processes in \pPb~\cite{Morsch:2013cfq,ALICE:2014xsp} and peripheral \PbPb\ collisions~\cite{Loizides:2017sqq}.
In both cases a detailed study of possible nucleon--nucleon impact parameter biases was triggered by unexpected deviations from the number of binary collisions ($\Ncoll$) scaling. In pp collisions, an equivalent scaling factor would be $\Nmpi$, however at present it cannot be extracted from data. Hence, in the absence of a standard scaling expectation, one has to be particularly careful with the interpretation of multiplicity-dependent measurements in pp collisions.

Assuming that a \pp\ event can be described by a superposition of independent parton--parton scatterings, it is reasonable to expect that particle multiplicity is approximately proportional to \Nmpi . 
For central collisions ($b = 0$), \Nmpi\ reaches about 3.5 times the value of \Nmpi\ averaged over all impact parameters.
Higher values are only accessed by rather improbable statistical (e.g.\ Poissonian) fluctuations. 
Hence, one can expect a roughly linear dependence between multiplicity and \Nmpi\ until 3.5 times the mean multiplicity, above which multiplicity fluctuations and not impact parameter variations dominate particle production. 
In such a model, both the yields of soft particles and those related to hard scatterings would be proportional to the event multiplicity over a wide range of multiplicities. 
However, one can expect important differences at low and high multiplicities for the following reasons:
\begin{enumerate}
\item Particles originating from hard scatterings have associated particle production contributing to the event multiplicity both close to the scattered particle in azimuth~($\varphi$) and pseudorapidity~($\eta$) (near-side) and back-to-back ($\varphi \approx \pi$) over a wide $\eta$-range. Particles are also produced from the fragmentation of strings spanned between partons and the beam remnants, also over a wide $\eta$-range.
The near-side contribution can be experimentally excluded by measuring the multiplicity in an $\eta$-region well separated from the region in which the signal particle yield is measured~\cite{Weber:2018ddv}.

\item Low-multiplicity events have a large contribution from soft collisions in which no hard scatterings occur and also from the soft underlying event.
\end{enumerate}

\begin{figure}[t!]
\begin{center}
\includegraphics[width=0.48\linewidth]{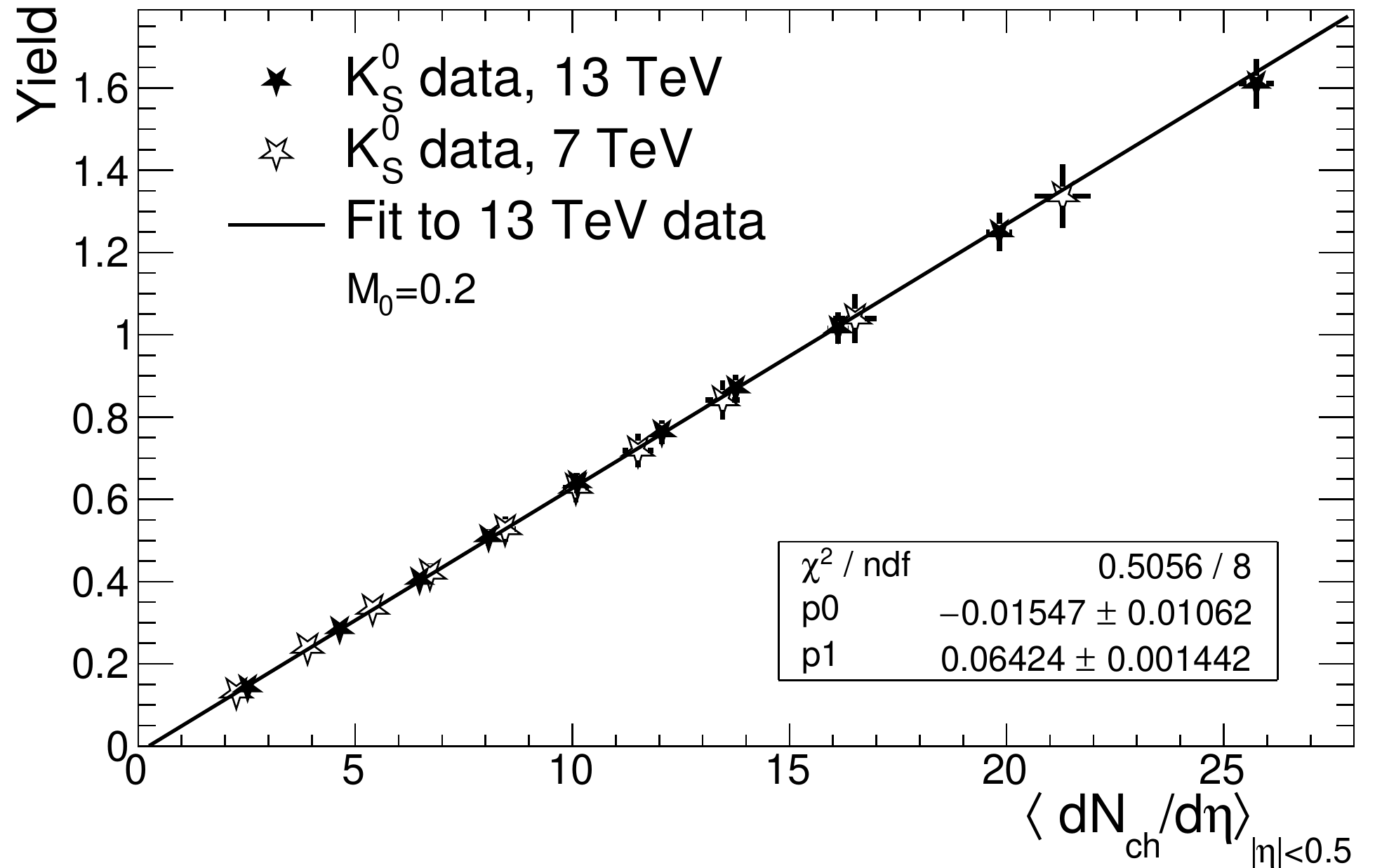}
\hspace{0.1cm}
\includegraphics[width=0.48\linewidth]{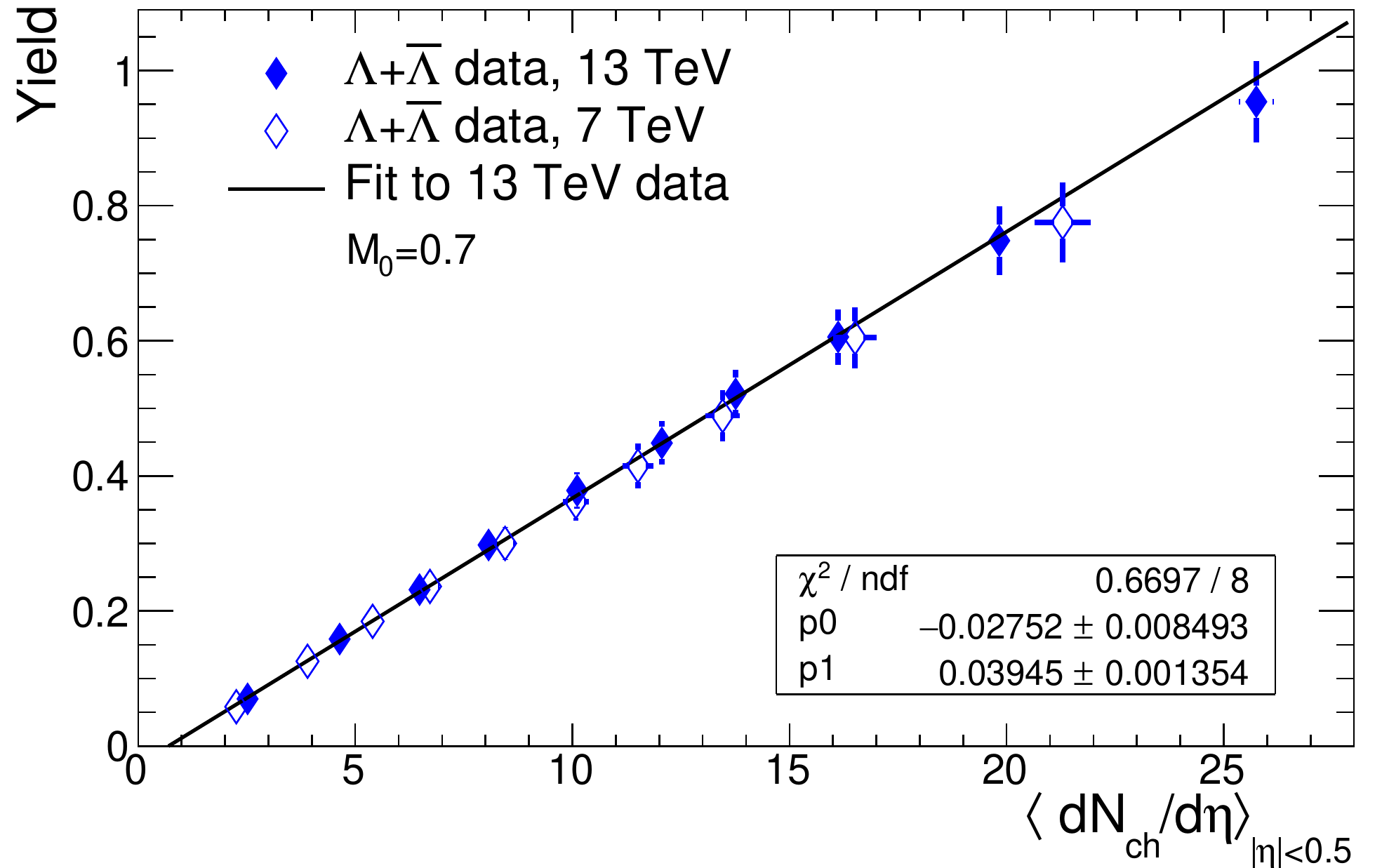} 
\includegraphics[width=0.48\linewidth]{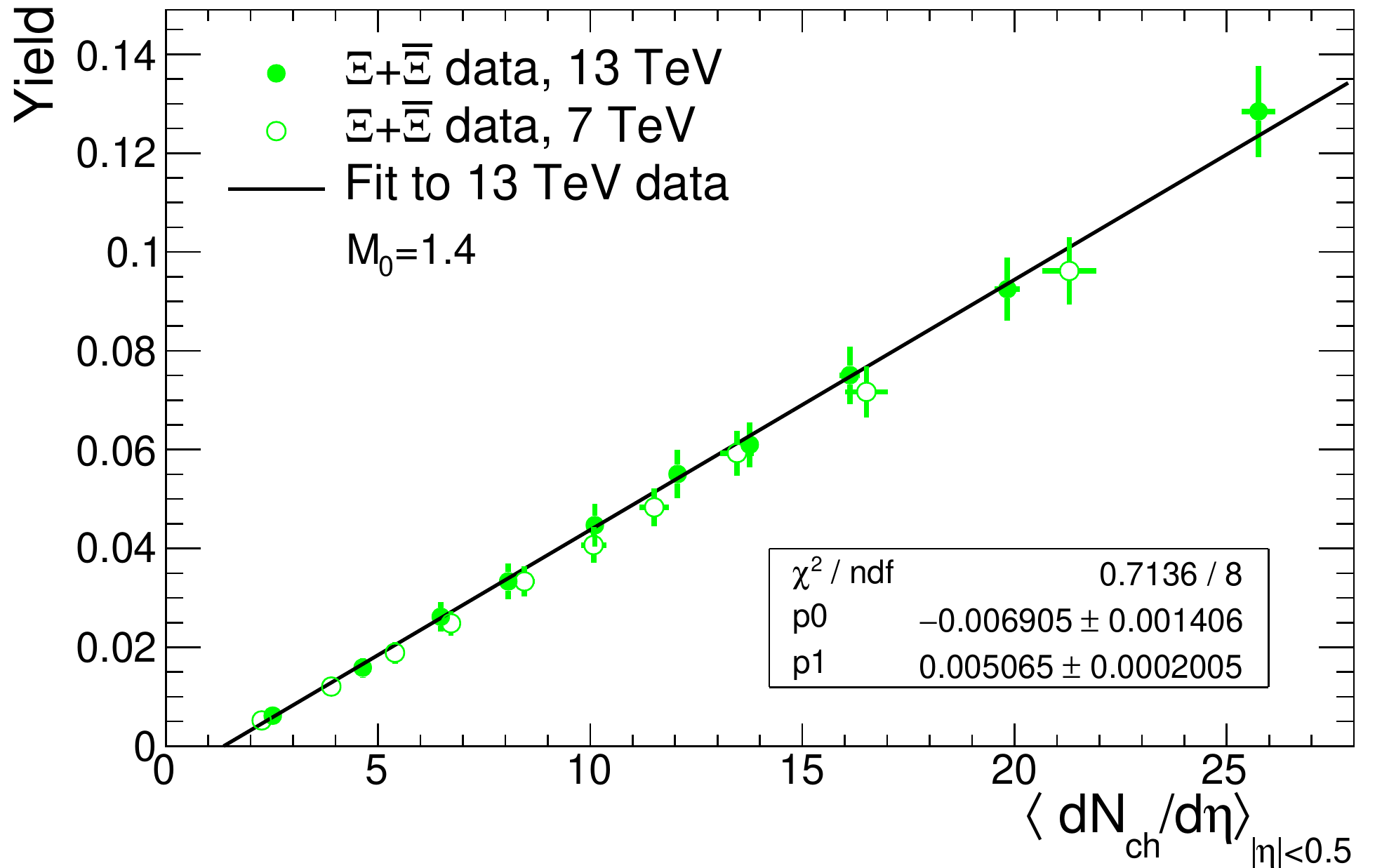} 
\hspace{0.1cm}
\includegraphics[width=0.48\linewidth]{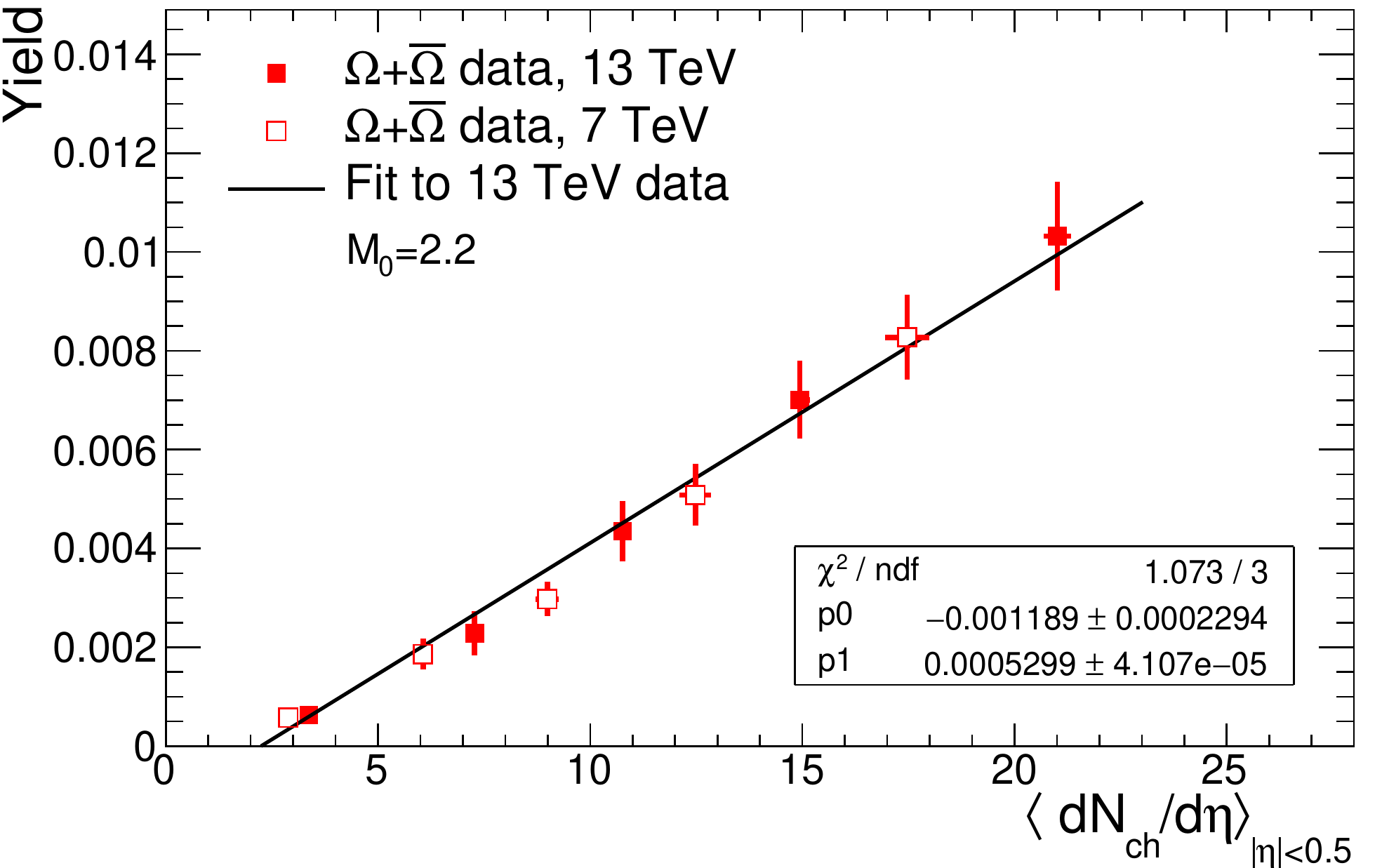}   
\caption{Measured yields of strange particles~($K_{\rm S}^0$, $\Lambda$, $\Xi$, $\Omega$) versus charged-particle multiplicity at midrapidity~($|\eta|<0.5$) in \pp\ collisions at $\s=7$ and 13 TeV~\cite{ALICE:2017jyt,Acharya:2019kyh}.
A straight-line fit to the 13 TeV data is shown in each case.
} %macro from ~/papers/alipap/enhancement/macros/data/pd13.c
\label{fig:strangedata}
\end{center}
\end{figure}
\begin{figure}[t!]
\begin{center}
\includegraphics[width=0.48\linewidth]{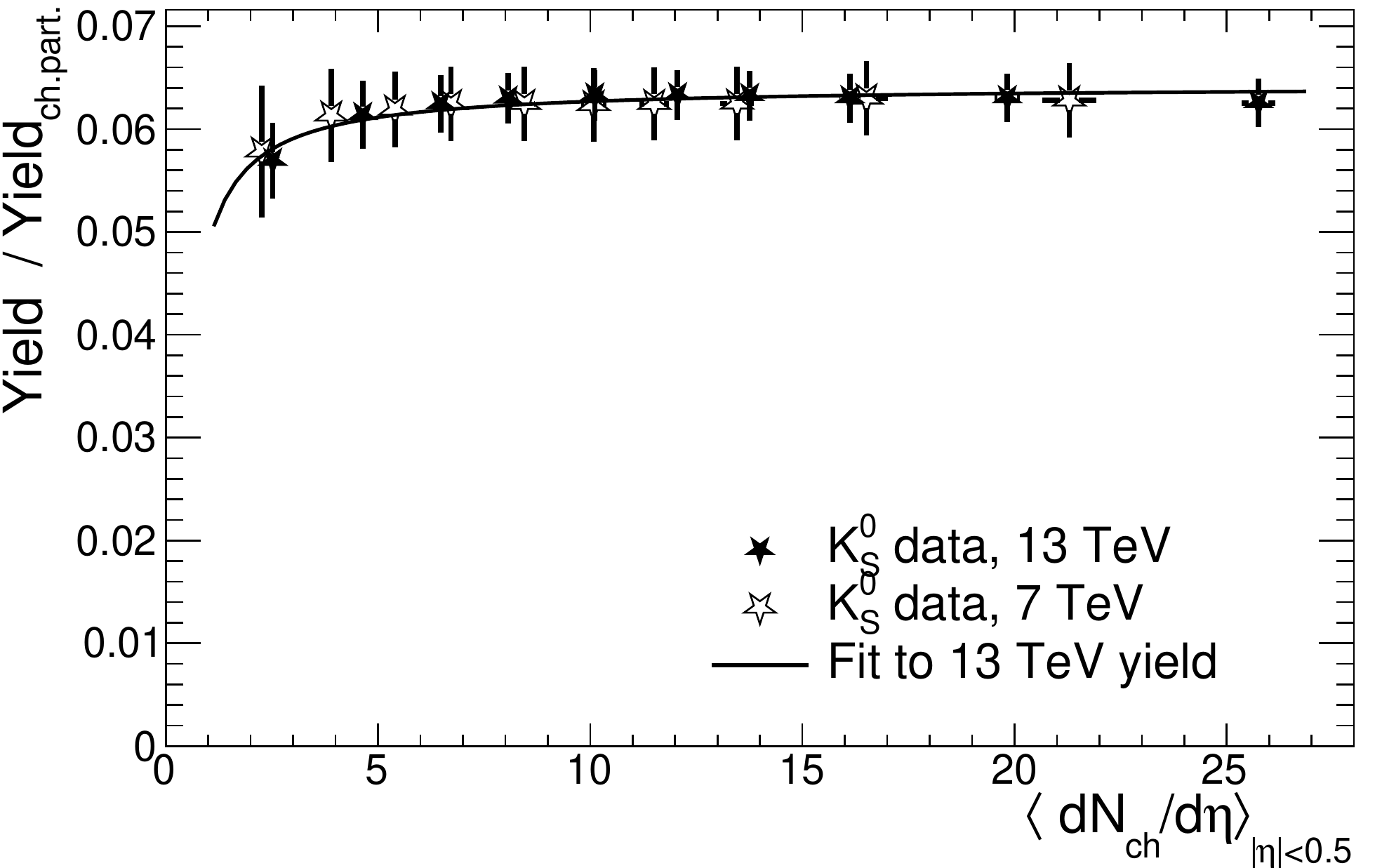}
\hspace{0.1cm}
\includegraphics[width=0.48\linewidth]{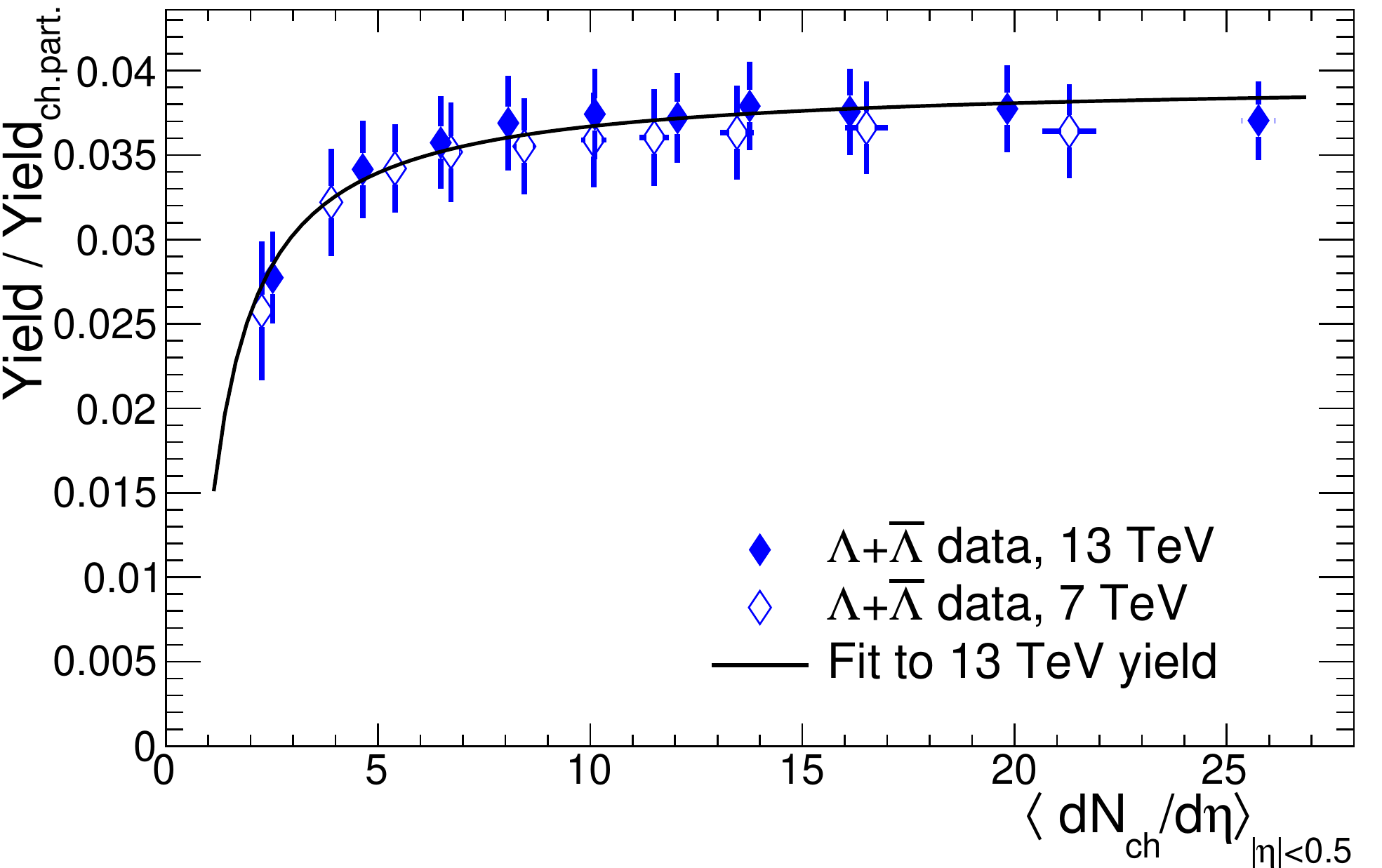} 
\includegraphics[width=0.48\linewidth]{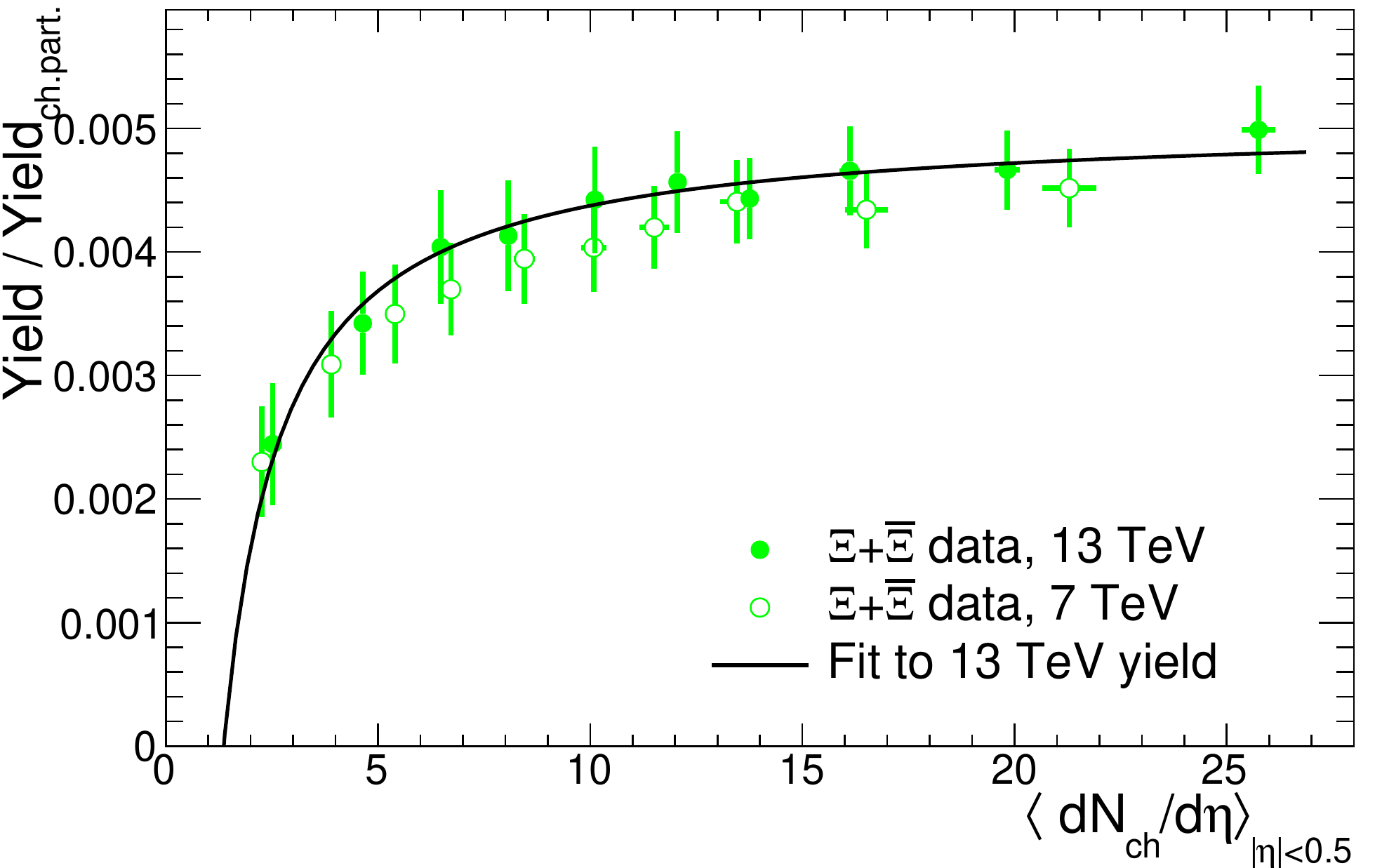} 
\hspace{0.1cm}
\includegraphics[width=0.48\linewidth]{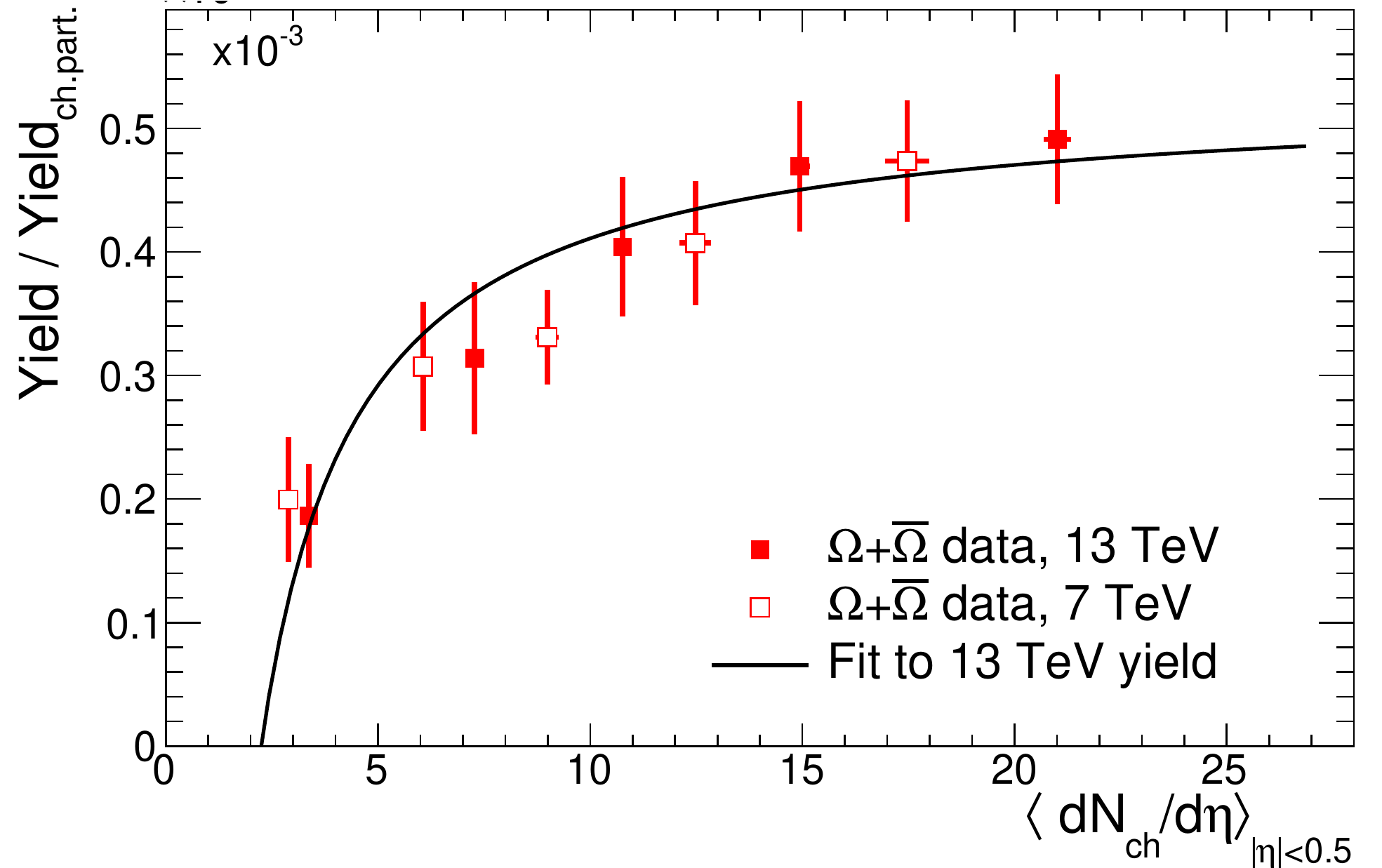}   
\caption{Same data and fits as in \Fig{fig:strangedata}, except that both data and fits were normalized normalized by the charged-particle multiplicity at midrapidity~($|\eta|<0.5$)~(i.e., normalized by the $x$-axis values).
} %macro from ~/papers/alipap/enhancement/macros/data/pd13.c
\label{fig:strangedatanorm}
\end{center}
\end{figure}

\ifcomment
The largest modification to the simplified assumption of proportionality between multiplicity and the yield of hard processes will be at low multiplicities, since the presence of the hard process will strongly increase the multiplicity. 
Moreover, since small multiplicities correspond to large impact-parameter \pp\ collisions soft processes are enhanced and hard processes suppressed.
A small change of the multiplicity in events with a hard process will only have small consequences for the pedestal effect. 
However, yield ratios will be strongly modified, leading to an apparent enhancement of hard processes in high multiplicity events.
Nevertheless several measurements of so called self-normalized yields of hard probes (yields normalised by their mean value) as a function of multiplicity have been performed and compared to models.

The observed enhancement in the strange particle yields versus multiplicity~($M$) shown in \Fig{fig:strangedata} can be understood to be the result of a pedestal or threshold effect:
The strange particle yields depend approximately linearly on multiplicity~($M$) but only above some minimum multiplicity~($M_0$) as demonstrated by the pol1 fits, while the pion~(and kaon) yields do not exhibit any apparent threshold~($p_0$ is close to 0).
Hence, normalizing the strange particle yields with the pion yields creates an enhancement $\propto (1-M_0/M)$, which leads to a rise with $M$ that is most apparent at low multiplicities, as shown in \Fig{fig:strangedatanorm}.
In microscopic models one can interpret the observed threshold in several ways:
i)~The yield increases with the number of overlapping strings, i.e.\ as a multi-string effect related to the string density driving the increase of strangeness production that however quickly saturates at higher multiplicity.
This is explored in works on modifications of string dynamics due to the increased density~\cite{Bierlich:2014xba,Bierlich:2015rha,Fischer:2016zzs,Pirner:2018ccp}.
ii)~There is minimum associated multiplicity necessary for multi-strange particle production and/or different scalings of soft and (semi-)hard processes, i.e.\ a single string effect that leads to suppression at low multiplicity and whose importance decreases with increasing multiplicity.

The largest modification to the simplified assumption of proportionality between multiplicity and the yield of hard processes is expected to be at low multiplicities, since its selection will favor more central than peripheral \pp\ collisions, which have a larger underlying event.
Moreover, small multiplicities correspond to large impact-parameter \pp\ collisions, where soft processes are enhanced and hard processes suppressed.
Hence, the \MPI\ model naturally predicts a larger intercept for a harder process as a consequence of the pedestal effect.
\fi

Associated particle production can lead to a stronger than linear increase of yields (auto-correlation bias) at high multiplicities (above 3.5 times the average multiplicity) where multiplicity fluctuations dominate. 
Both effects mentioned above lead to a threshold multiplicity above which a proportionality between yields and multiplicity sets in. 

Indeed, as demonstrated by the straight-line fits in \Fig{fig:strangedata}, the measured yields of strange particles have an approximate linear dependence on the charged-particle multiplicity, but only above some minimum multiplicity~($M_0$). 
The multiplicity reaches 3.5 (3.8) times the minimum bias multiplicity for $\s=7$~TeV (13~TeV), i.e.\ within the expected range of proportionality. 
The intercept increases with mass and strangeness, indicative of a higher importance of hard processes for multi-strange baryon production.
Using the intercept, the functional relation of the yields on multiplicity can be expressed as $\propto (M-M_0)$.
\footnote{
Here and in the following the charged-particle multiplicity is always expressed as average charged-particle pseudorapidity density at midrapidity, i.e.\ $\avg{\dNdeta}_{|\eta|<0.5}$, in event classes which were obtained by dividing the multiplicity distributions measured or calculated with the \Pythia\ generator in the forward ranges of $-3.7<\eta<-1.7$ and $2.8<\eta<5.1$~(ALICE V0 detector acceptance) to suppress the auto-correlation bias.
}

The reported strangeness enhancement~\cite{ALICE:2017jyt,Acharya:2019kyh} is reproduced in \Fig{fig:strangedatanorm} by normalizing the strange particle yields with the charged-particle multiplicity~(as a proxy for the charged pion yields). %, which are $\propto M$ trivially.
Due to the normalization, which is $\propto M$ trivially, the resulting functional dependence is $\propto (1-M_0/M)$, i.e.\ a hyperbolic decrease with multiplicity towards the intercept, which leads to a rise with $M$ that is most apparent at low multiplicities. 
However, the rise is still visible above the mean multiplicity ($\dNdeta = 6.9$) leading to the impression of an enhancement effect.
%In other words, the yields of strange particles~(like that of other rare particles) are strongly suppressed in collisions with multiplicities below the average.
Hence, in the original \MPI\ picture, the data can be interpreted to result from a minimum associated multiplicity necessary for multi-strange particle production and/or different scaling of soft and (semi-)hard processes, i.e.\ a single string effect leading to suppression at low multiplicity that is reduced at higher multiplicity.

\begin{figure}[t!]
\begin{center}
\ifplotpold    
\includegraphics[width=0.6\linewidth]{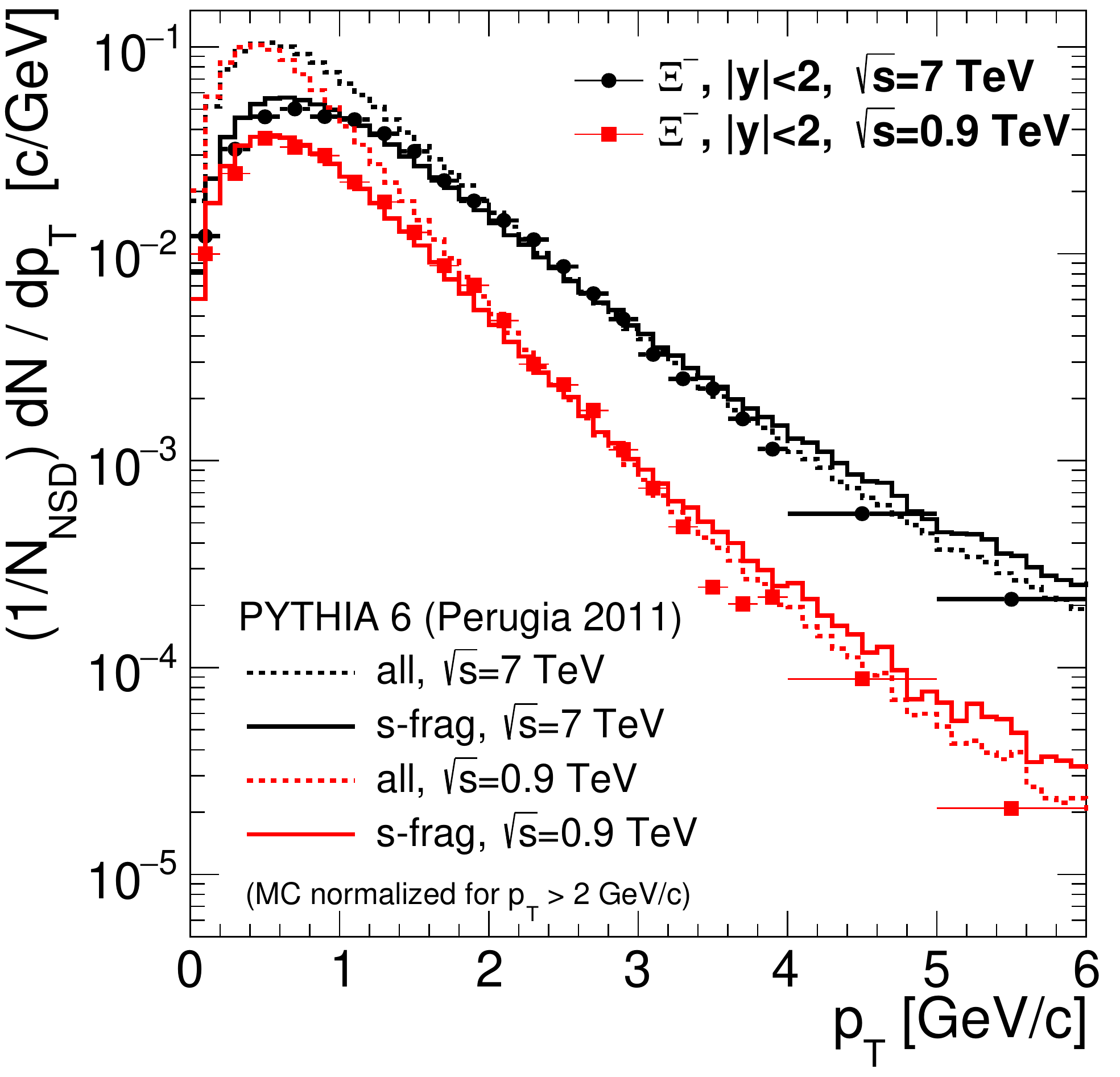}
\else
\includegraphics[width=0.6\linewidth]{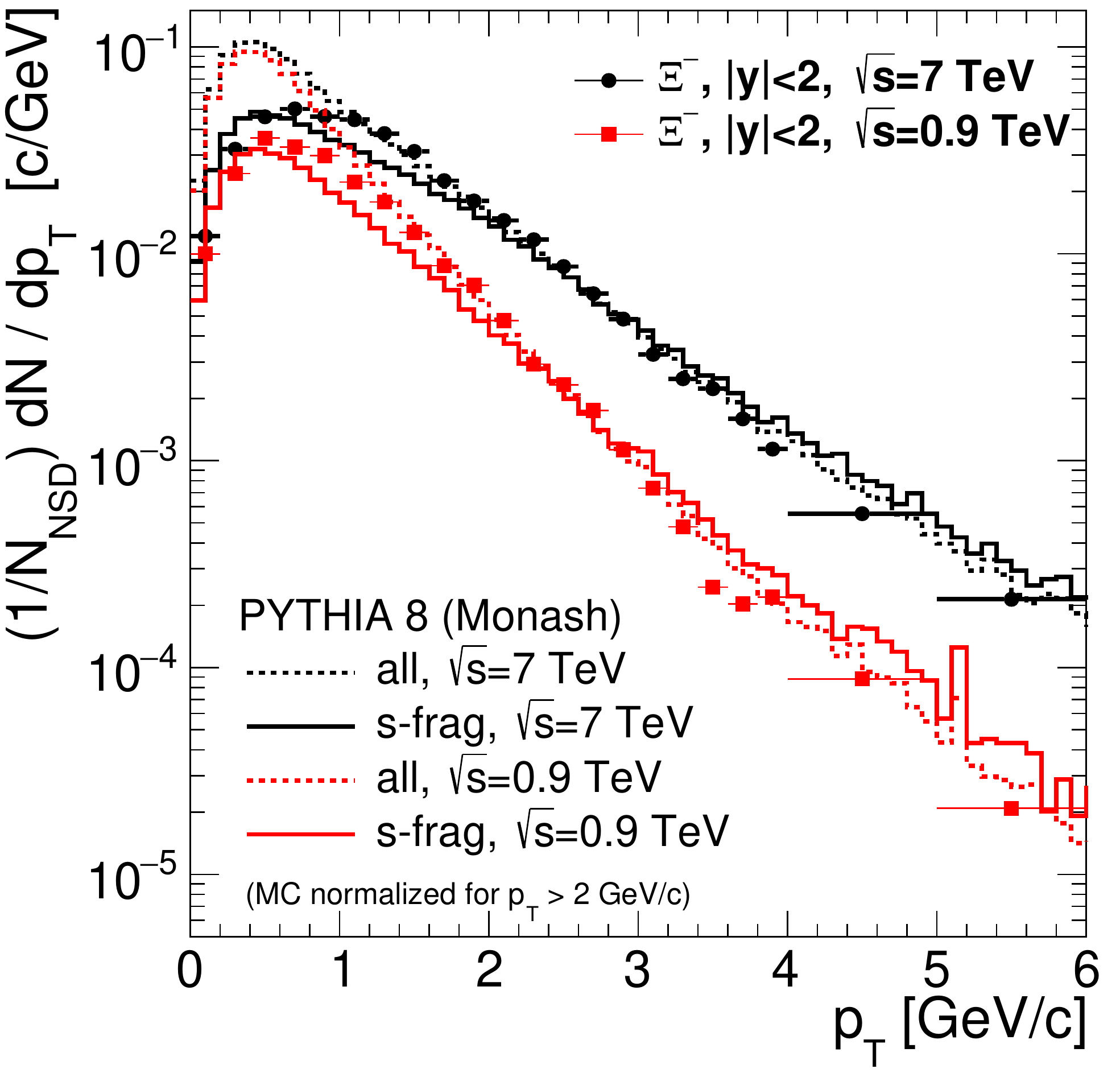}
\fi
\caption{$\Xi^{-}$ spectra in \pp\ collisions at 0.9 and 7~TeV measured by CMS~\cite{Khachatryan:2011tm} down to nearly zero $\pt$ compared with \pytstr\ calculations. The calculated spectra, which are shown for all produced $\Xi^{-}$ as well as for those produced by s-quark fragmentation alone, are normalized to the data in the region $\pt>2$~GeV/$c$.
}
\label{fig:cmscomparison}
\end{center}
\end{figure}

However, the \Pythia\ generator, in its original form, does not describe the strangeness data, and does not even exhibit a rising trend of the strange particle yield ratios with multiplicity~(see Fig.\ 2 in~\cite{ALICE:2017jyt}).
This is in particular surprising because it generally describes the multiplicity-dependence of particle production at high $\pT$ well~(see for e.g.\ \cite{ALICE:2019dfi}). 

For light-flavor particle yield measurements, it is experimentally challenging, if not impossible, to separate the yields into soft and hard production based on \pt . Hence, it is a minimum requirement for meaningful comparisons of multiplicity dependent yield measurements to MC calculations, that the corresponding \pt\ spectra are well described. 
However, $\Xi$ and $\Omega$ baryon production measured above a \pT\ of about 1~GeV/$c$ in \pp\ collisions at 7 TeV, is significantly harder~(by factor 2--4) above 2~GeV/$c$ in data compared to \Pythia\ calculations~(see Fig.\ 2 in \cite{ALICE:2012yqk}).
Hence, in \Fig{fig:cmscomparison} we compare $\Xi^{-}$ spectra in non-single diffractive \pp\ collisions at 0.9 and 7~TeV, which were measured by CMS~\cite{Khachatryan:2011tm} down to nearly zero $\pt$,  with calculations of
\ifplotpold
\Pythia~6~(Perugia 2011 tune~\cite{Skands:2010ak}).
\else
\Pythia~8~(Monash tune~\cite{Skands:2014pea}).
\fi
The calculated spectrum, which is normalized to the data for $\pt>2$~GeV/$c$, is found to well describe the measured spectral shape at high-$\pt$, but significantly overshoots the data at low $\pt$.
To understand the low-$\pt$ excess in \Pythia, we separate the produced strange particles into two categories:
These are strange particles that result from string fragmentation of a string containing:
\begin{itemize}
\item only $u$-quarks, $d$-quarks or gluons, called ``$udg$-hadronization'';
\item at least one $s$-quark produced by flavor creation or excitation in the parton shower, called ``$s$-fragmentation''.
\end{itemize}
\ifcomment
\begin{figure}[t!]
\begin{center}
  \includegraphics[width=0.37\linewidth]{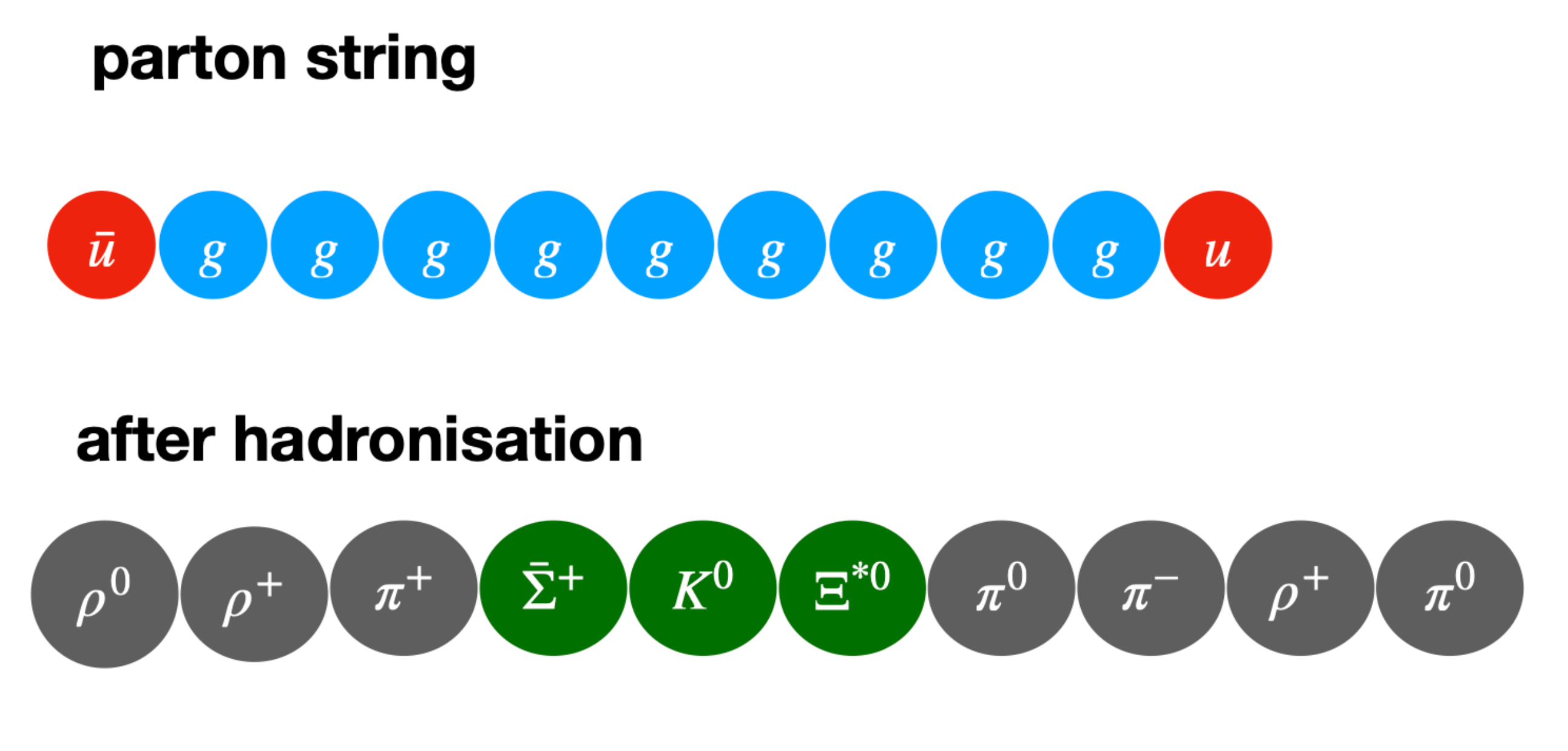}
  \hspace{0.2cm}
  \includegraphics[width=0.5\linewidth]{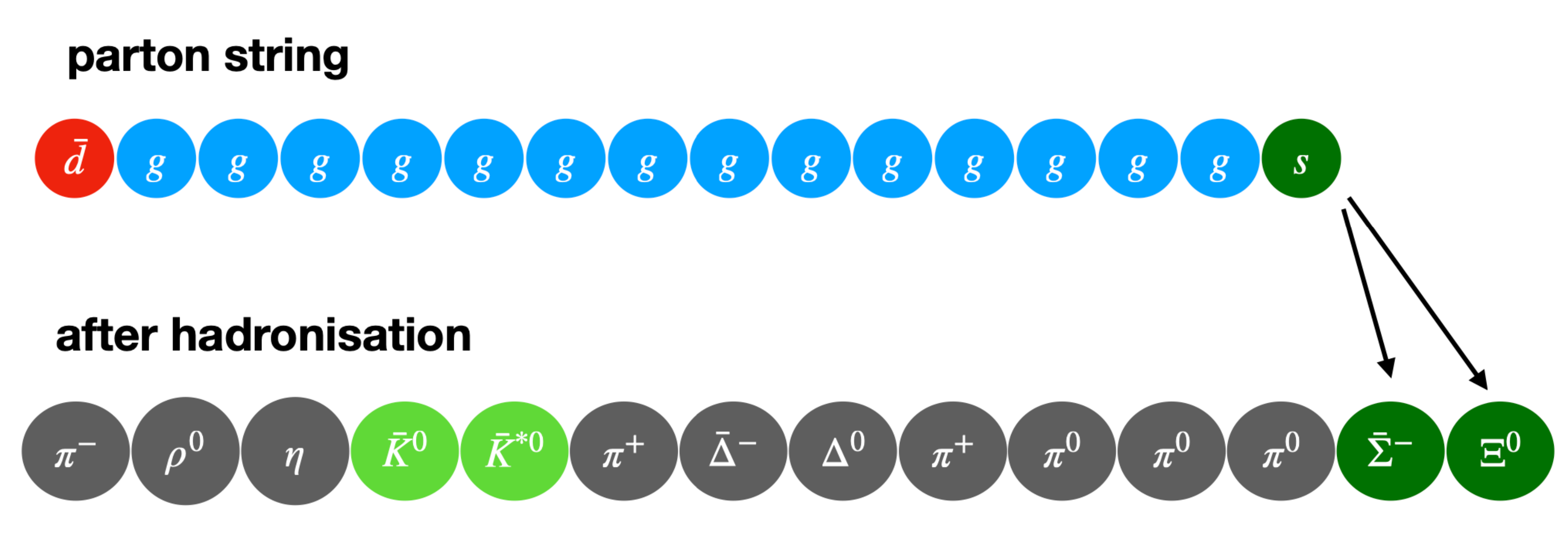}
  \caption{Illustration of $udg$-hadronization~(left) and $s$-fragmentation~(right) in \Pythia .
} 
\label{fig:}
\end{center}
\end{figure}
\fi
As can be seen in \Fig{fig:cmscomparison}, the spectra of $\Xi^{-}$ produced by $s$-quark fragmentation alone in \Pythia\ well describe the spectral shape of the data, even down to low $\pt$.
In general, particle production by $udg$-hadronization dominates the \Pythia\ calculation, in particular of course for the soft component.
In the normalization region (above 2~GeV/$c$), the yield produced by $s$-fragmentation is about half of the total calculated yield.

\begin{figure}[t!]
\begin{center}
\ifplotpold
\includegraphics[width=0.32\linewidth]{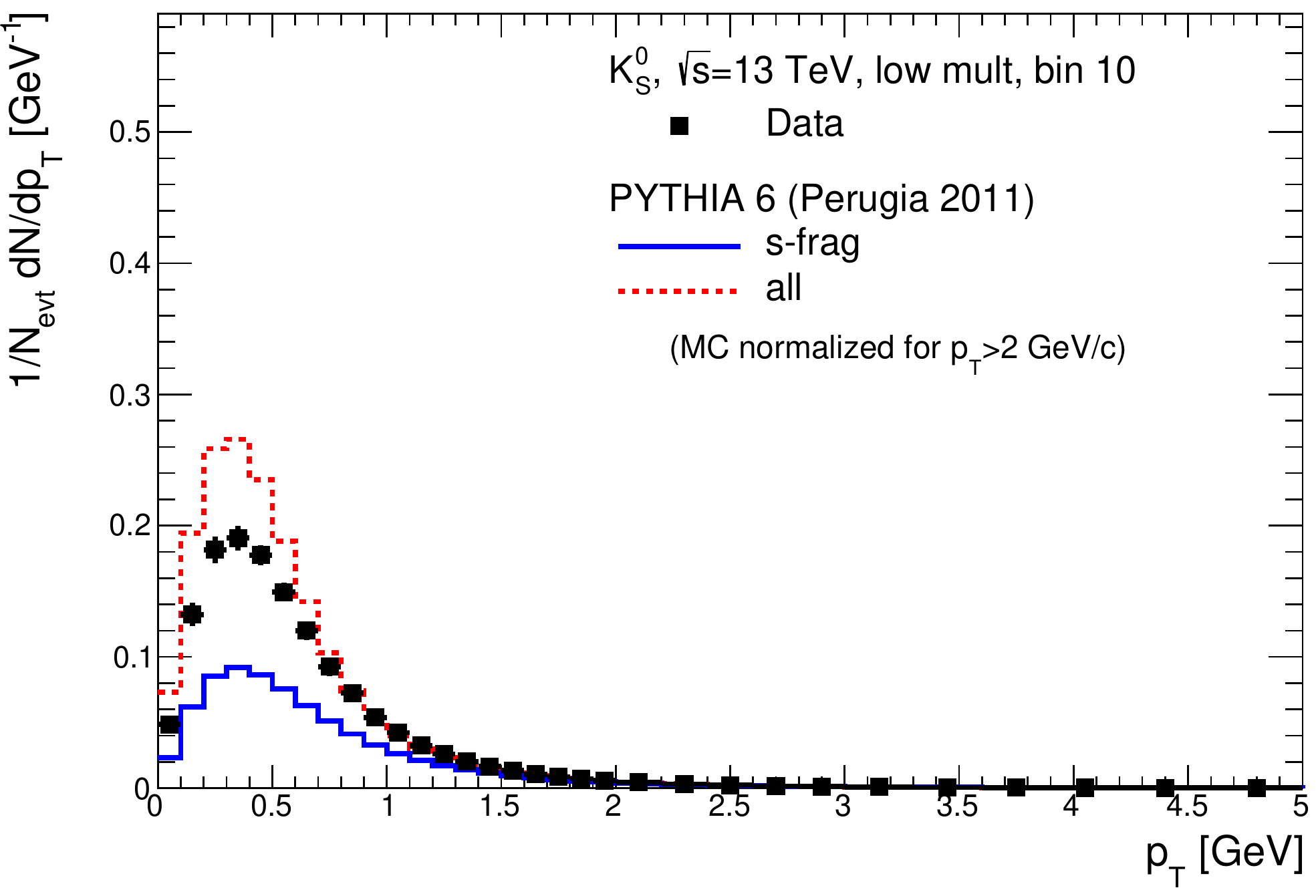}
\includegraphics[width=0.32\linewidth]{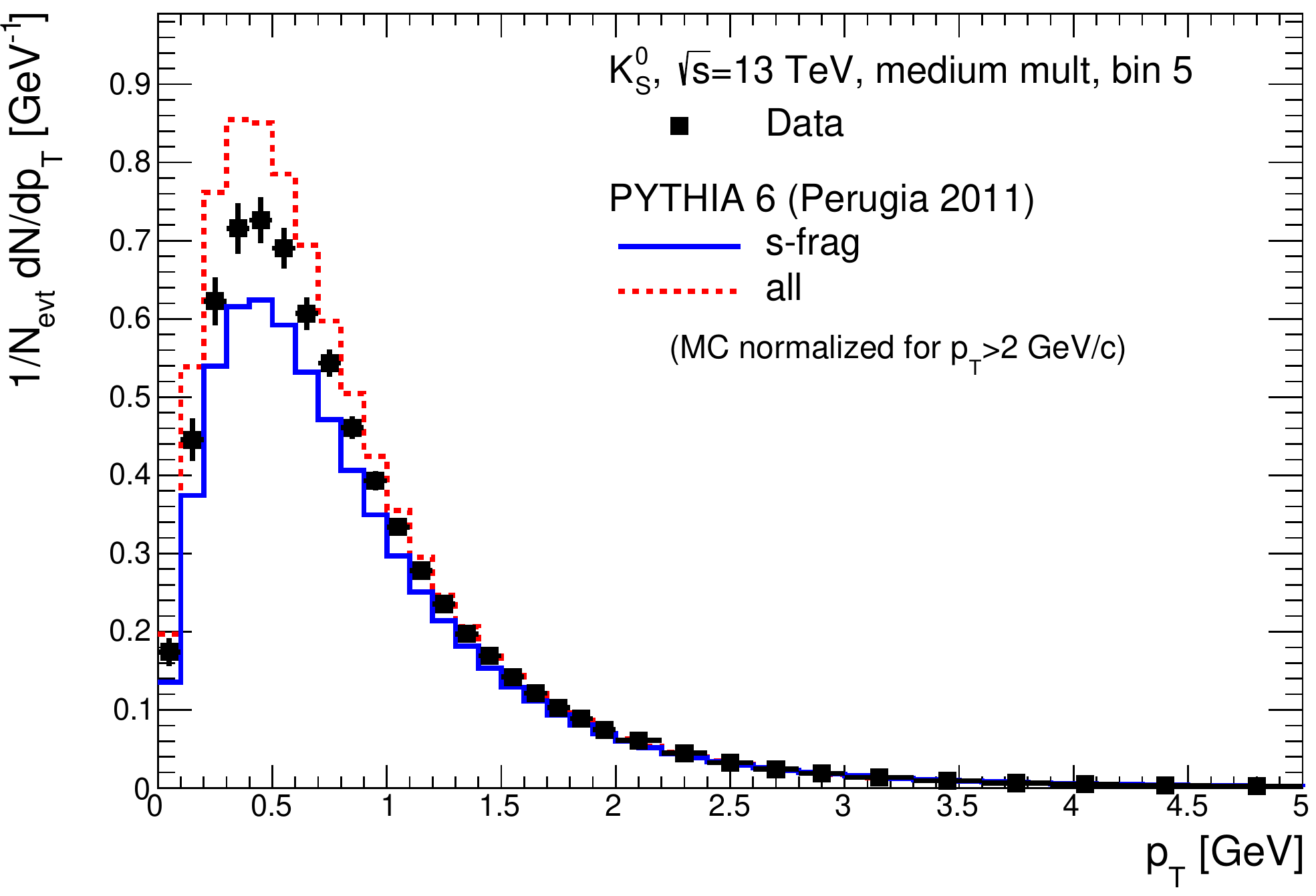}   
\includegraphics[width=0.32\linewidth]{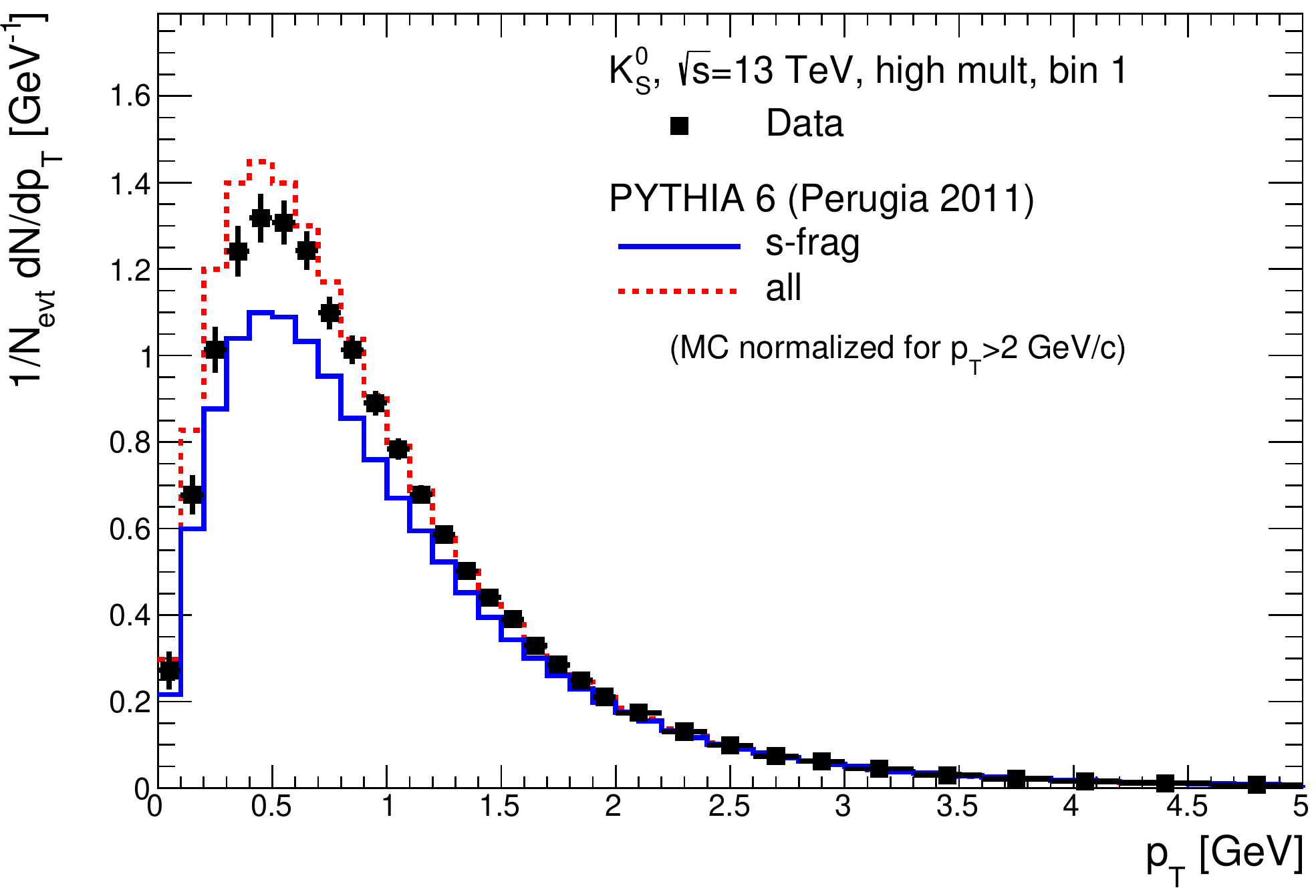}       
\includegraphics[width=0.32\linewidth]{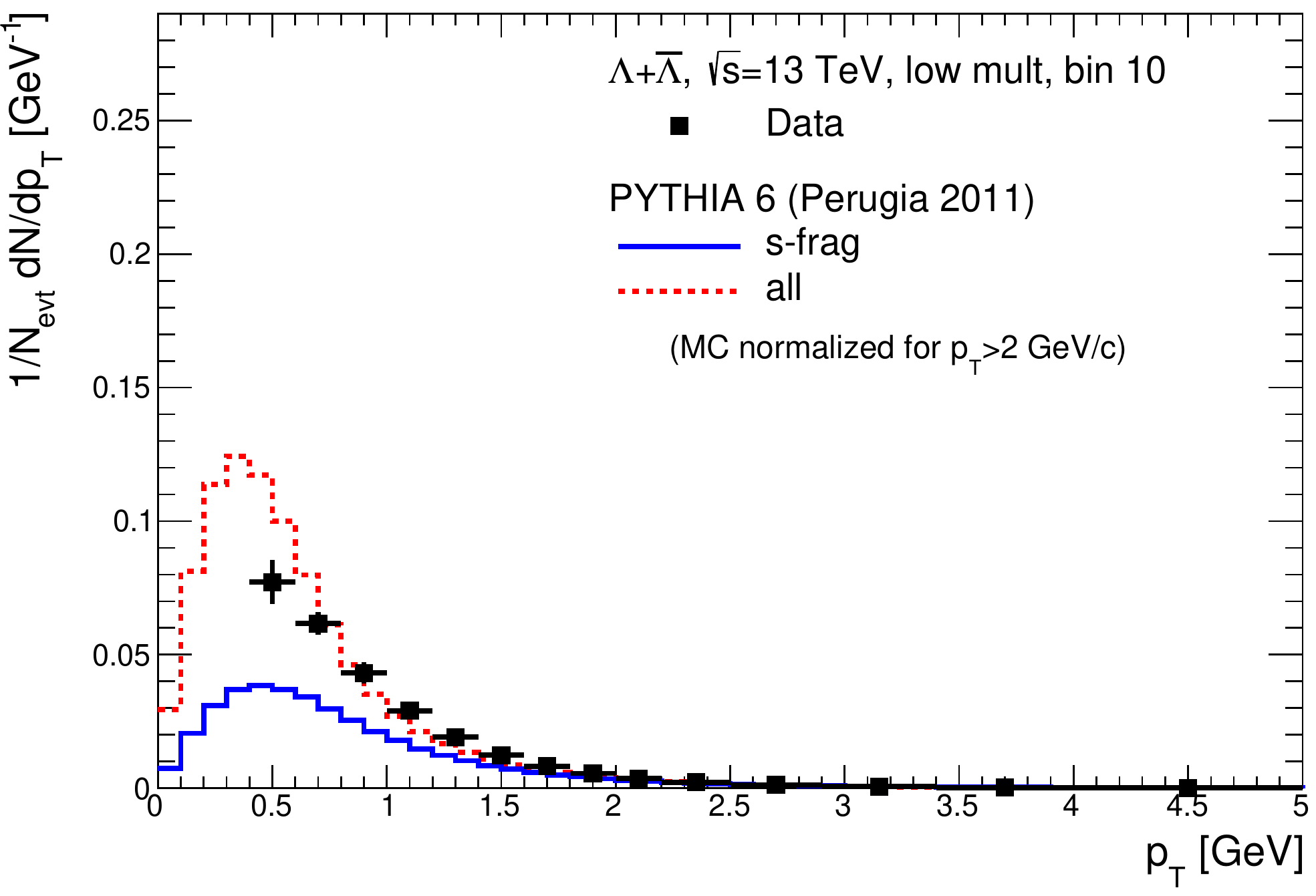}
\includegraphics[width=0.32\linewidth]{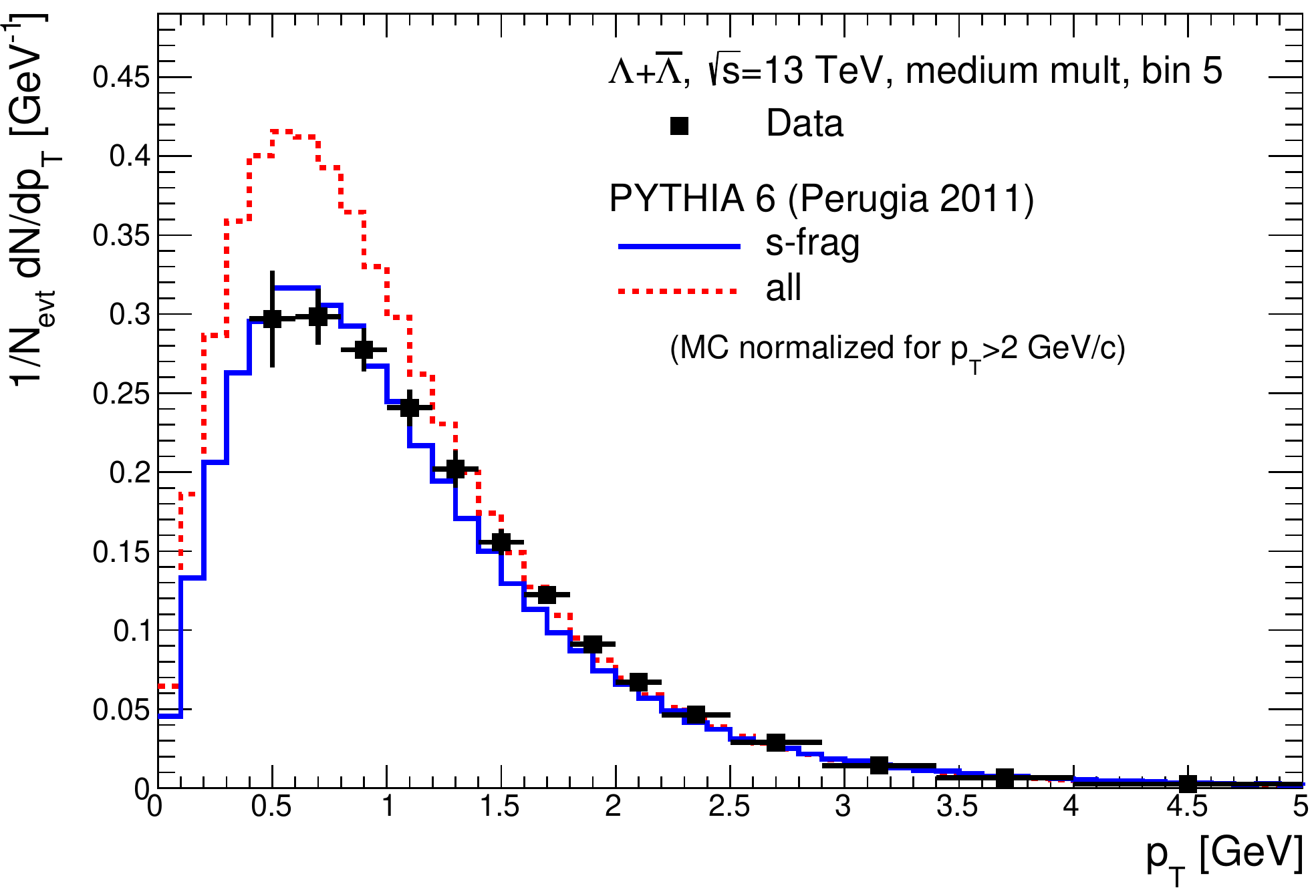}   
\includegraphics[width=0.32\linewidth]{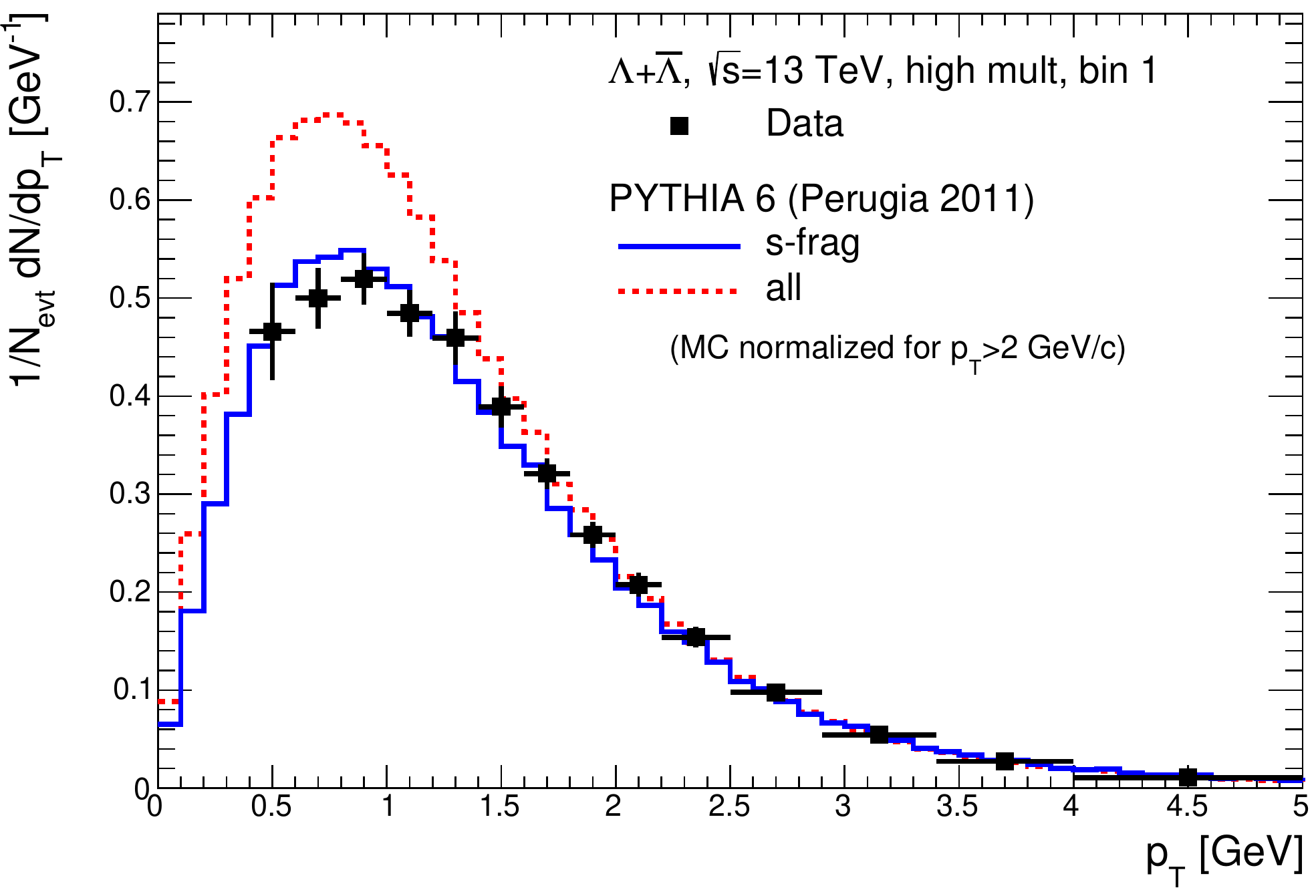}
\includegraphics[width=0.32\linewidth]{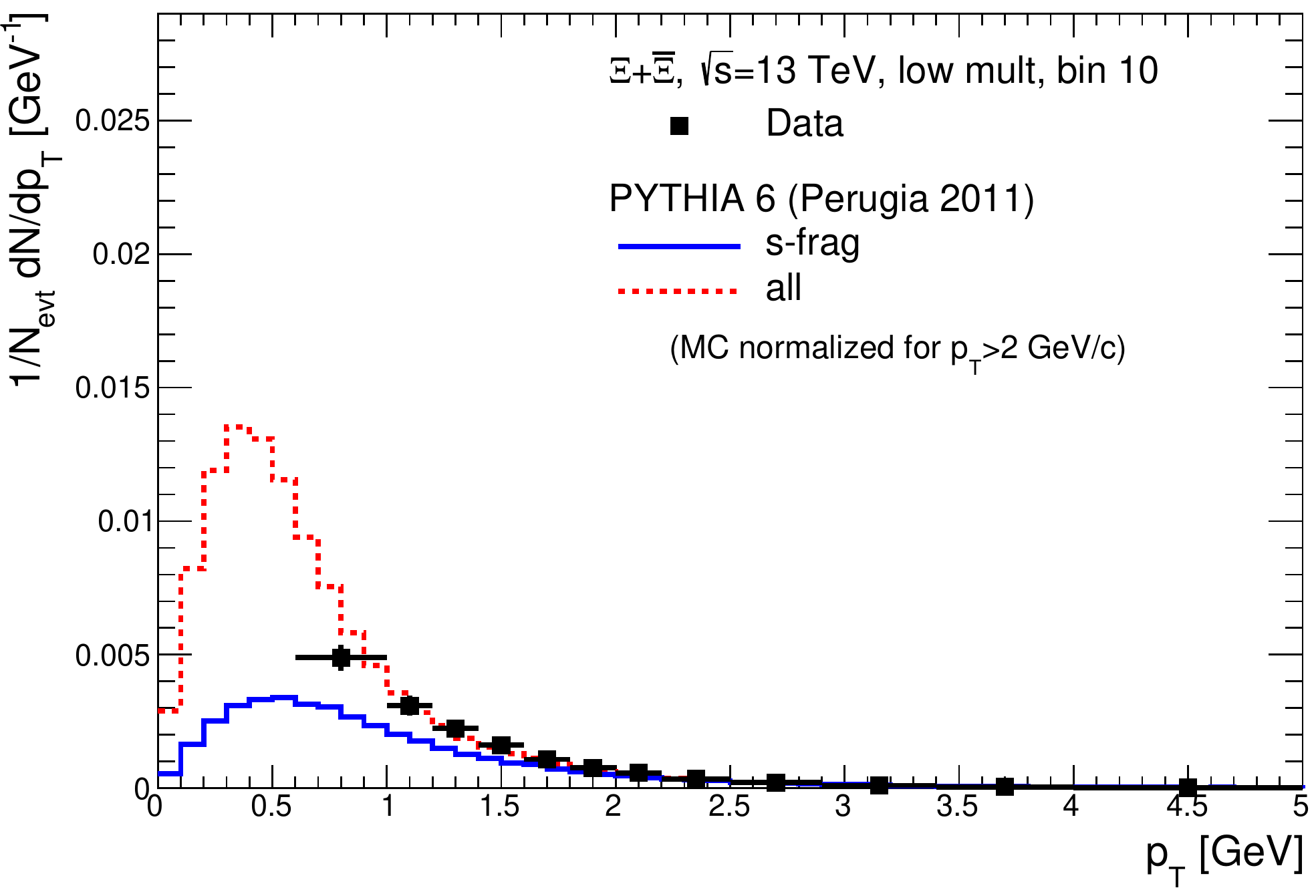}
\includegraphics[width=0.32\linewidth]{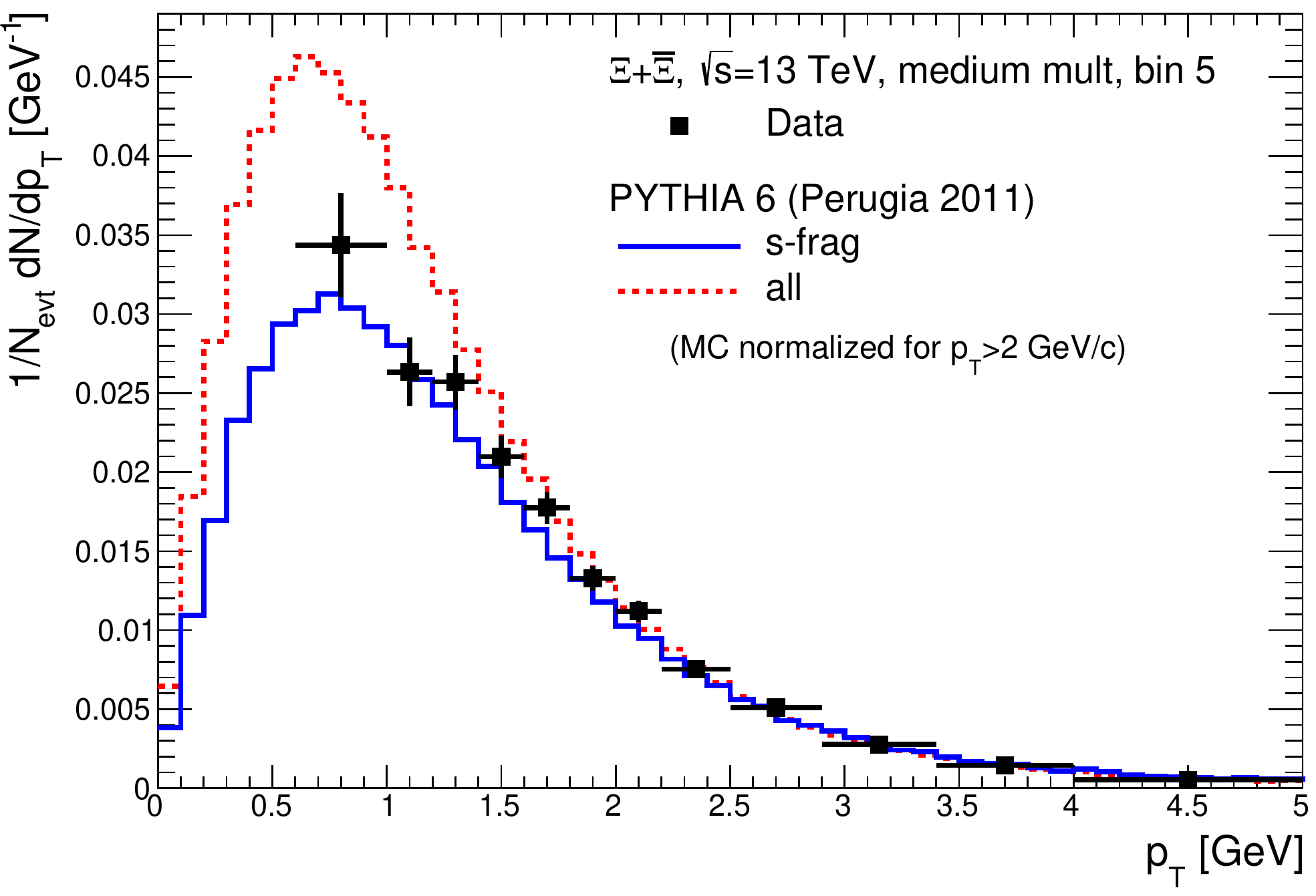}   
\includegraphics[width=0.32\linewidth]{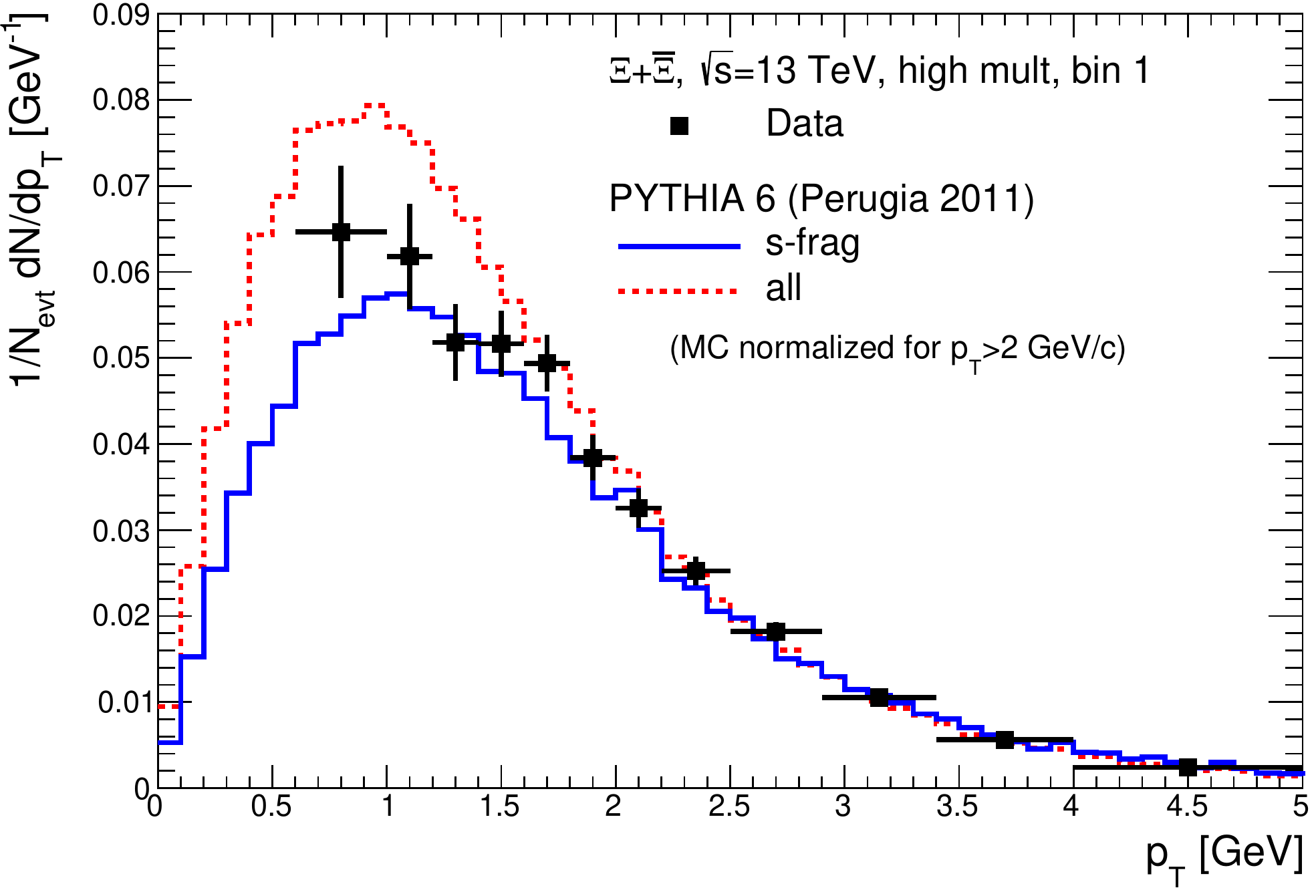}       
\includegraphics[width=0.32\linewidth]{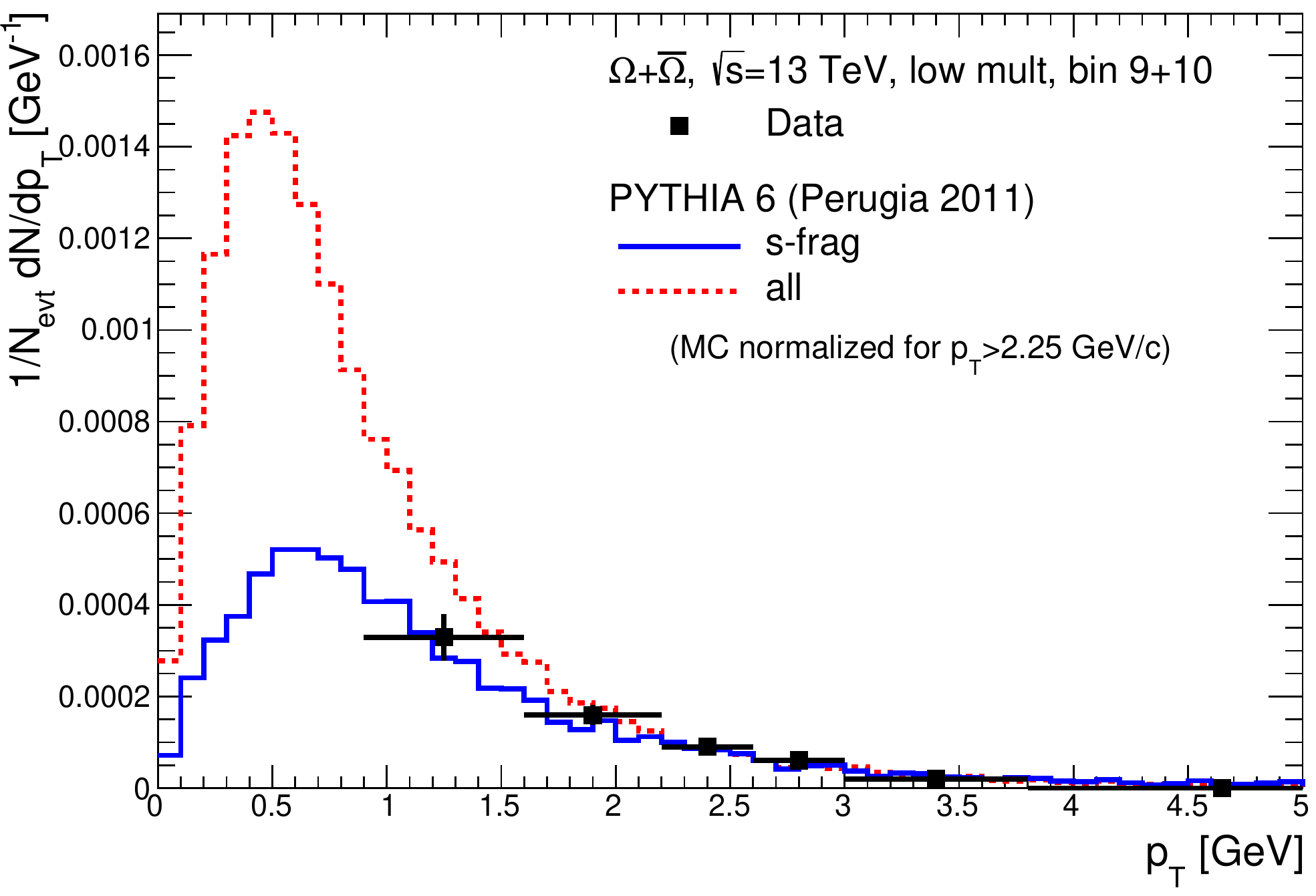}
\includegraphics[width=0.32\linewidth]{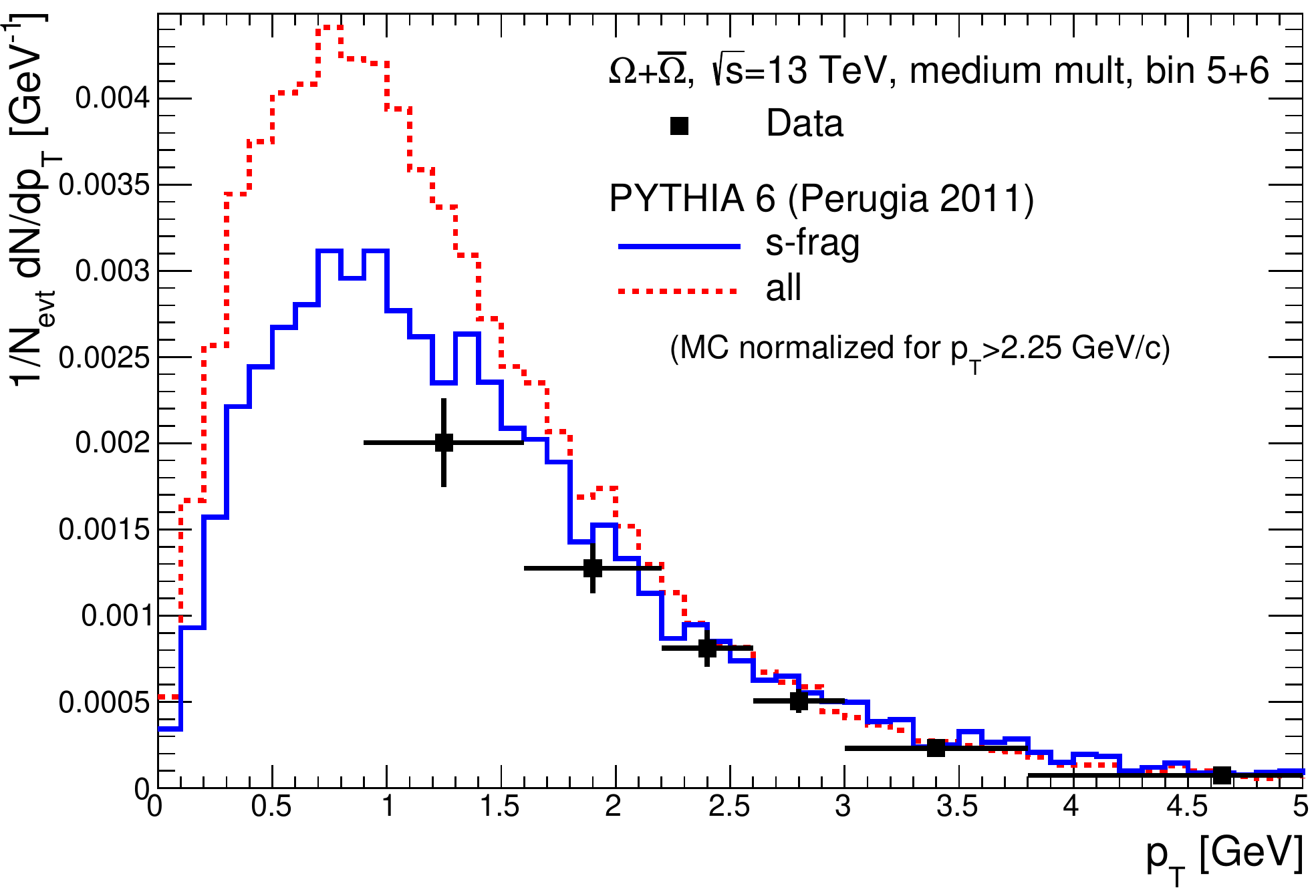}   
\includegraphics[width=0.32\linewidth]{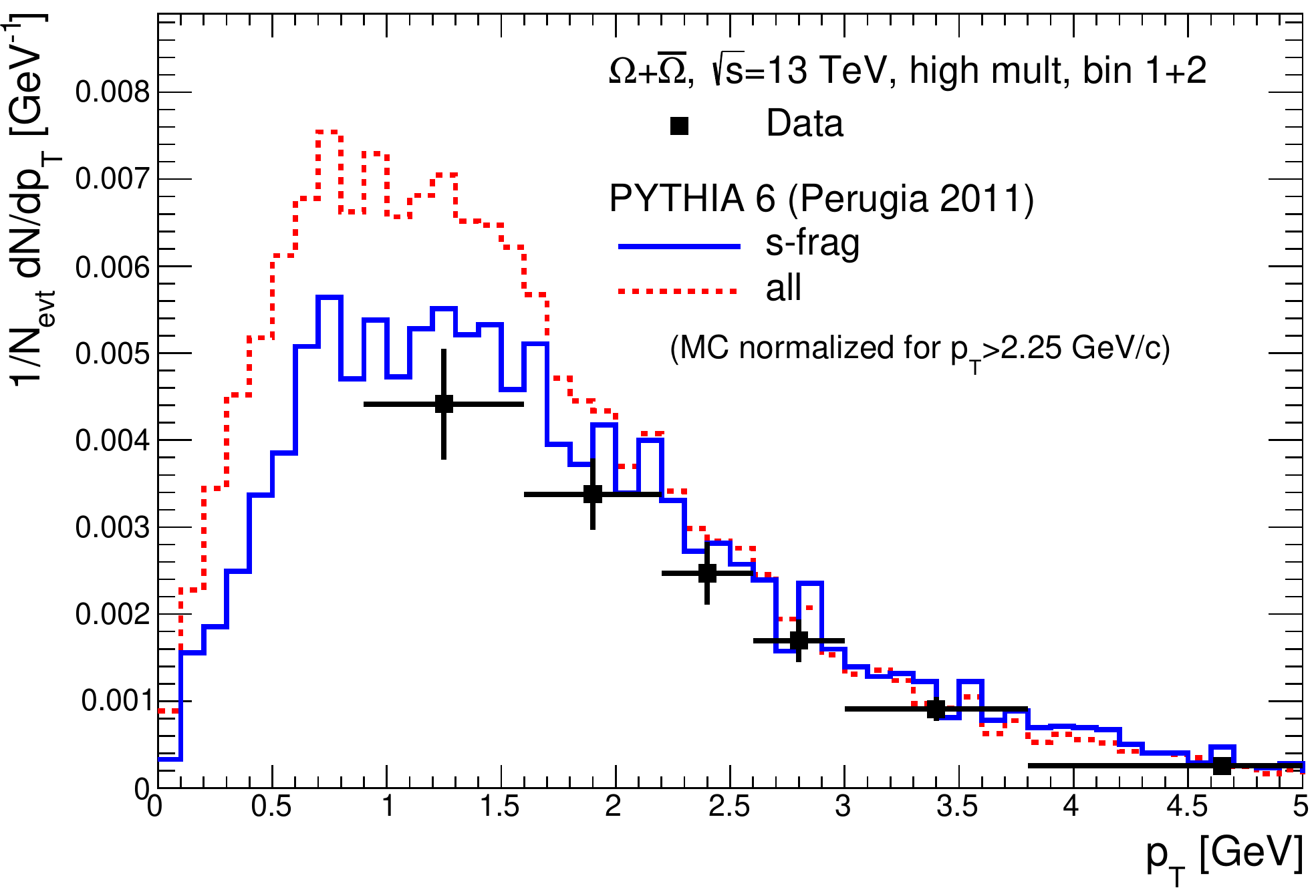}
\else
\includegraphics[width=0.32\linewidth]{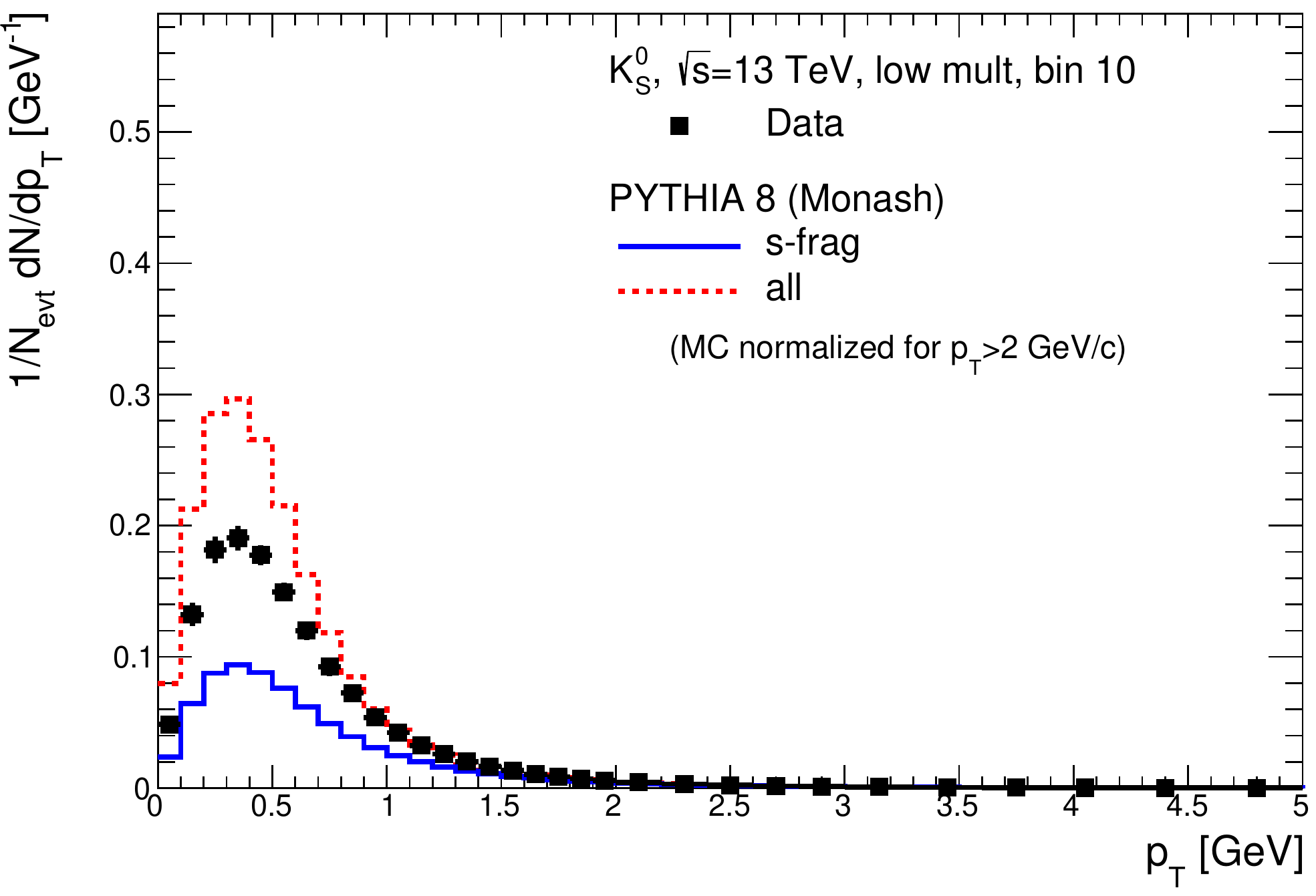}
\includegraphics[width=0.32\linewidth]{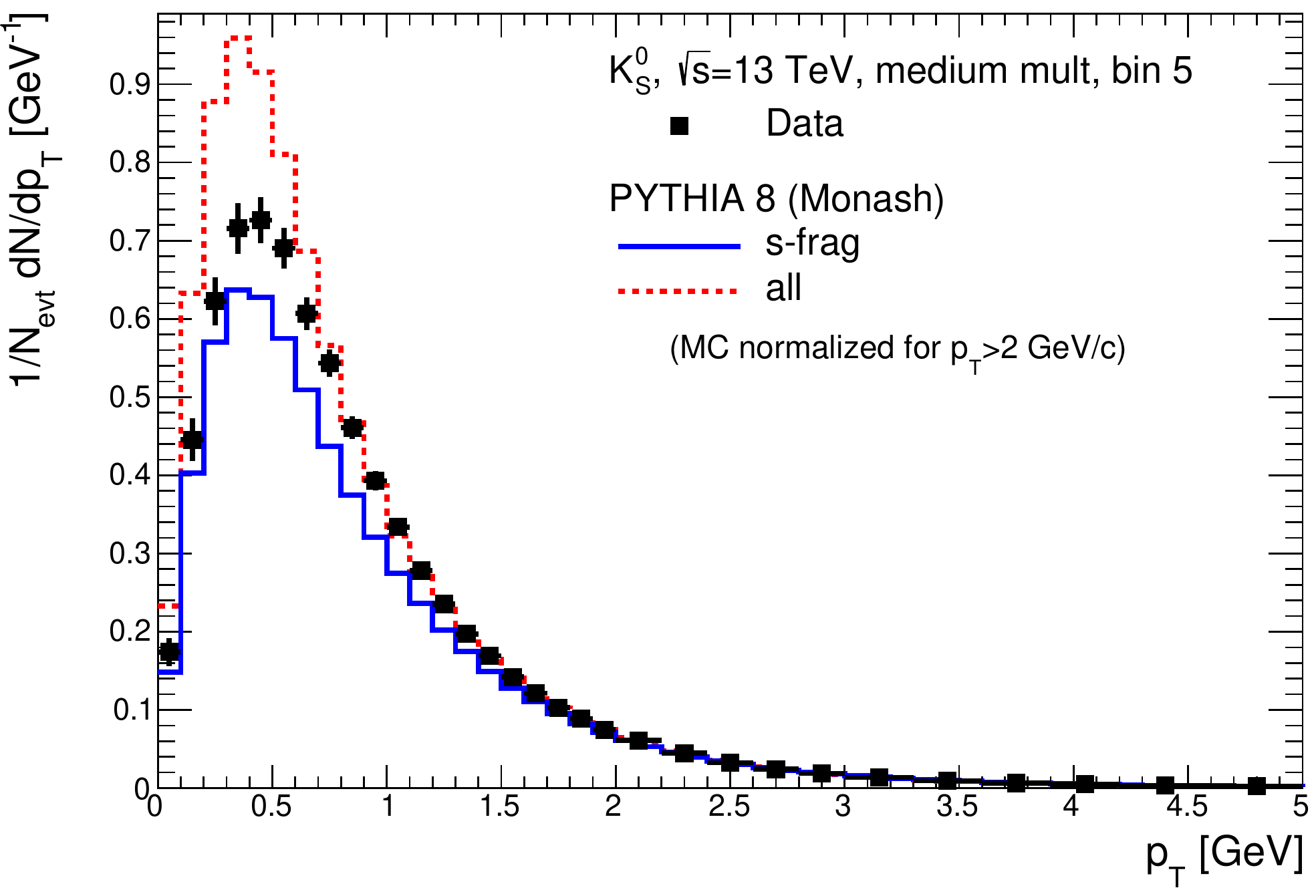}   
\includegraphics[width=0.32\linewidth]{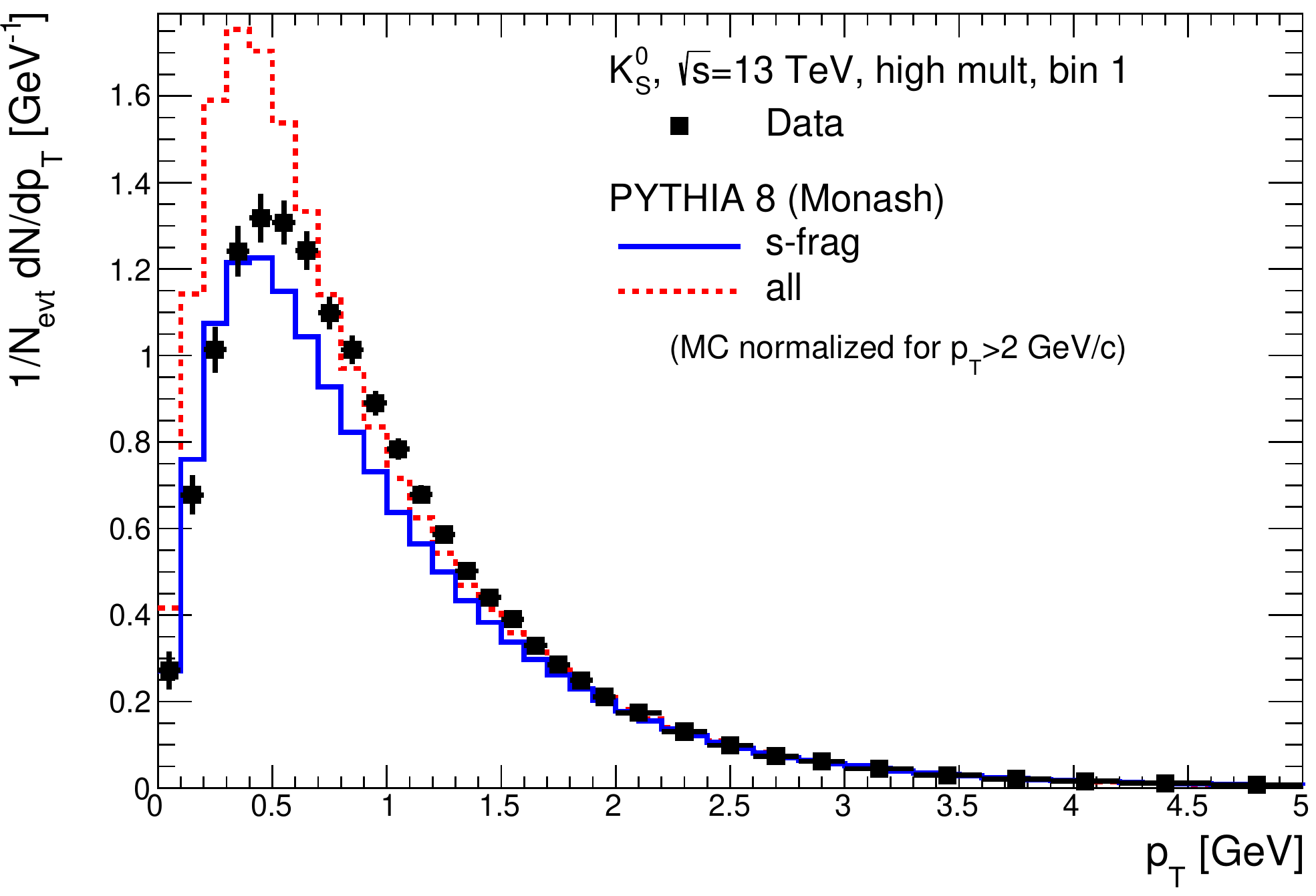}       
\includegraphics[width=0.32\linewidth]{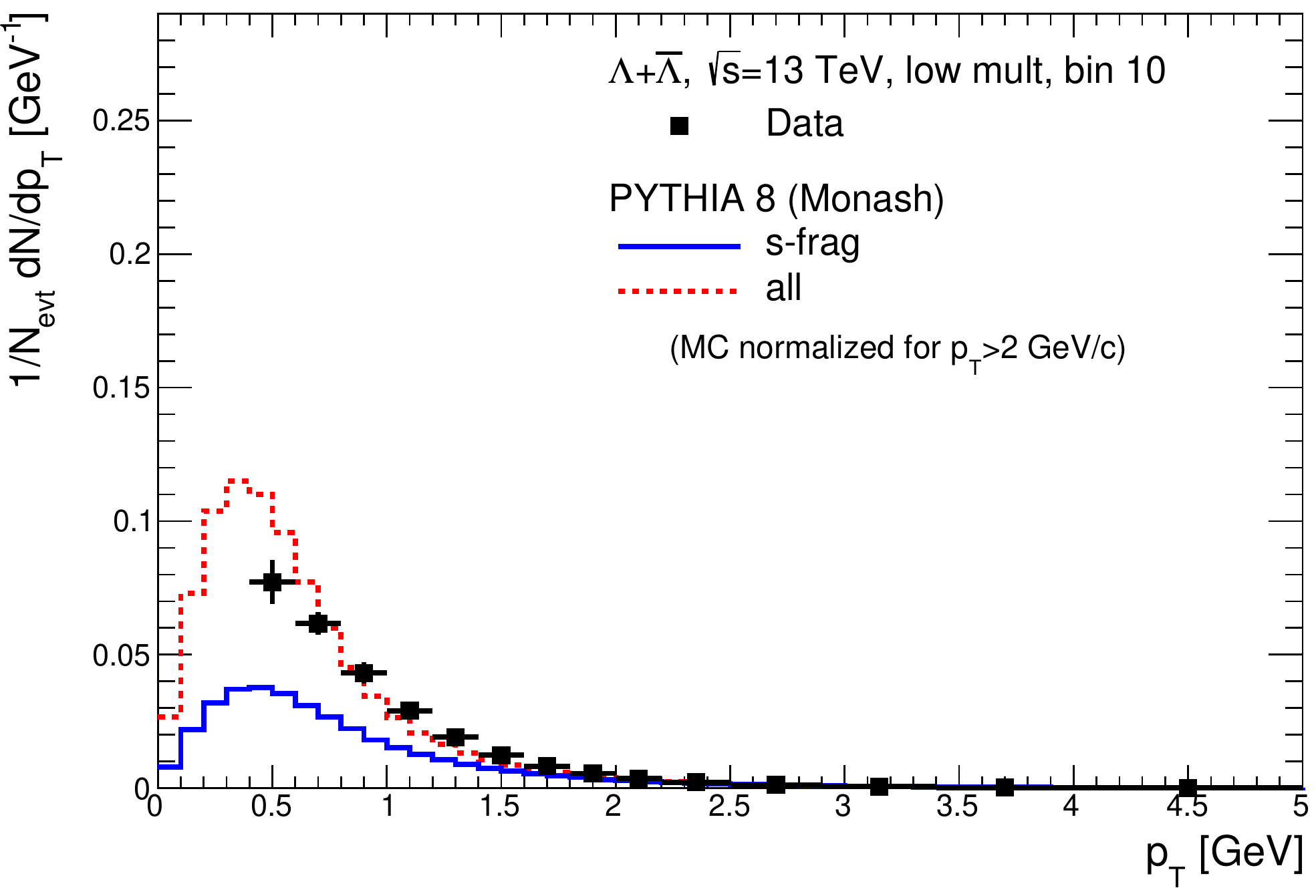}
\includegraphics[width=0.32\linewidth]{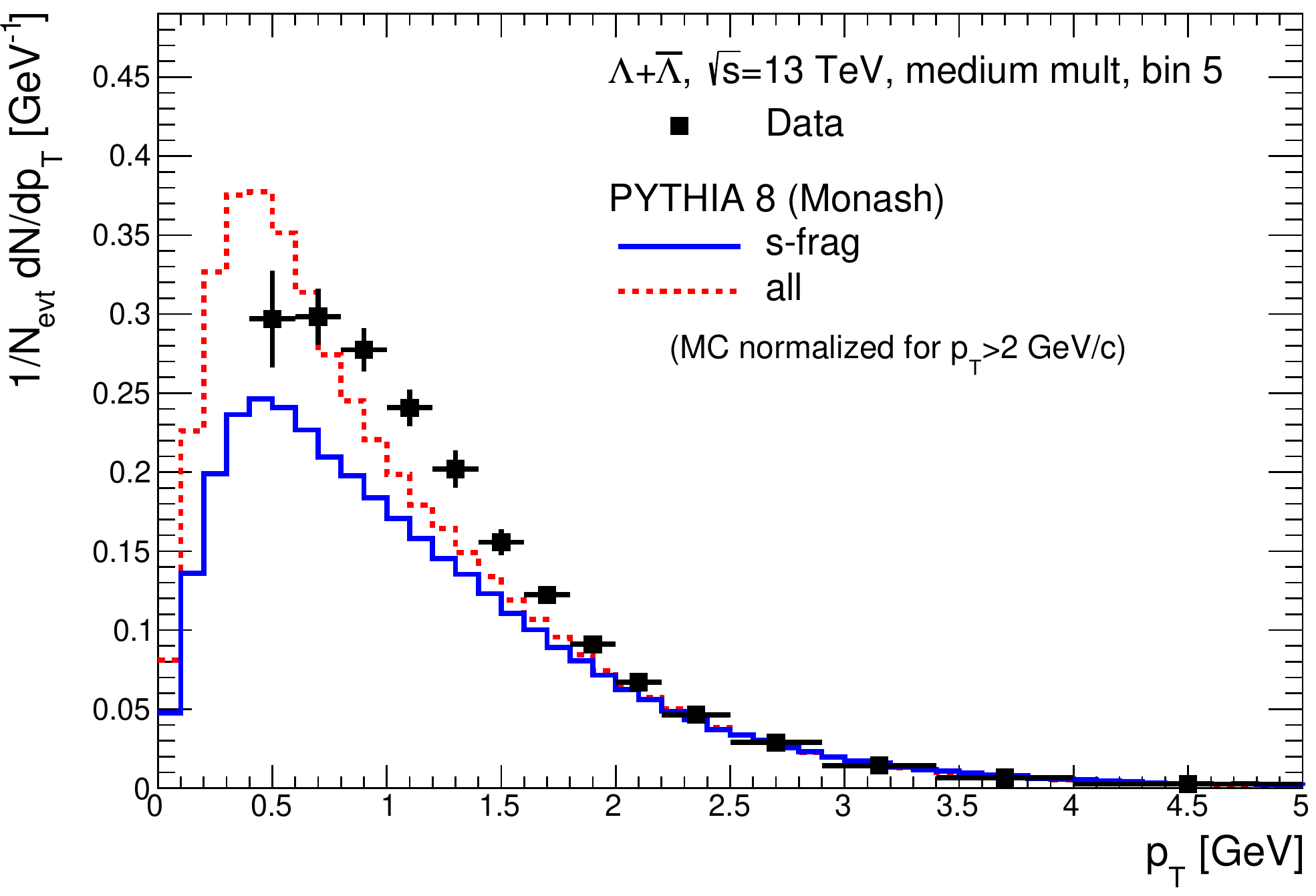}   
\includegraphics[width=0.32\linewidth]{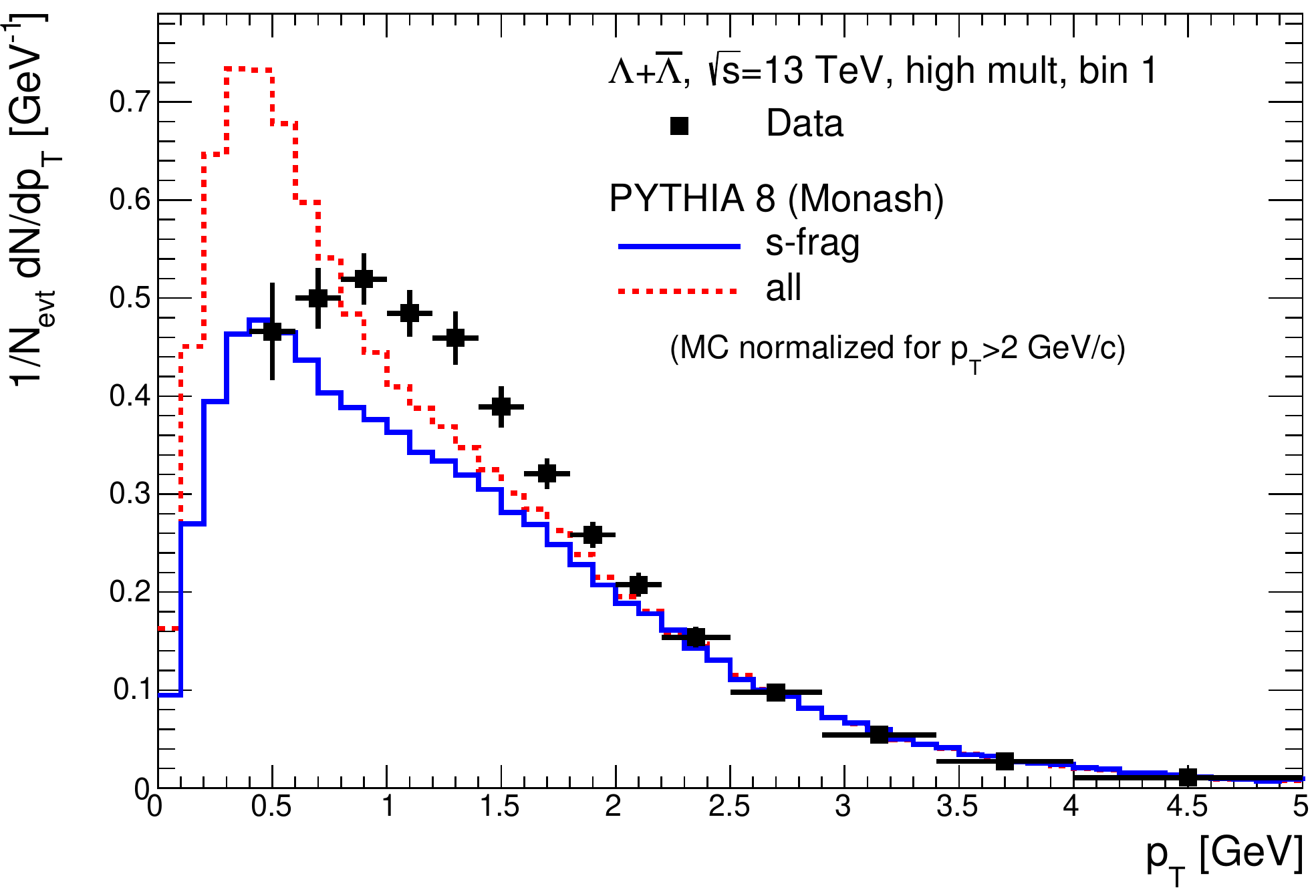}
\includegraphics[width=0.32\linewidth]{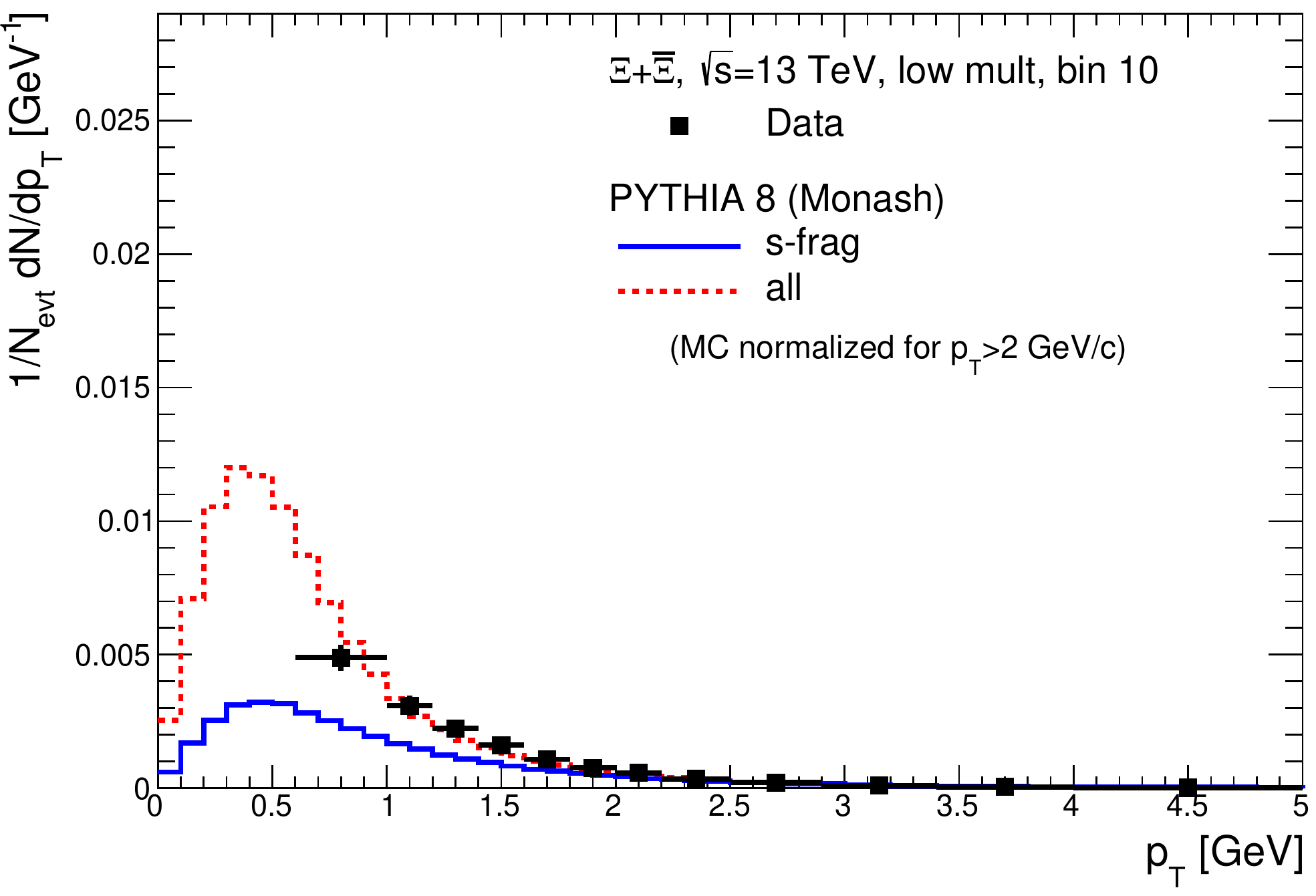}
\includegraphics[width=0.32\linewidth]{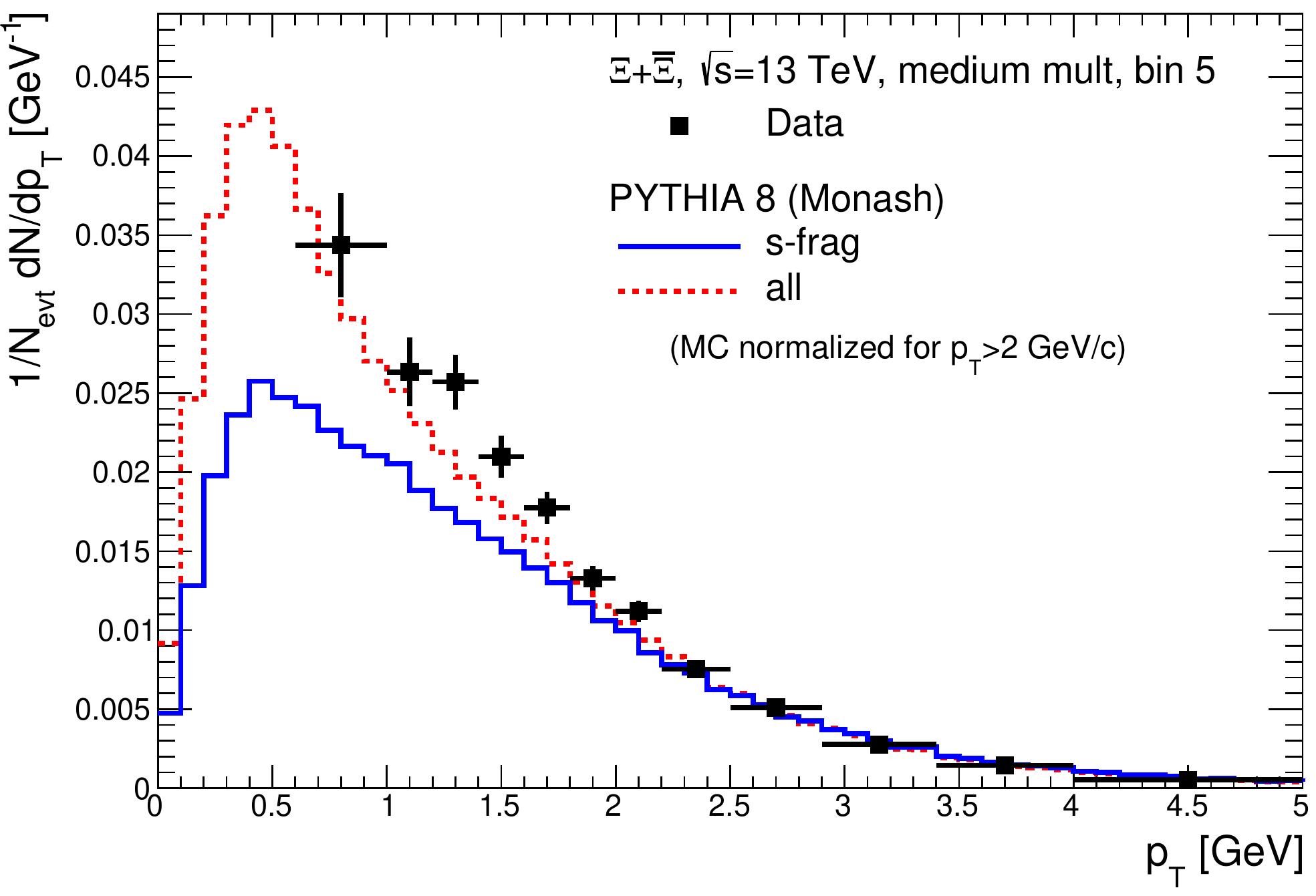}   
\includegraphics[width=0.32\linewidth]{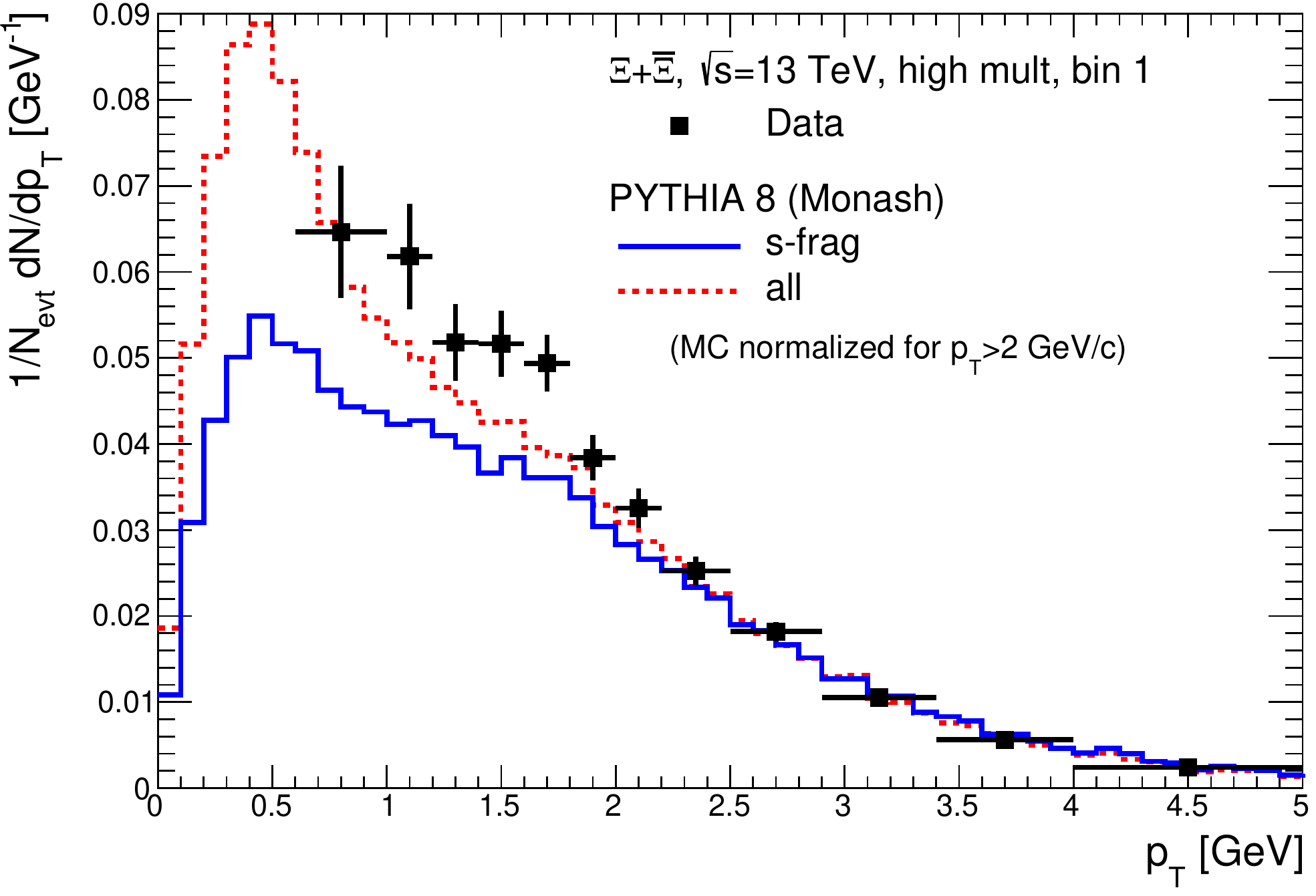}       
\includegraphics[width=0.32\linewidth]{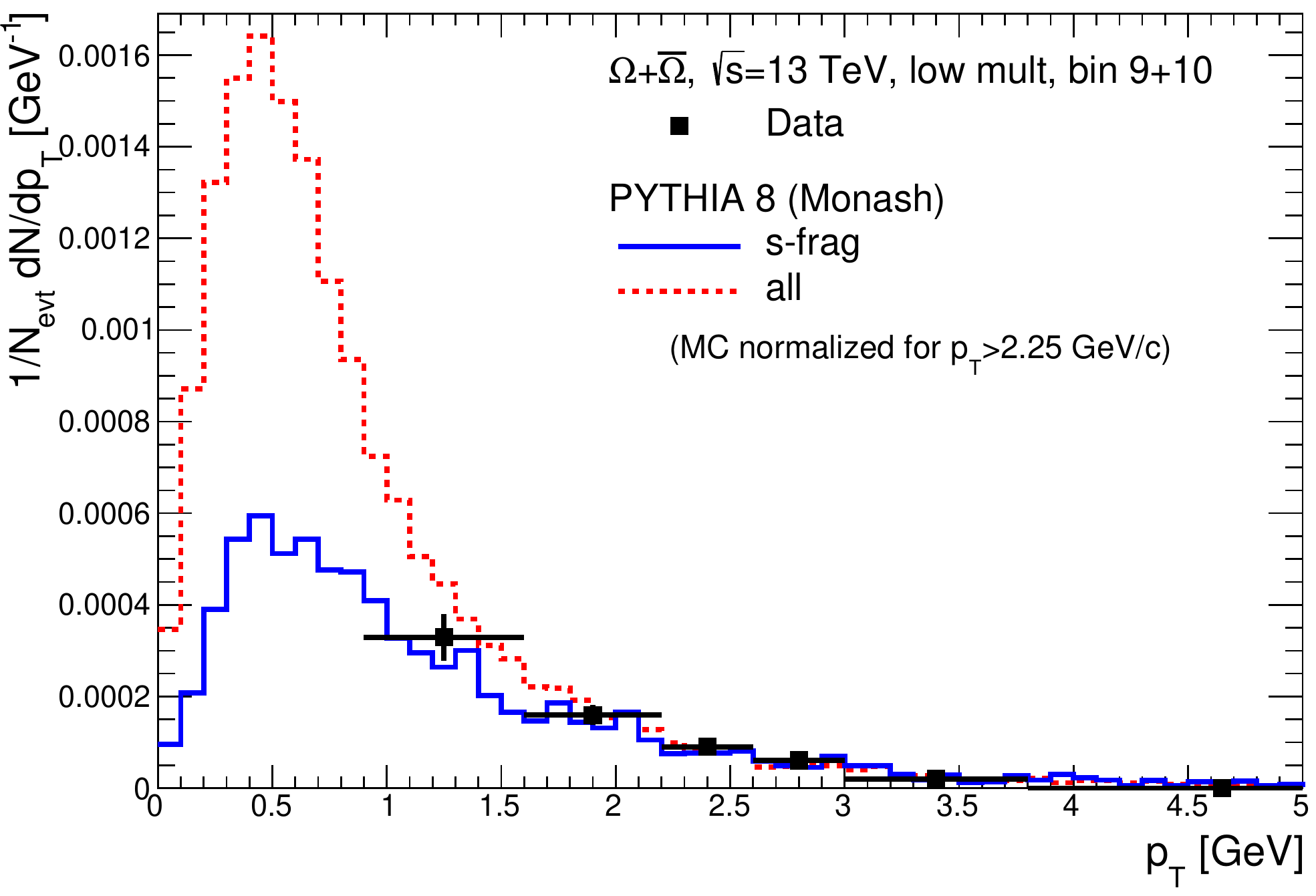}
\includegraphics[width=0.32\linewidth]{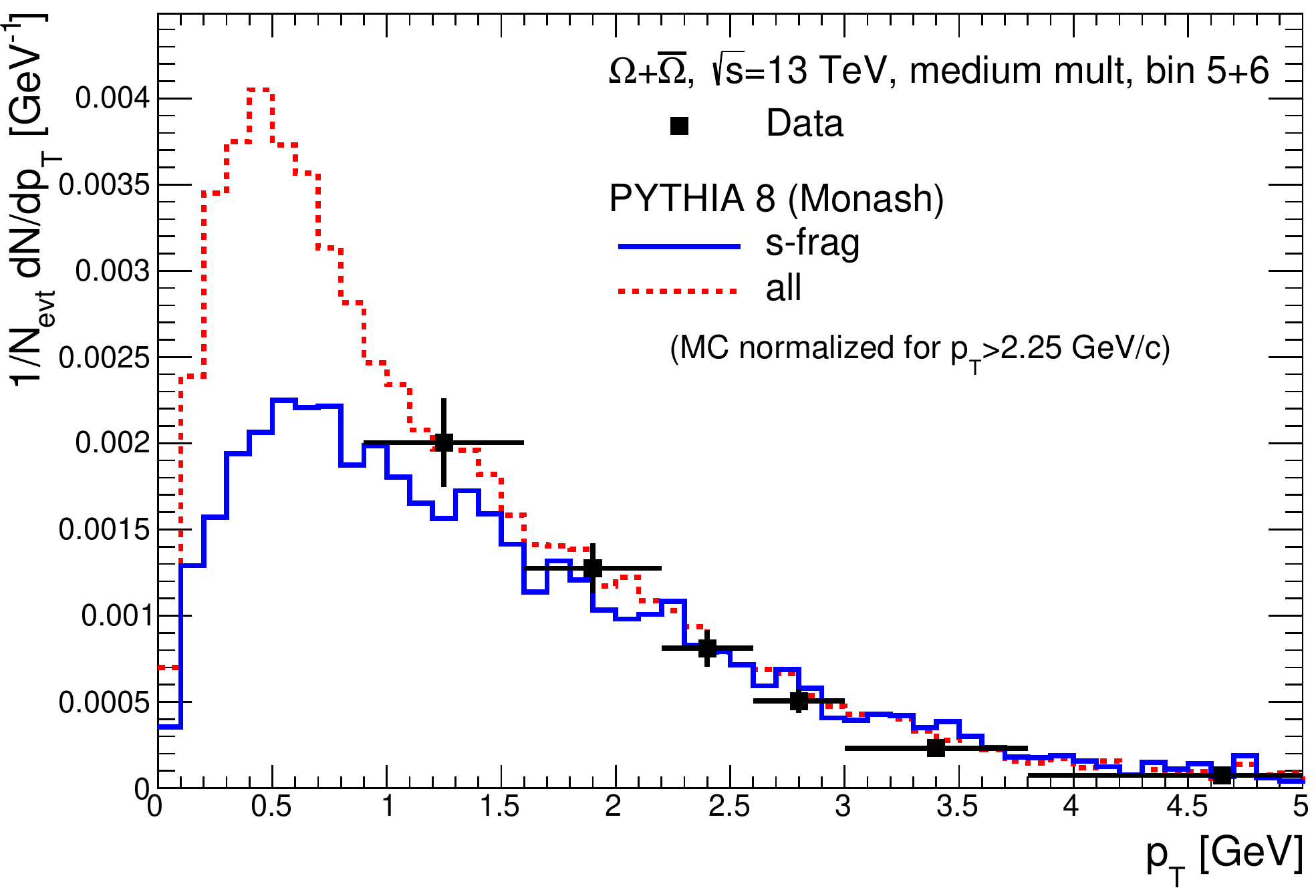}   
\includegraphics[width=0.32\linewidth]{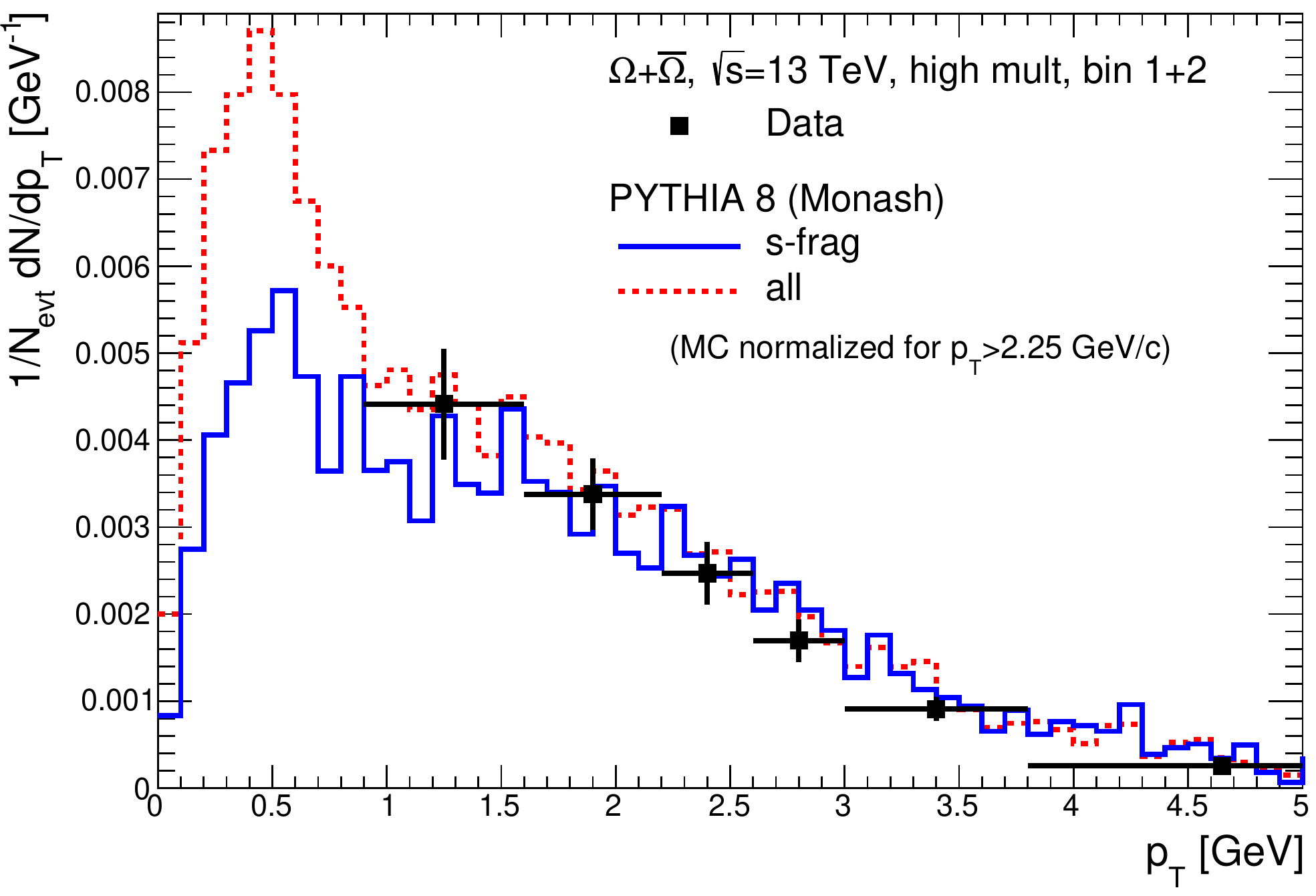}
\fi
\caption{Strange particle $\pt$-spectra~($K_{\rm S}^0$, $\Lambda$, $\Xi$, $\Omega$) measured in low~(left panels), medium~(middle panels) and high~(right panels) multiplicity \pp\ collisions at $\s=13$ by ALICE TeV~\cite{Acharya:2019kyh} compared with \pytstr\ calculations. The calculated distributions, which show all produced particles of a given type, as well as those produced by $s$-quark fragmentation alone, are normalized to the data in the region $\pt\gtrsim2$~GeV/$c$.
} 
\label{fig:multdep}
\end{center}
\end{figure}
In \Fig{fig:multdep}, we compare calculated spectra using \pytstr\ with the measured $\pt$-spectra~($K_{\rm S}^0$, $\Lambda$, $\Xi$, $\Omega$) in low, medium and high multiplicity \pp\ collisions at $\s=13$ TeV.
As before, the calculated spectra are normalized to the data in the region $\pt\gtrsim2$~GeV/$c$.
In data, the soft component appears in particular for the low-multiplicity interval as well as for the $K_{\rm}^{0}$, while otherwise the spectral shapes are well described by the $s$-fragmentation component.

\begin{figure}[t!]
\begin{center}
\ifplotpold
\includegraphics[width=0.75\linewidth]{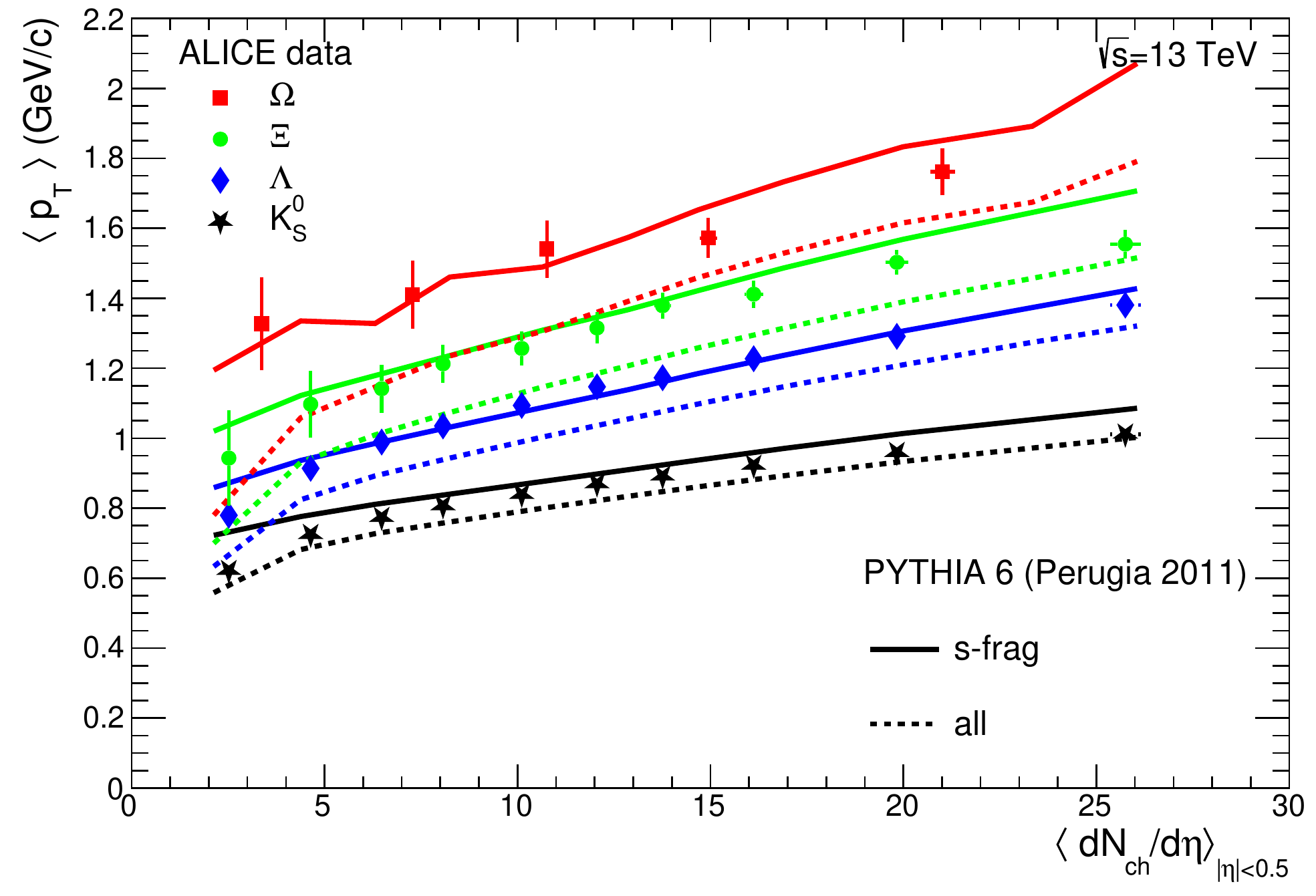}
\else
\includegraphics[width=0.75\linewidth]{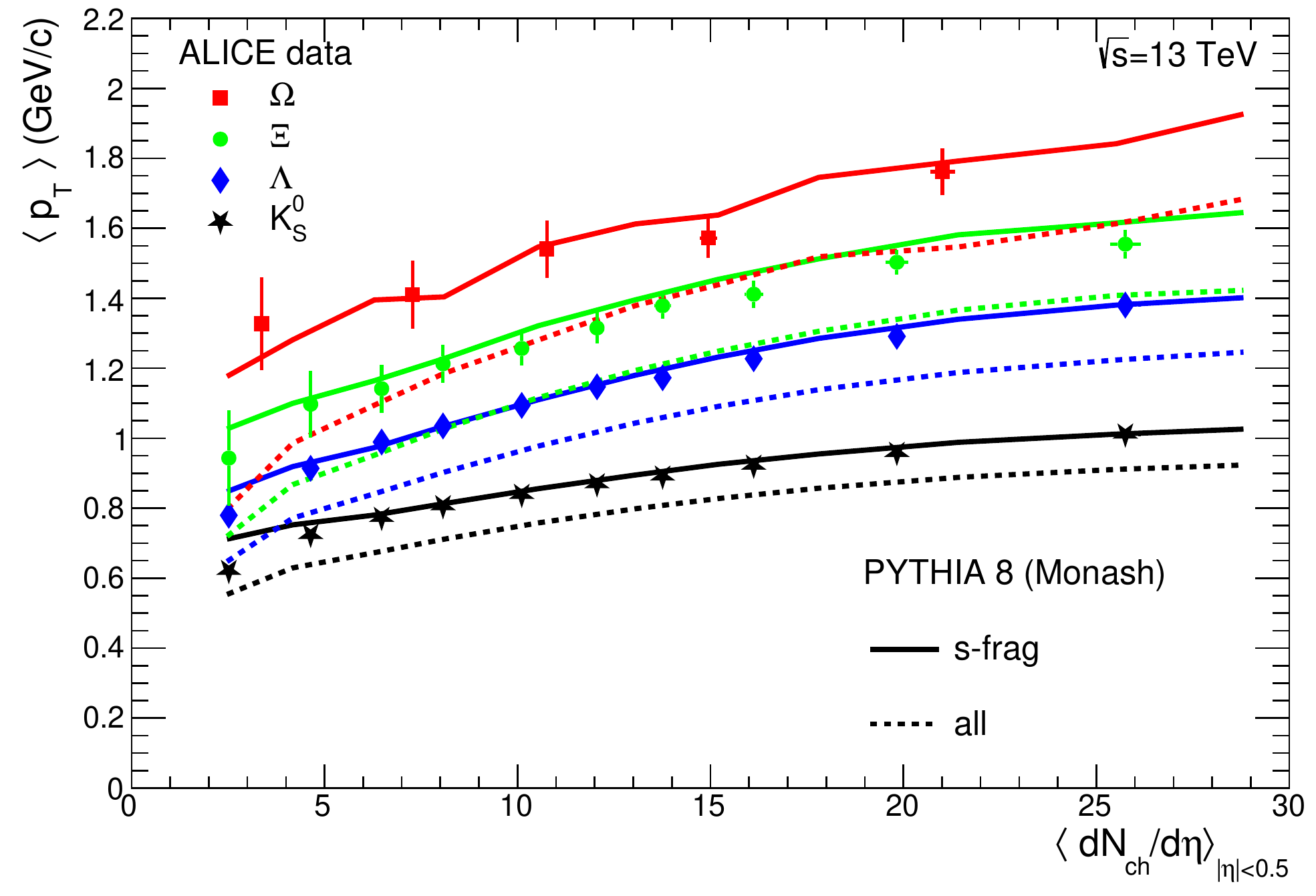}
\fi
\caption{Average transverse momenta of strange particles~($K_{\rm S}^0$, $\Lambda$, $\Xi$, $\Omega$) versus charged-particle multiplicity at midrapidity in \pp\ collisions at $\s=13$ TeV measured by ALICE~\cite{Acharya:2019kyh} compared with  \pytstr\ calculations. The calculation shows the $\avg{\pt}$ of all produced particles of a given type, as well as the $\avg{\pt}$ of those produced by $s$-quark fragmentation alone.
} 
\label{fig:meanpt}
\end{center}
\end{figure}
In \Fig{fig:meanpt}, we compare the measured $\avg{\pt}$ of strange particles~($K_{\rm S}^0$, $\Lambda$, $\Xi$, $\Omega$) versus multiplicity in \pp\ collisions at $\s=13$ TeV~\cite{Acharya:2019kyh} with calculations of \pytstr.
As expected from the comparison of the $\pT$-spectra, the data are generally well reproduced by the calculations using $s$-quark fragmentation alone\ifplotpold, except in the case of the $K_{\rm S}^0$\fi.

\begin{figure}[t!]
\begin{center}
\ifplotpold
\includegraphics[width=0.75\linewidth]{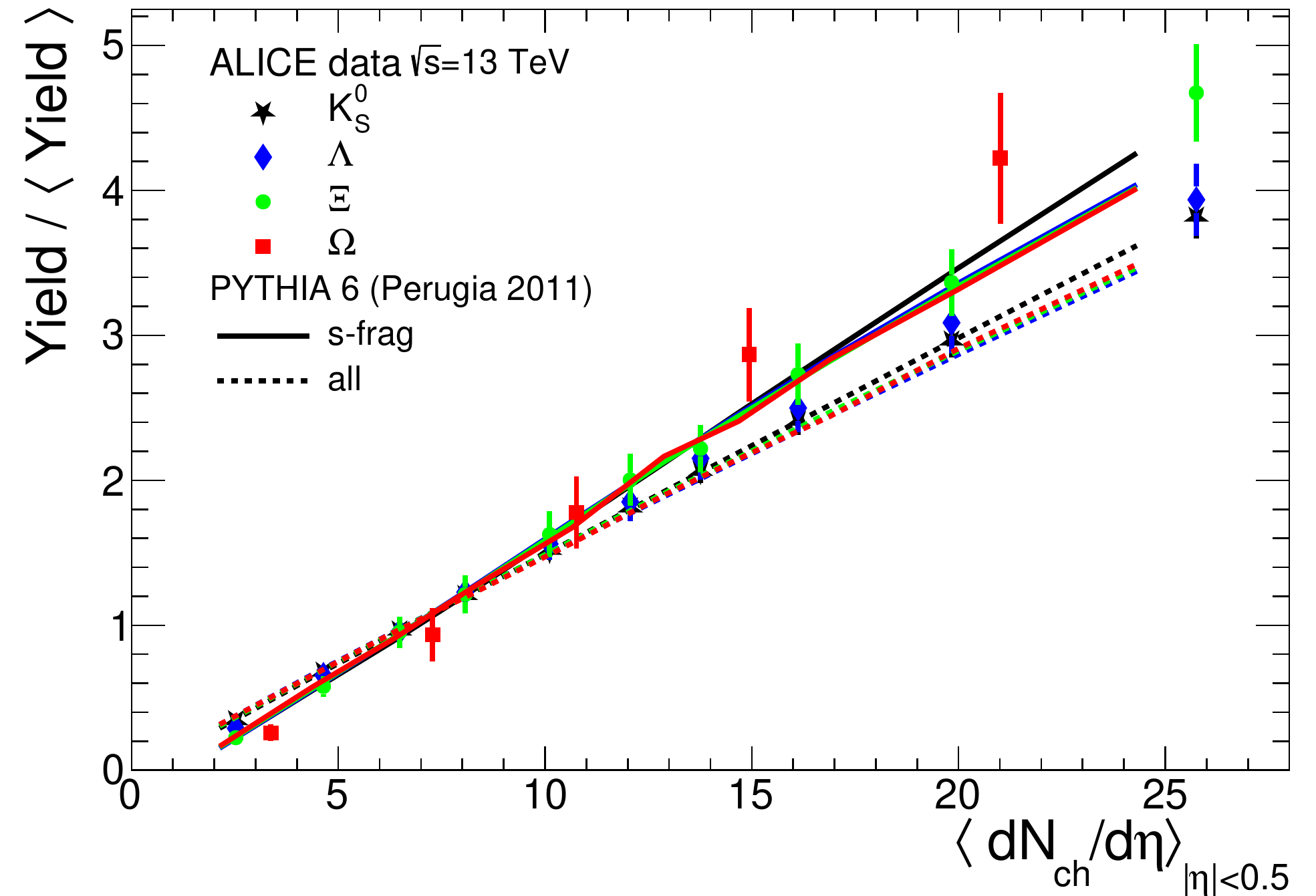}
\else
\includegraphics[width=0.75\linewidth]{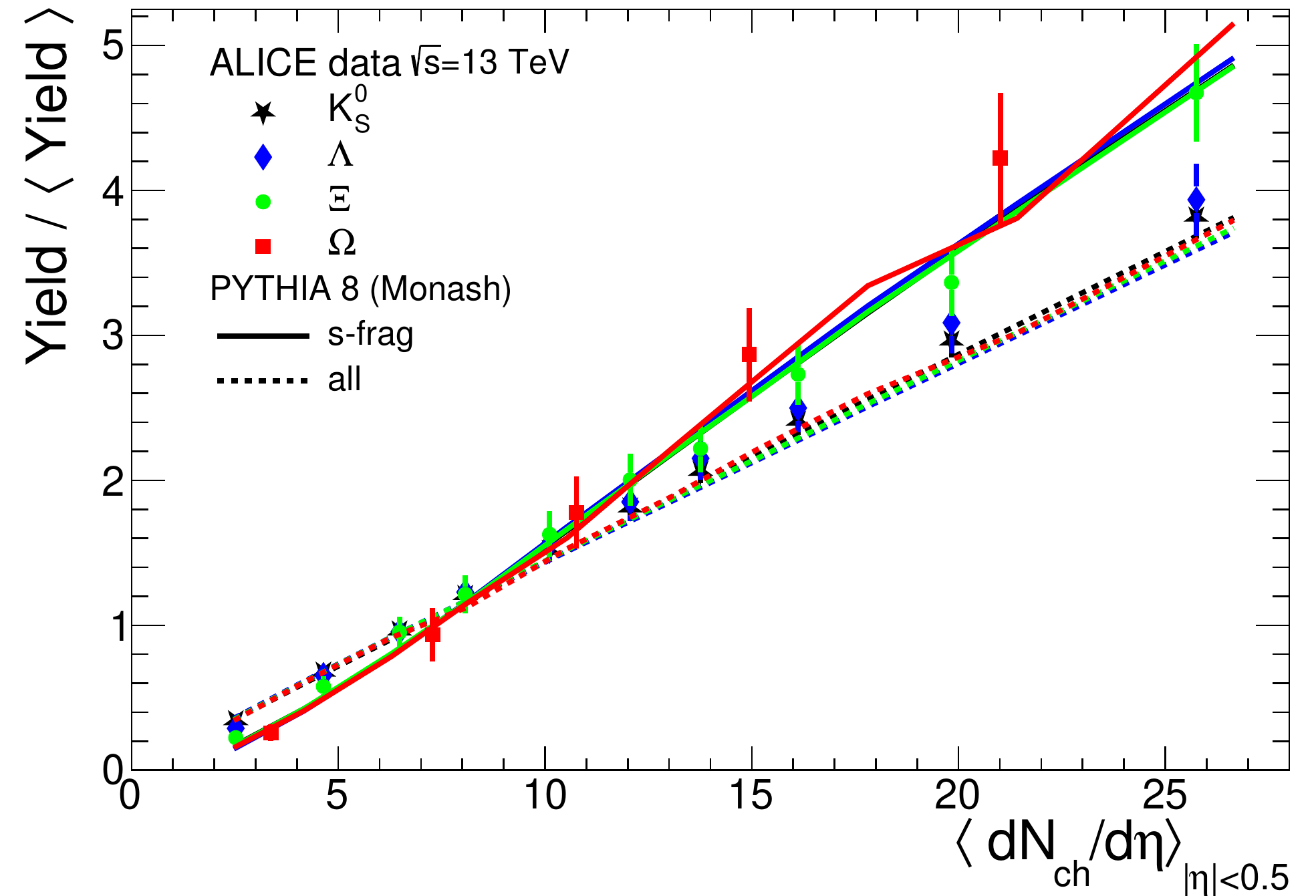}
\fi
\caption{Self-normalized yields of strange particles~($K_{\rm S}^0$, $\Lambda$, $\Xi$, $\Omega$) versus charged-particle multiplicity at midrapidity in \pp\ collisions at $\s=13$ TeV~\cite{Acharya:2019kyh} compared with \pytstr\ calculations. The calculation shows the yield of all produced particles of a given type, as well as the yield of those produced by $s$-quark fragmentation alone.
} 
\label{fig:selfnormyields}
\end{center}
\end{figure}

Since the absolute yield is typically not well reproduced in calculations, we compare the self-normalized yields~(i.e.\ the yield in a given multiplicity interval normalized to its average yield) as a function of multiplicity, in \Fig{fig:selfnormyields}, between data and \pytstr\ calculations.
The self-normalized yields for all produced particles which are dominated by $udg$-hadronization and by $s$-quark fragmentation alone enclose the data points.
The $udg$-dominated yields pass through (0,0) while the $s$-fragmentation component has a finite intercept with the $x$-axis.
The comparison of the data with the calculations suggests that the $K_{\rm S}^0$ and $\Lambda$ yields are dominated by $udg$-hadronization, while $\Xi$ and $\Omega$ yields by $s$-fragmentation. 
The production of $\Lambda$ particles may also have a sizeable contribution from the $s$-fragmentation component.
However, these interpretations rely on 
the shape of the $\pt$-spectra as modelled by PYTHIA. It cannot be excluded that the $udg$-hadronization dynamics for 
multi-strange baryons is not implemented correctly and that with a correct implementations both mechanisms would show the threshold behavior.
A way to avoid the interference of the soft and semi-hard components is to compare to self-normalized strange-particle yields extracted above a certain $\pt$ threshold~(Fig.\ 6 of \cite{Acharya:2019kyh}), which were found to be well described by \Pythia.

\begin{figure}[t!]
\begin{center}
\ifplotpold
\includegraphics[width=0.49\linewidth]{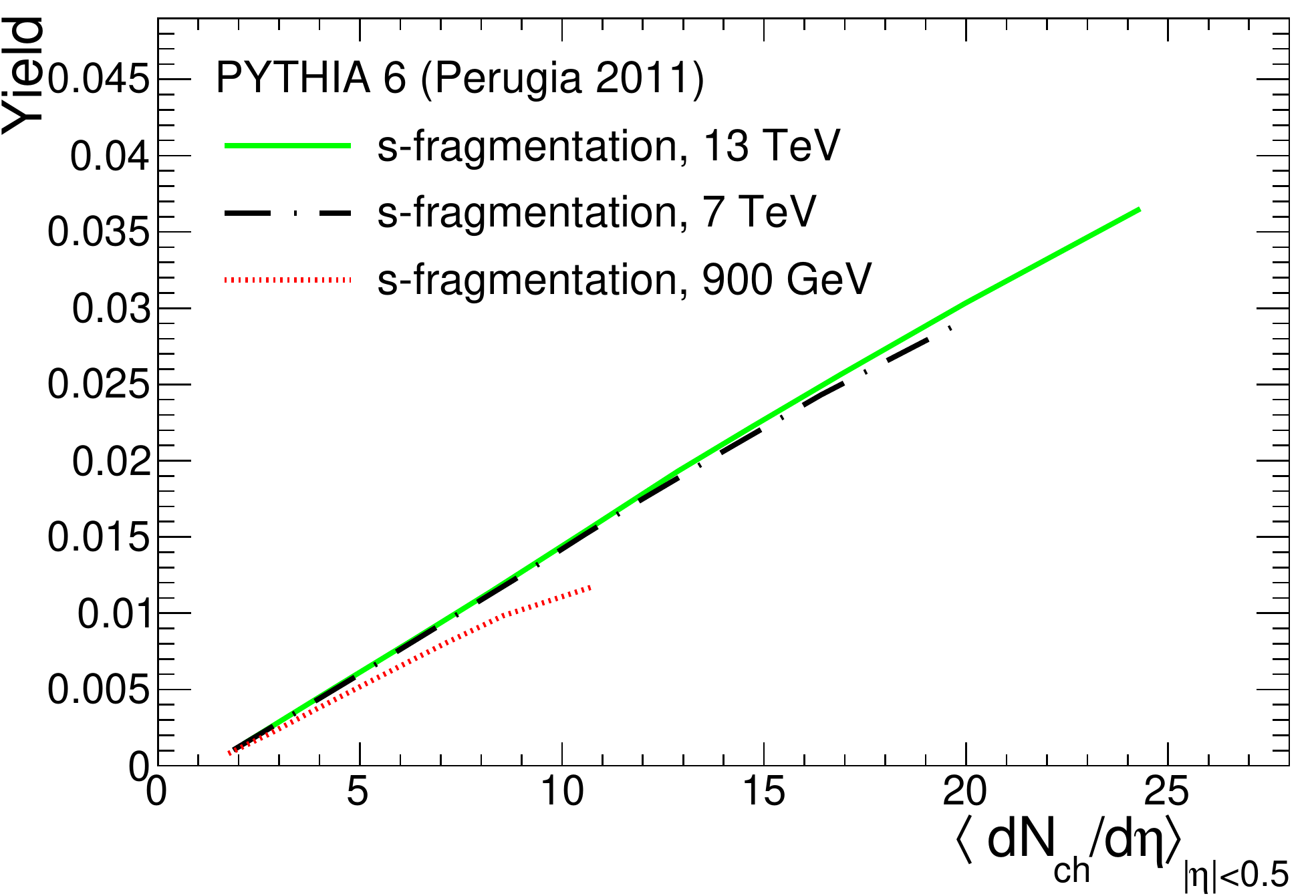}
\else
\includegraphics[width=0.49\linewidth]{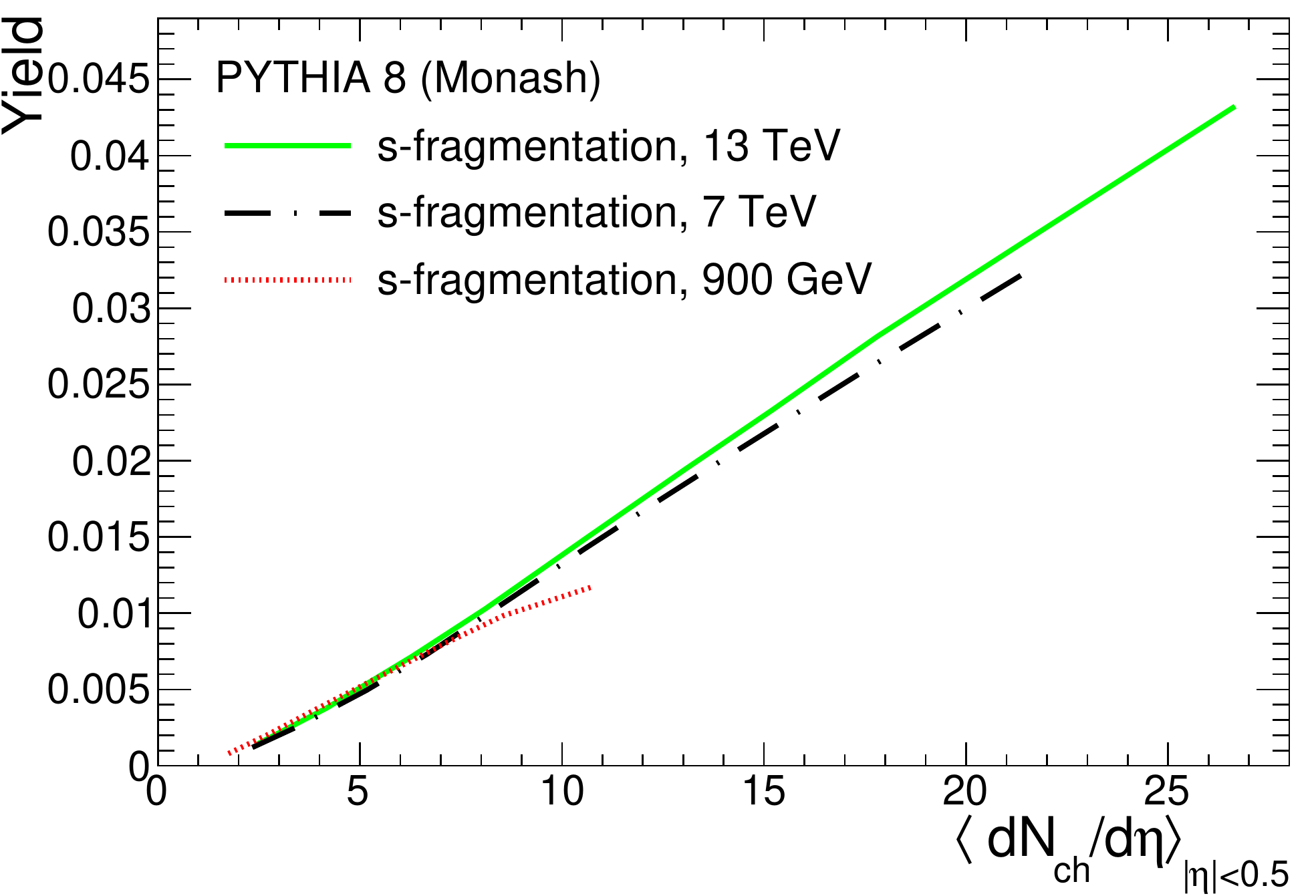}
\fi
\caption{Comparison of $\Xi$ yield versus charged-particle multiplicity at midrapidity in \pp\ collisions at $\s=0.9$, 7 and 13 TeV for $s$-quark fragmentation computed with \pytstr.
} 
\label{fig:energyscaling}
\end{center}
\end{figure}
\begin{figure}[t!]
\begin{center}
\ifplotpold
\includegraphics[width=0.49\linewidth]{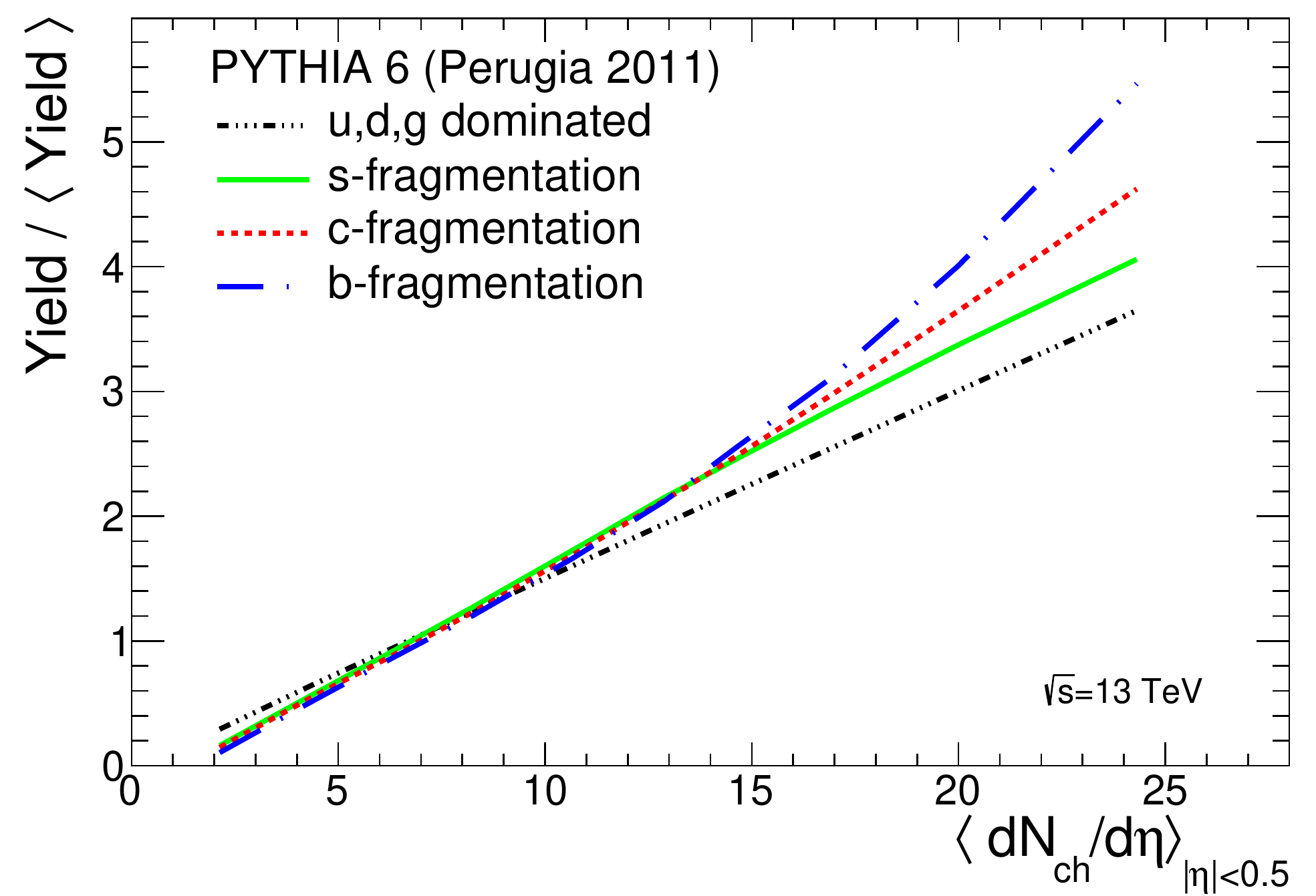}
\includegraphics[width=0.49\linewidth]{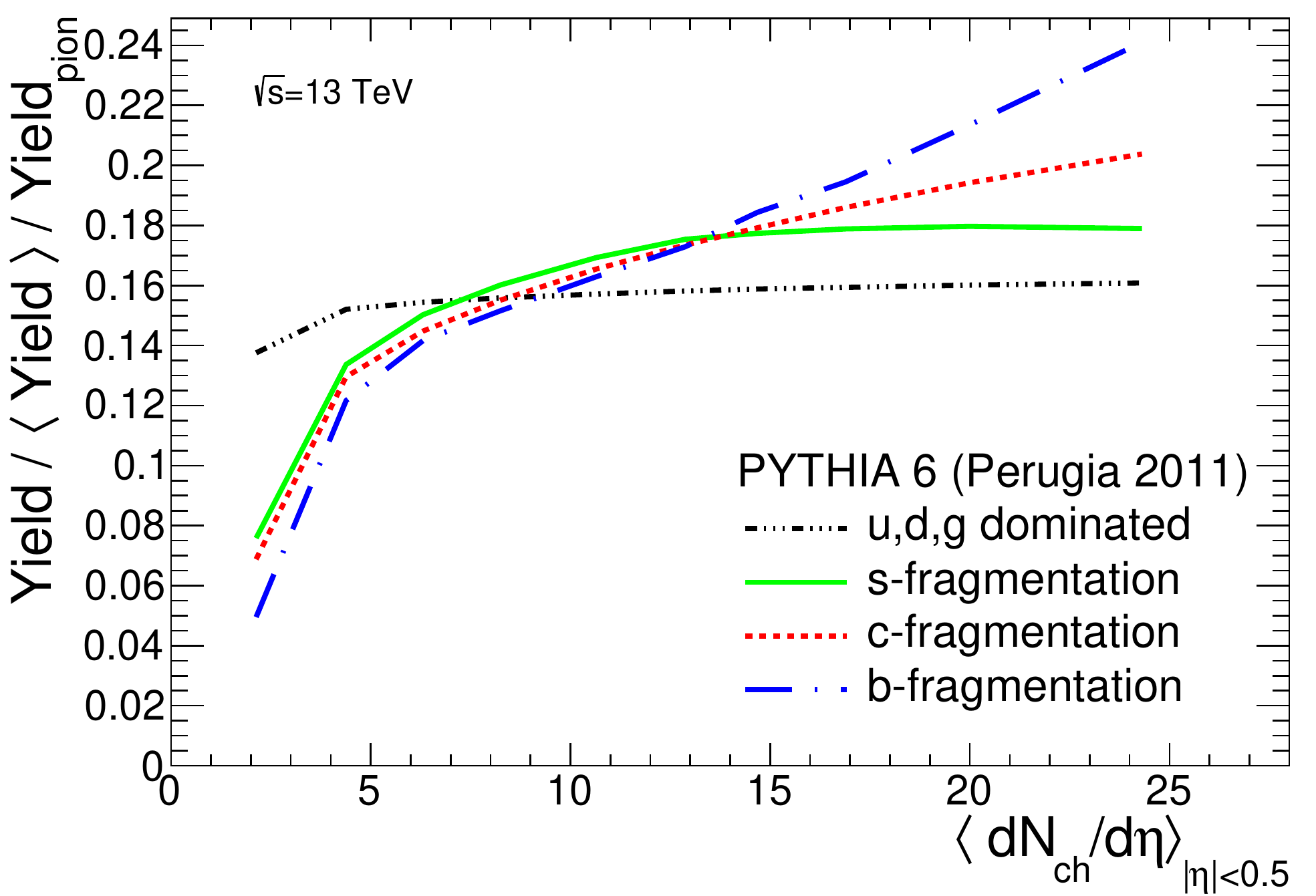}
\else
\includegraphics[width=0.49\linewidth]{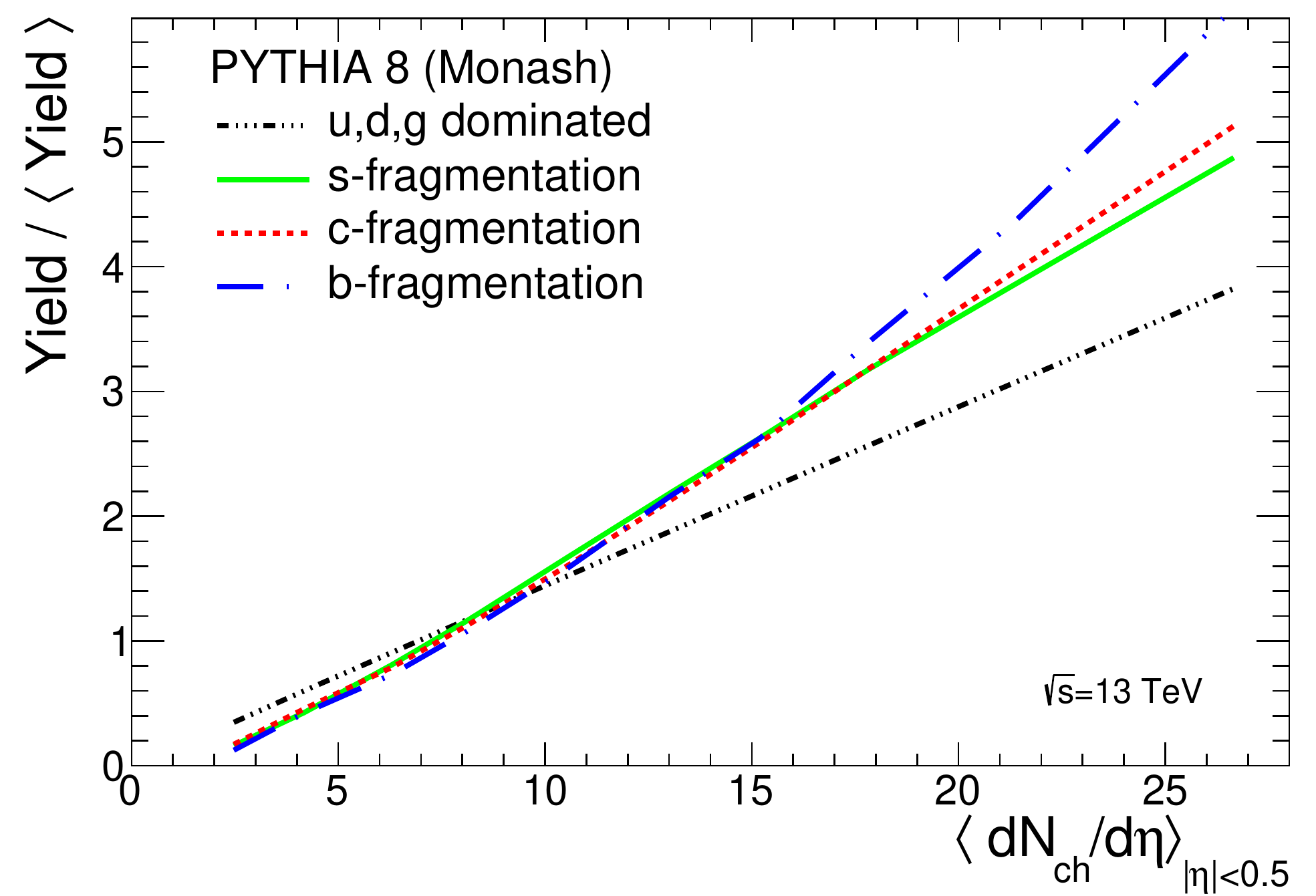}
\includegraphics[width=0.49\linewidth]{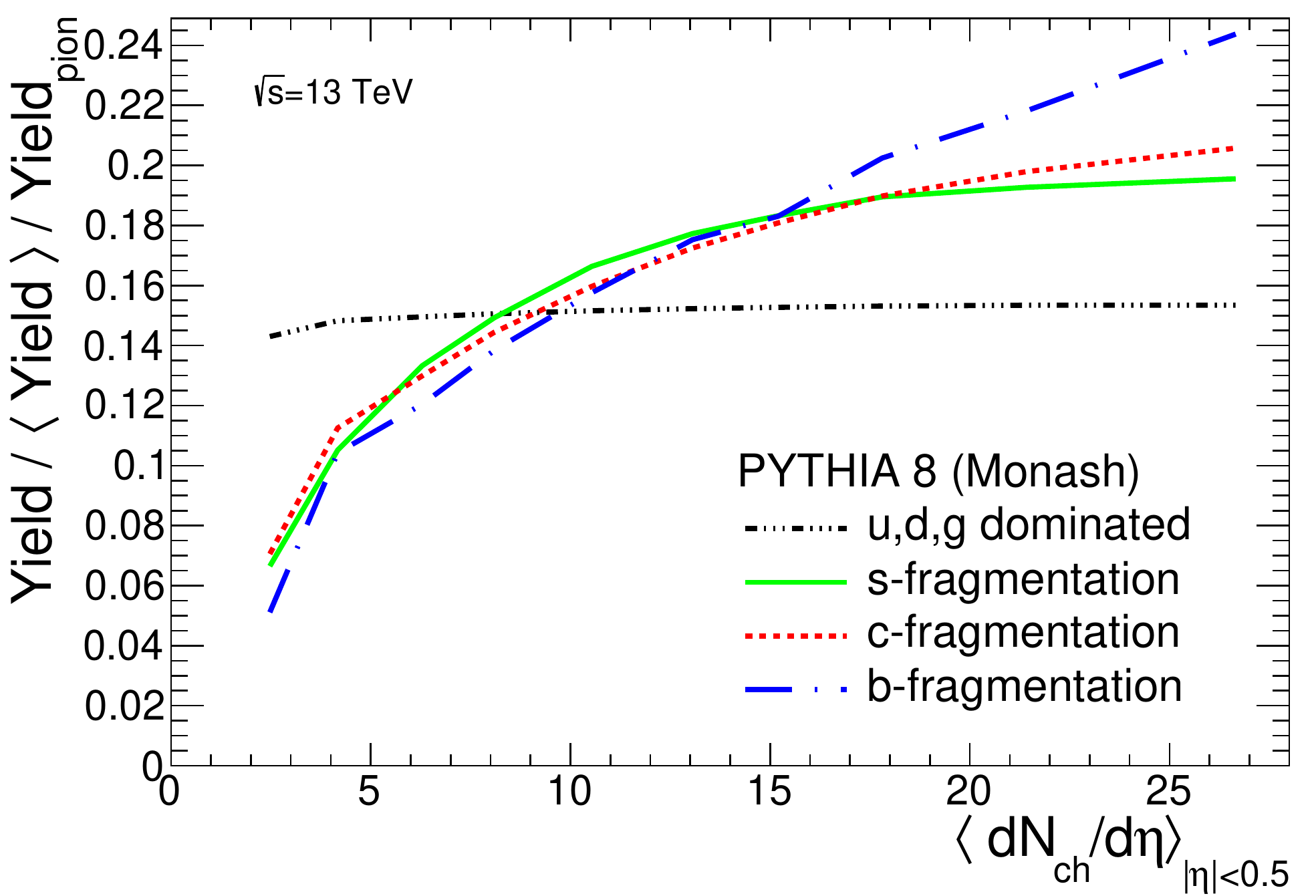}
\fi
\caption{Left:~Self-normalized yields versus charged-particle multiplicity at midrapidity in \pp\ collsions at $\s=13$ TeV for different flavors computed with \pytstr. Right:~Same as on the left panel, except that the yields are further normalized by the charged pion yield.
}
\label{fig:flavordep}
\end{center}
\end{figure}

The $\sqrt{s}-$independence of the strange-particle yields versus multiplicity between $\s= 7$ and 13 TeV~(see \Fig{fig:strangedata} and \Fig{fig:strangedatanorm}) is reproduced in \Pythia\ as shown in \Fig{fig:energyscaling}.
Hence, ``multiplicity scaling'' in-itself cannot be interpreted as a consequence of final state effects. A significant deviation from this scaling is, however, expected at lower energies ($\s= 900$ GeV in \Fig{fig:strangedatanorm}). The reason is that the number of charged particles produced per \MPI\ decreases only logarithmically with $\s$.

The quark-flavor evolution of both the self-normalized yields and self-normalized yields per charged-pion yields computed with \pytstr\ are shown in \Fig{fig:flavordep}.
A smooth transition from linear to quadratic dependence from the $udg$-hadronization and $s$-fragmentation components to $c$-and $b$-fragmentation components emerges, which in the \MPI\ picture, results as a consequence of the significant contribution of associated particle production to the measured multiplicity~(auto-correlation bias)~\cite{Weber:2018ddv}.

\begin{figure}[t!]
\begin{center}
\includegraphics[width=0.49\linewidth]{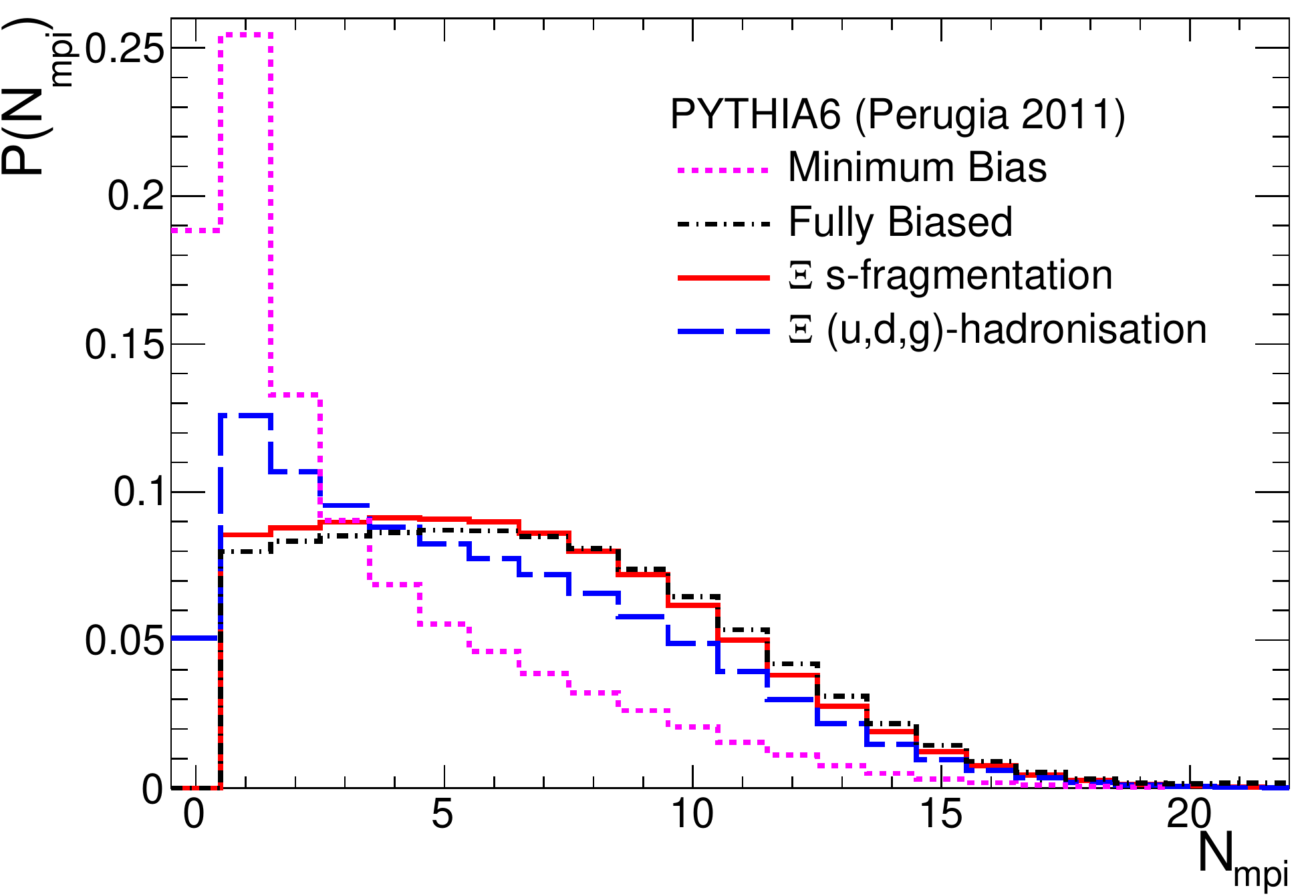}
%\hspace{0.1cm}
\includegraphics[width=0.49\linewidth]{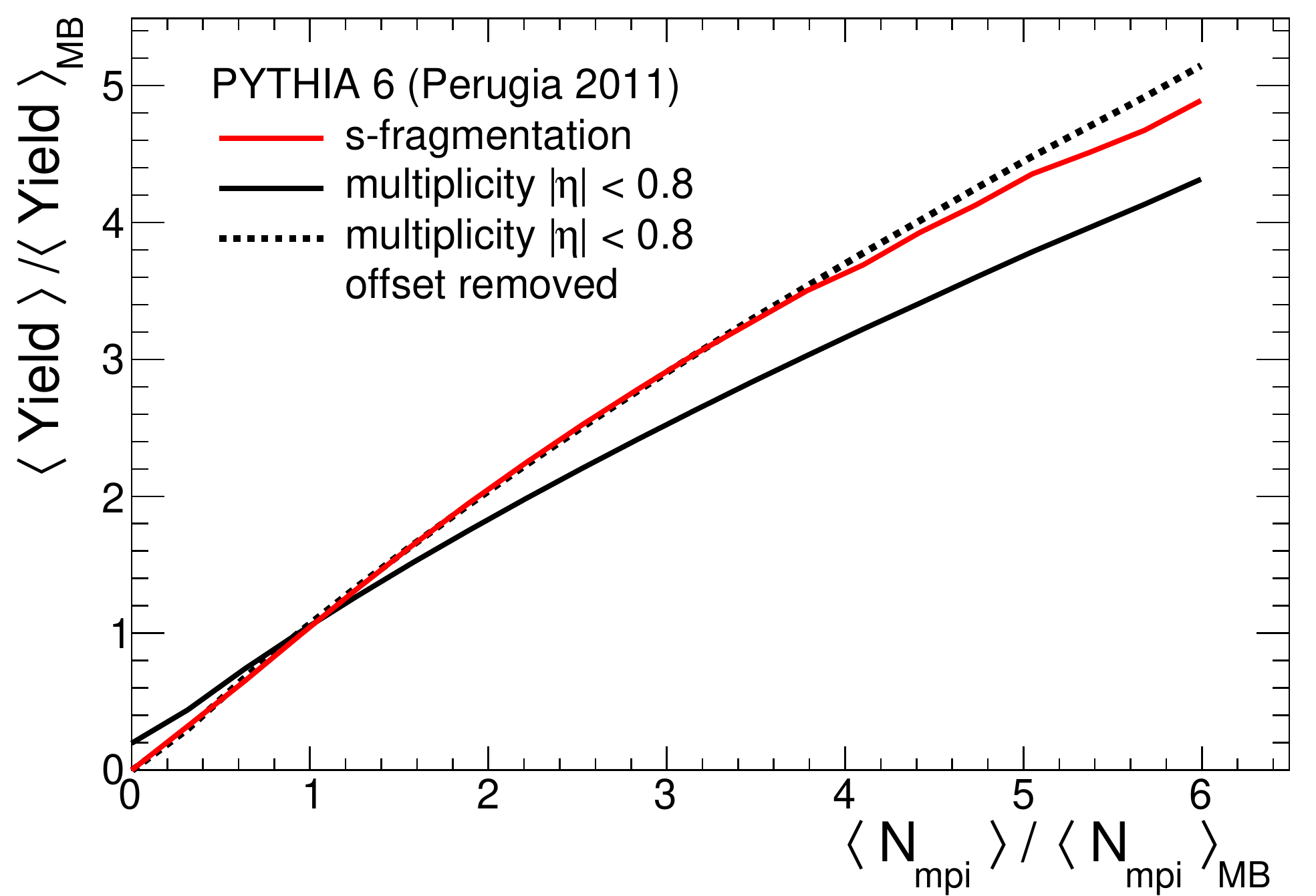}
\caption{ Left: The \MPI\ probability distribution for $\Xi$ baryons from ($u,d,g$)-hadronization~(long-dashed blue) is compared to the one from $s$-fragmentation~(solid red). As references we show also the minimum-bias distribution (fine-dashed) and the fully biased distributions~($\Nmpi P_{\rm MinBias}(\Nmpi)$) expected for hard processes~(short-long dashed black).
Right: Self-normalised $\Xi$ yields from $s$-fragmentation~(red) and charged particle multiplicity (black solid) as a function of the self-normalised number of \MPI. The dotted line shows the multiplicity with the soft contribution ($N_{\rm mpi} = 0$) removed.
} %macro from macros/plotD.C
\label{fig:OriginOffset_MPI}
\end{center}
\end{figure}

In order to elucidate the origin of the multiplicity threshold we investigate the effect from three different angles:

i)~In the left panel of \Fig{fig:OriginOffset_MPI} we compare the $\Nmpi$ probability distribution for $\Xi$ baryons from ($u,d,g$)-hadronization to the those for $s$-fragmentation. 
Both processes bias the $\Nmpi$-distribution towards the ``fully biased'' distribution expected for rare processes.
However, the $s$-fragmentation tag biases the MPI distribution to higher values, i.e.\ more central collisions, compared to ($u,d,g$)-hadronization. 
The difference is particularly important below the unbiased~(minimum-bias) average $\Nmpi$ of 3.2 corresponding to the lowest multiplicity bins in data. 
It results from the presence of double-diffractive collisions which have $\Nmpi = 0$ (only seen for ($u,d,g$)-hadronization) and an important contribution to soft particle production in peripheral collisions. 
Note that this interpretation is further corroborated by the shape of the multiplicity dependence of the average $\pt$ at low multiplicity~(\Fig{fig:meanpt}, "ledge effect" \cite{Wang:1988bw}). 

ii)~In the right panel of \Fig{fig:OriginOffset_MPI} we show the self-normalised yield of $\Xi$ baryons from $s$-fragmentation versus the self-normalised number of \MPIs\ and compare it to the self-normalised multiplicity versus the same quantity. 
While $s$-fragmentation exhibits a strict proportionality at low $\Nmpi$, charged particle multiplicity stays finite down to $\Nmpi = 0$. 
The reason is that multiplicity also receives contributions from soft processes. 
If one subtracts this soft offset the multiplicity follows the $s$-fragmentation yield over a wide range of $\Nmpi$ which explains the linear behaviour of $s$-fragmentation versus multiplicity above the threshold. 
In particular, the weaker than linear increase at higher $\Nmpi$ results in PYTHIA from color-reconnection and affects multiplicity and $s$-fragmentation almost equally.

\begin{figure}[t!]
\begin{center}
\includegraphics[width=0.49\linewidth]{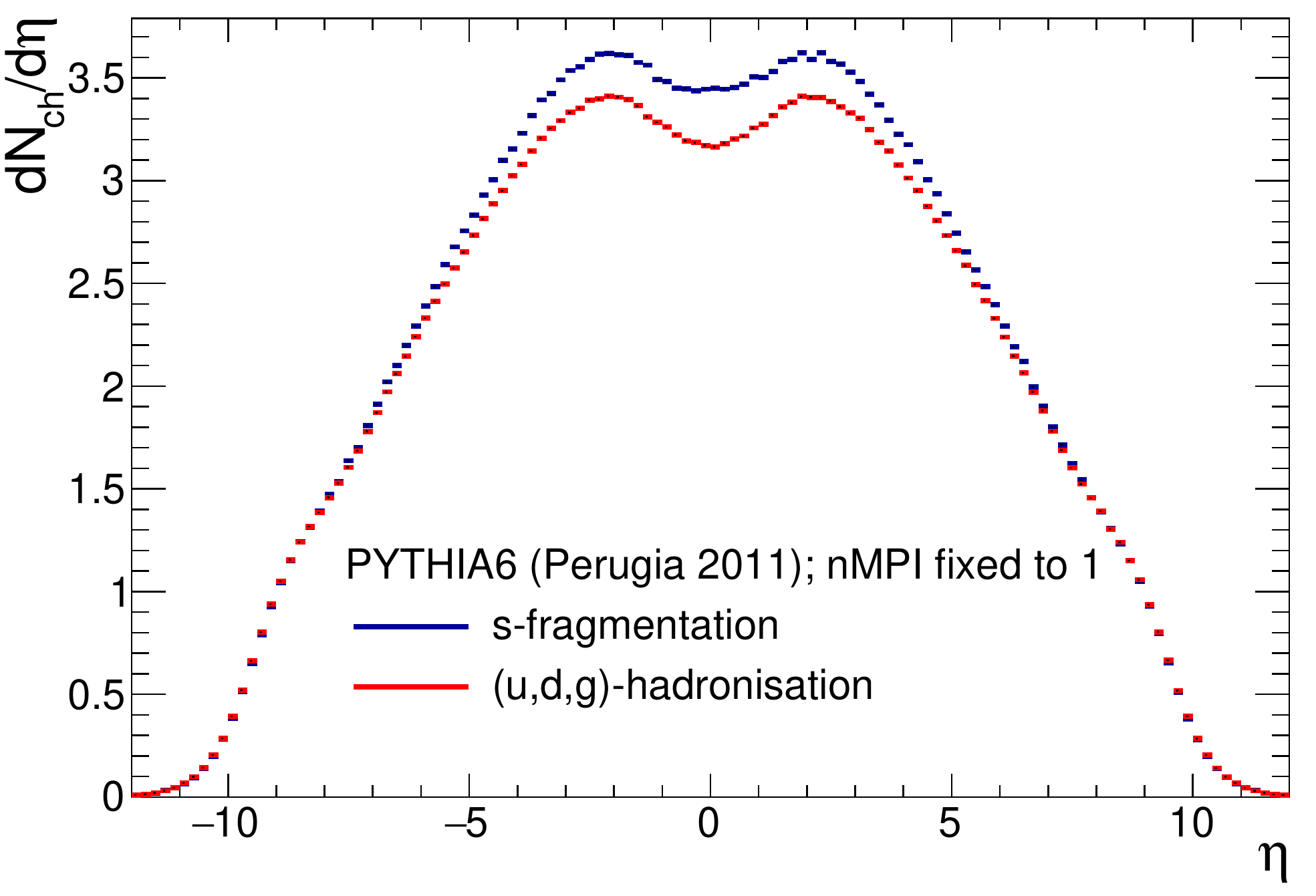}
%\hspace{0.1cm}
\includegraphics[width=0.49\linewidth]{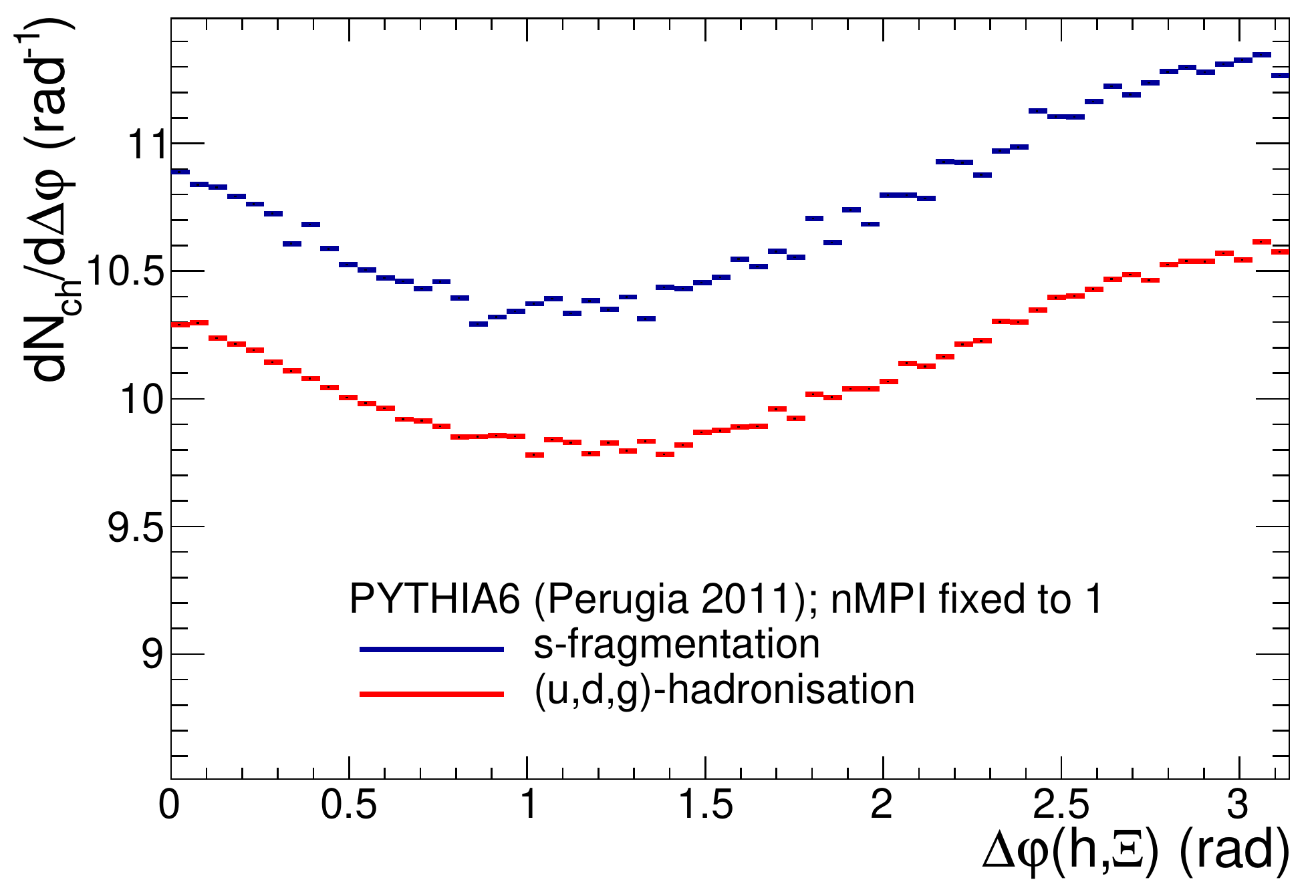}
\caption{Left: For a fixed $\Nmpi = 1$,
the charged particle pseudorapidity density for events with a $\Xi$  ($u,d,g$)-hadronization at midrapidity (blue) is compared to the one for $s$-fragmentation tagged events (red). 
Right: The corresponding azimuth angle difference distribution between charged hadrons and the $\Xi$ for $|\eta| < 5$.
} %macro from macros/plotD.C
\label{fig:OriginOffset_etaphi}
\end{center}
\end{figure}

iii)~In order to remove the centrality bias we compare in the left panel of \Fig{fig:OriginOffset_etaphi}, for a fixed $\Nmpi = 1$, the charged particle pseudorapidity density for events with a $\Xi$ from ($u,d,g$)-hadronization at midrapidity to the one for $s$-fragmentation tagged events. For $s$-fragmentation the density is about 0.2 units higher. 
The difference is also reflected in the azimuthal angle distribution between charged hadrons and the $\Xi$ baryon with a slightly more pronounced back-to-back structure for $s$-fragmentation~(\Fig{fig:OriginOffset_etaphi}(right)).

In summary, motivated by the importance of modelling the pp impact parameter dependency of \MPIs\ for the understanding of not only the multiplicity dispersion, underlying event properties and multiplicity dependence of hard processes in \pp\ but also the centrality dependence of hard processes in \pPb\ and peripheral \PbPb\ collisions, we investigated whether strangeness enhancement in \pp\ collisions at the LHC can be understood within the original \MPI\ picture implemented in the \Pythia\ event generator.
We started by showing that the measured yields can be parameterized by straight-line fits, where the intercept increases with increasing strangeness and mass of the particle~(\Fig{fig:strangedata}).
Normalizing the strange-particle yields by the charged-particle multiplicity leads to a hyperbolic decrease with multiplicity towards the  intercept~(\Fig{fig:strangedatanorm}), which in the original \MPI\ picture is naturally explained as a suppression of strange-particle yields at low multiplicity that is reduced for higher multiplicities. 
By comparing \pytstr\ calculations with data down to low $\pT$~(\Fig{fig:cmscomparison}) and in intervals of multiplicity~(\Fig{fig:multdep}), we argued that the strangeness enhancement in \Pythia\ is masked by a large excess of soft baryons, which largely results from the hadronization of $u$ quark, $d$ quark and gluon strings.
At higher $\pt$, strings formed from parton showers initiated by strange quarks describe the spectral shape of $\Xi$ and $\Omega$ baryons well. 
Furthermore, we demonstrate that the $s$-fragmentation component exhibits the transverse momentum~(\Fig{fig:meanpt}) and yield enhancements~(\Fig{fig:selfnormyields}) seen in the data, including the approximate multiplicity scaling for different collision energies~(\Fig{fig:energyscaling}). This suggests that $s$-quark fragmentation plays a dominant role for the production of these baryons. However, we cannot exclude that a ($udg$)-hadronization implementation which reproduces the $\pt$-spectra could also describe the enhancement.
The \MPI\ framework exhibits a smooth evolution~(\Fig{fig:flavordep}) from strict-proportional multiplicity scaling~($K_{\rm S}^0$, $\Lambda$ where the $udg$-hadronization component dominates) to linearity~($\Xi$, $\Omega$ where $s$-fragmentation dominates) and on to increasingly non-linear behavior~($c$  and $b$ quark fragmentation as well as high-$\pt$ particles),
and hence provides a unified approach for particle production across all particle species in pp collisions.
In \Pythia, the multiplicity threshold for $s$-fragmentation has two origins:
i)~a bias towards more central collisions compared to yields which also receive contributions from soft processes~(\Fig{fig:OriginOffset_MPI})
ii)~for a fixed centrality (in particular for low-multiplicity peripheral collisions) particle production is correlated with multi-strange baryons~(\Fig{fig:OriginOffset_etaphi}).
The centrality difference originates from the presence of double-diffractive collisions and an important contribution to soft particle production in peripheral collisions.

While all calculations for the paper were made with \pytstr\, the same qualitative and mostly also the same quantitative description was found with
\ifarxiv
\pytstrb~(as shown in the appendix~\ref{app:addfigs}).
In appendix~\ref{app:rope} we additionally show that, as expected by the authors, the current implementation of rope hadronization introduced in \cite{Bierlich:2014xba} does not describe the mean $\pt$. 
Note that this observation is at  variance with the claims in \cite{Nayak:2018xip}.
\else
\pytstrb.
\fi
We stress that while the multiplicity-dependencies of the measured strange-particle yields are described, their absolute yields are significantly underestimated within the original \Pythia\ model. %, like it is for other heavy-flavor baryons~\cite{todo}. 
Further insight in to the interplay of the underlying particle production mechanisms can be expected from two-particle angular correlation measurements between strange and (non-)strange particles.
These and other studies, in particular when they involve the $\Omega$ baryon, will greatly benefit from the large increase in the number of events provided by the $200$/pb \pp\ program planned for Run-3/4 at the LHC~\cite{Citron:2018lsq}.

\begin{figure}[htb!]
\begin{center}
\ifplotpold
\includegraphics[width=0.32\linewidth]{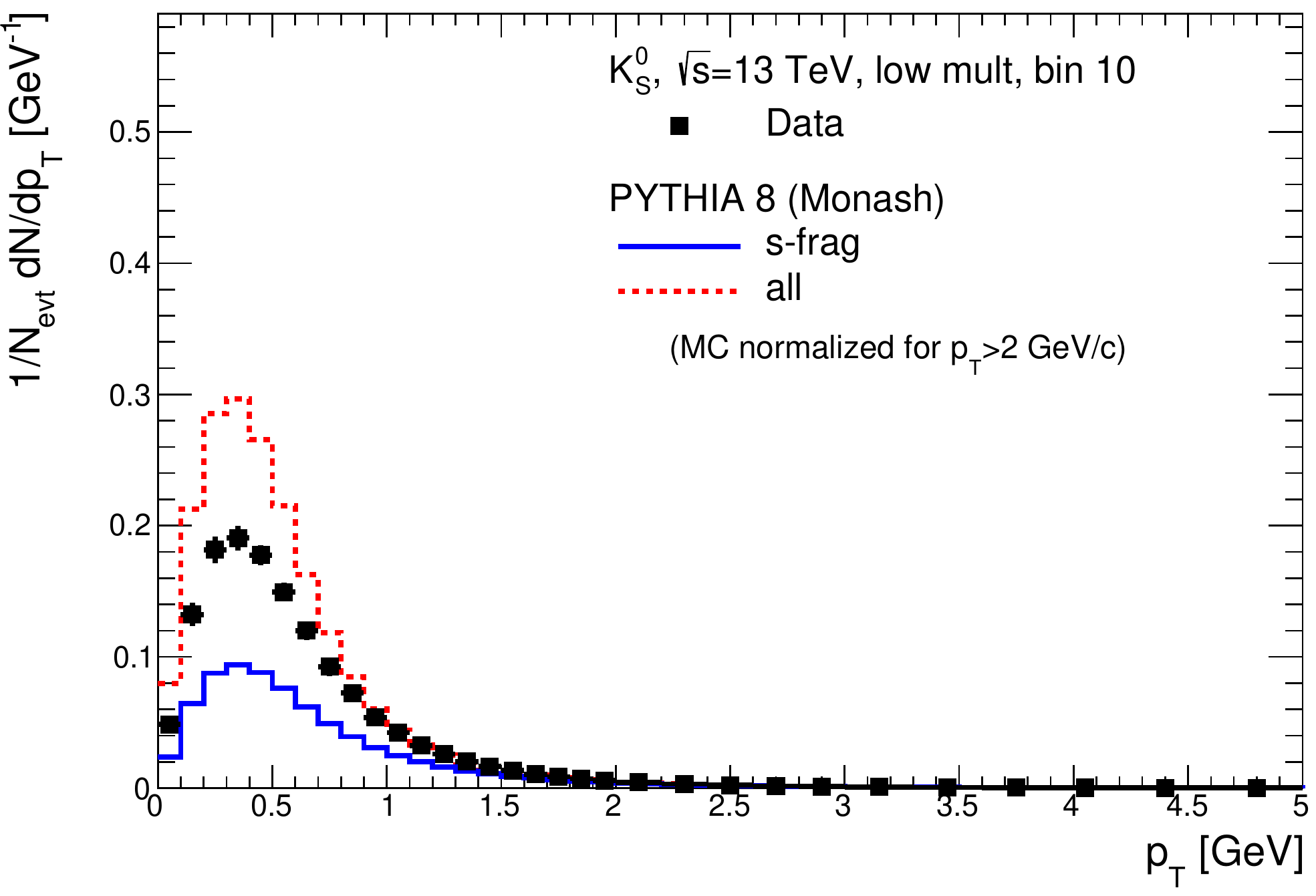}
\includegraphics[width=0.32\linewidth]{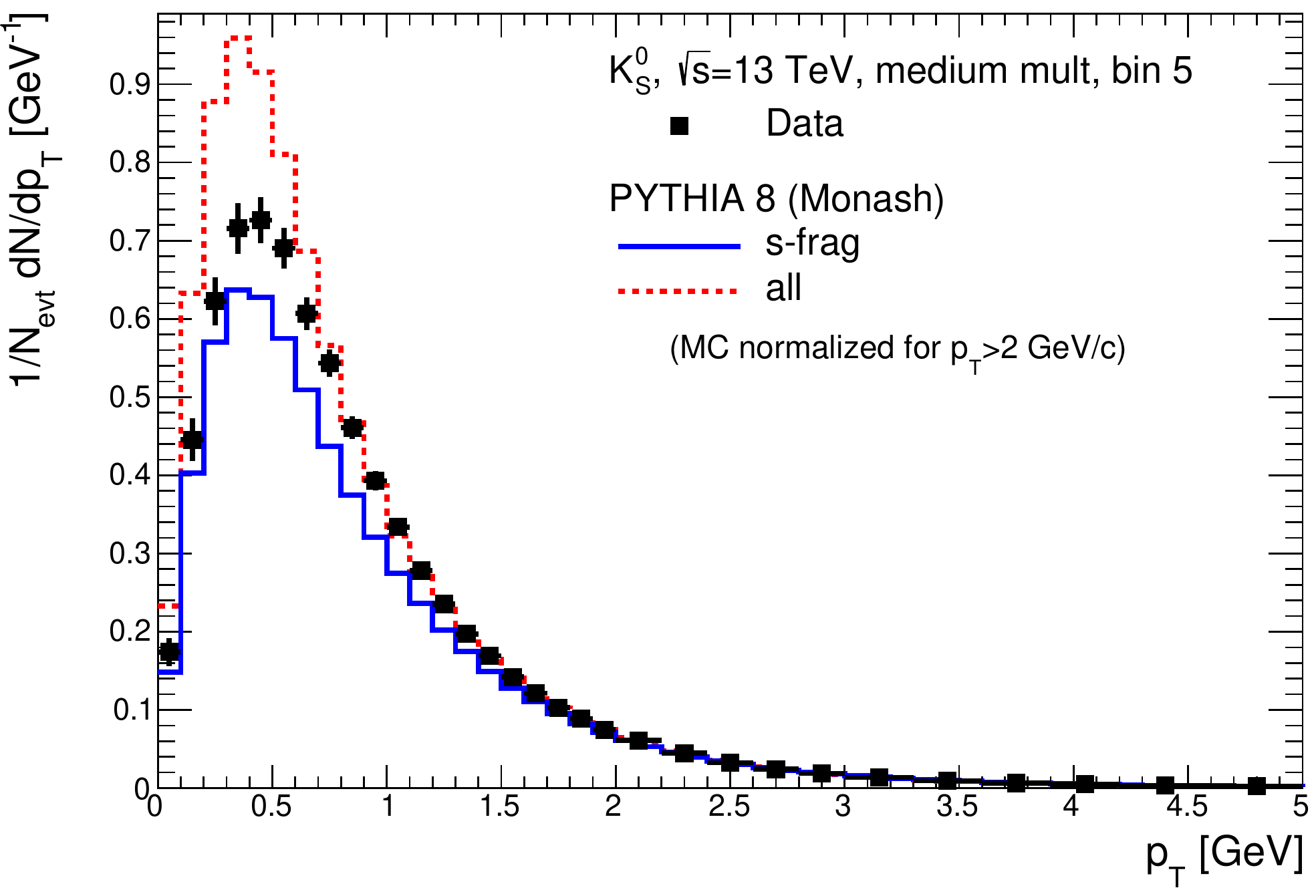}   
\includegraphics[width=0.32\linewidth]{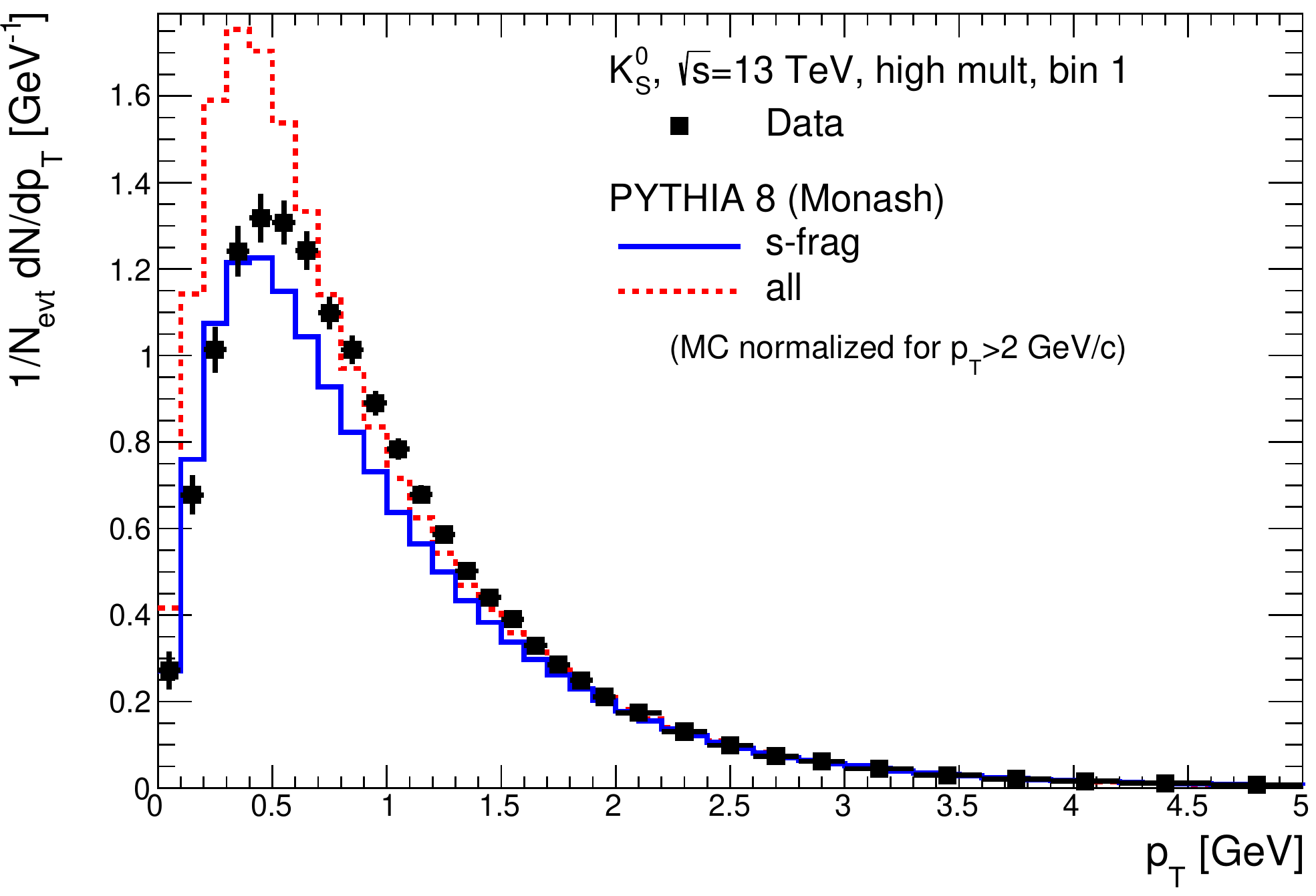}       
\includegraphics[width=0.32\linewidth]{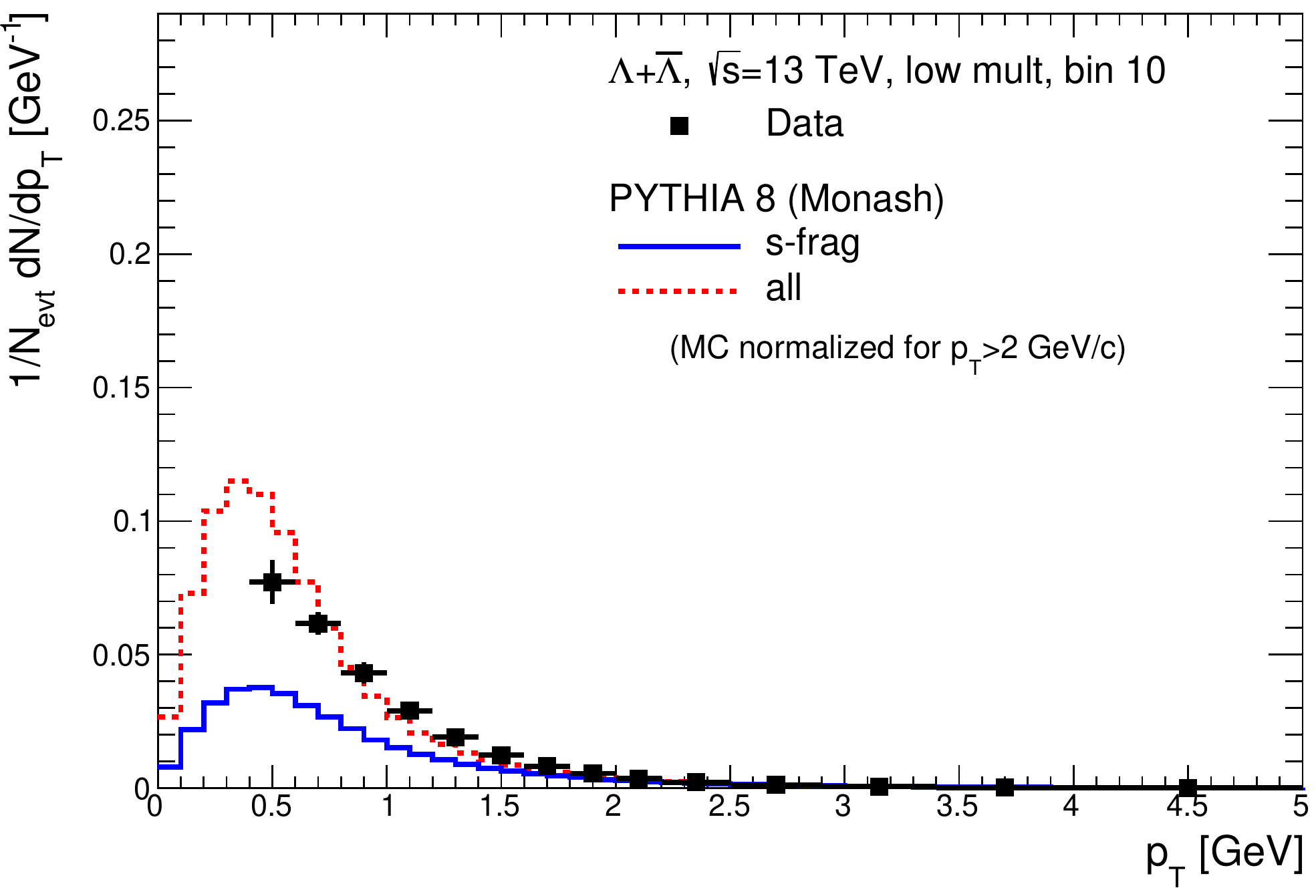}
\includegraphics[width=0.32\linewidth]{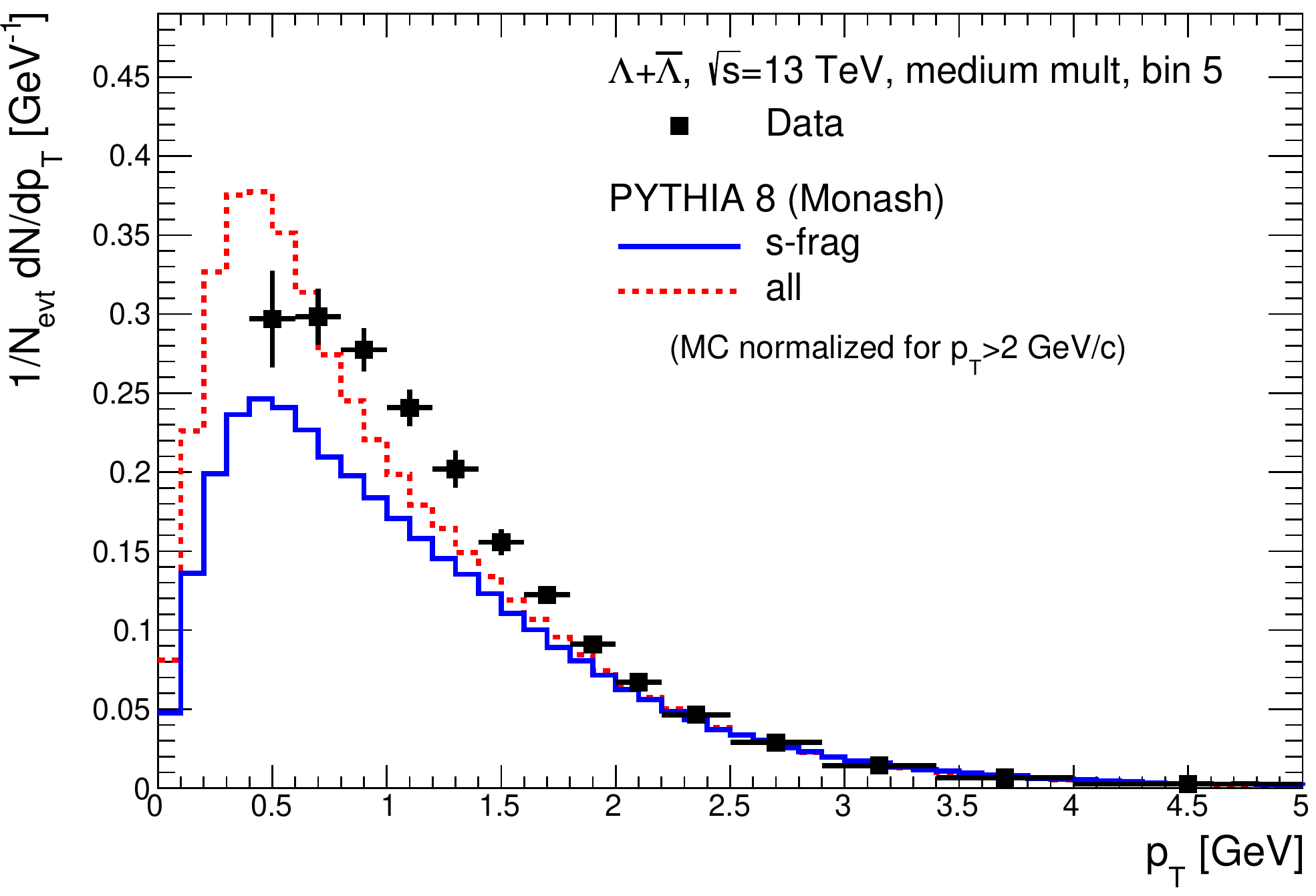}   
\includegraphics[width=0.32\linewidth]{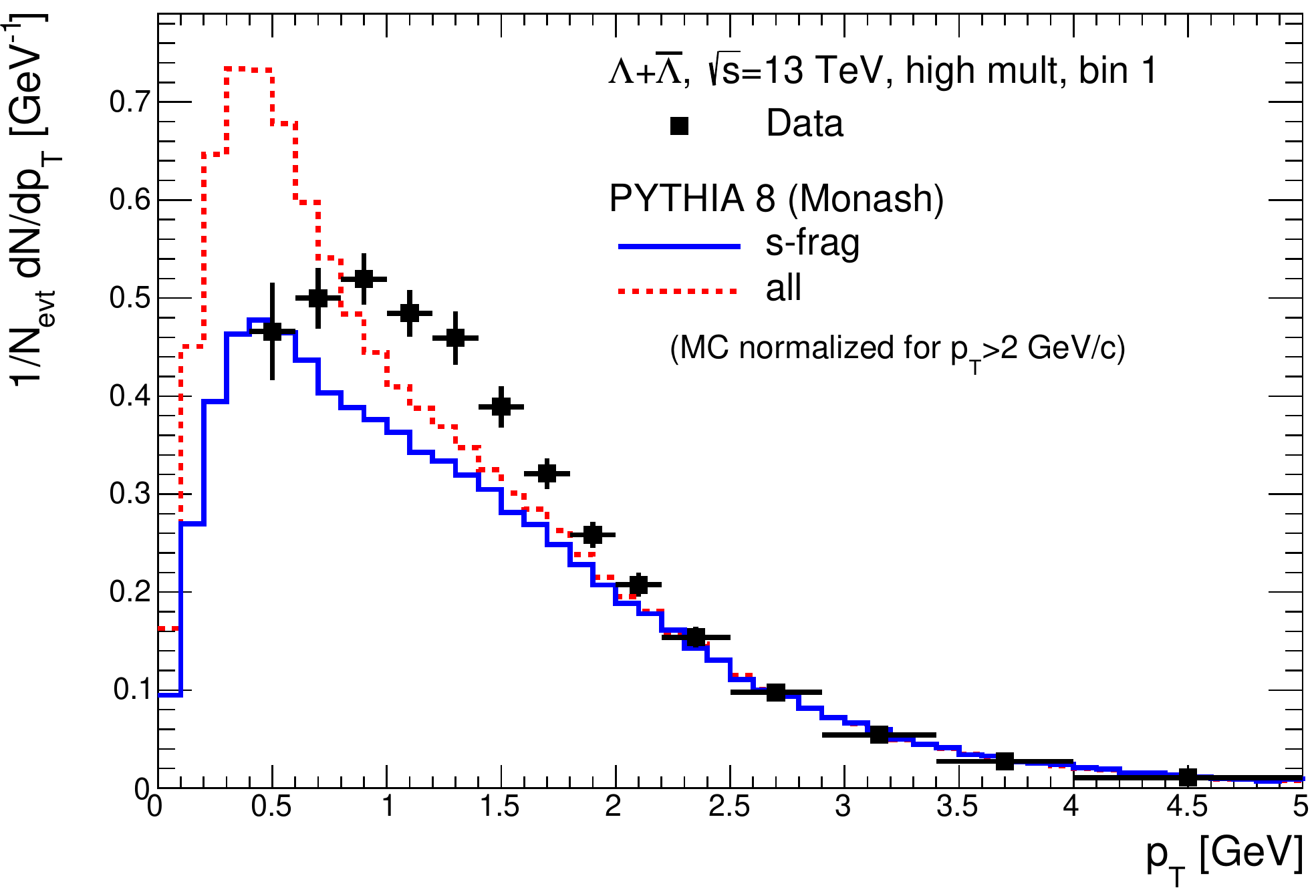}
\includegraphics[width=0.32\linewidth]{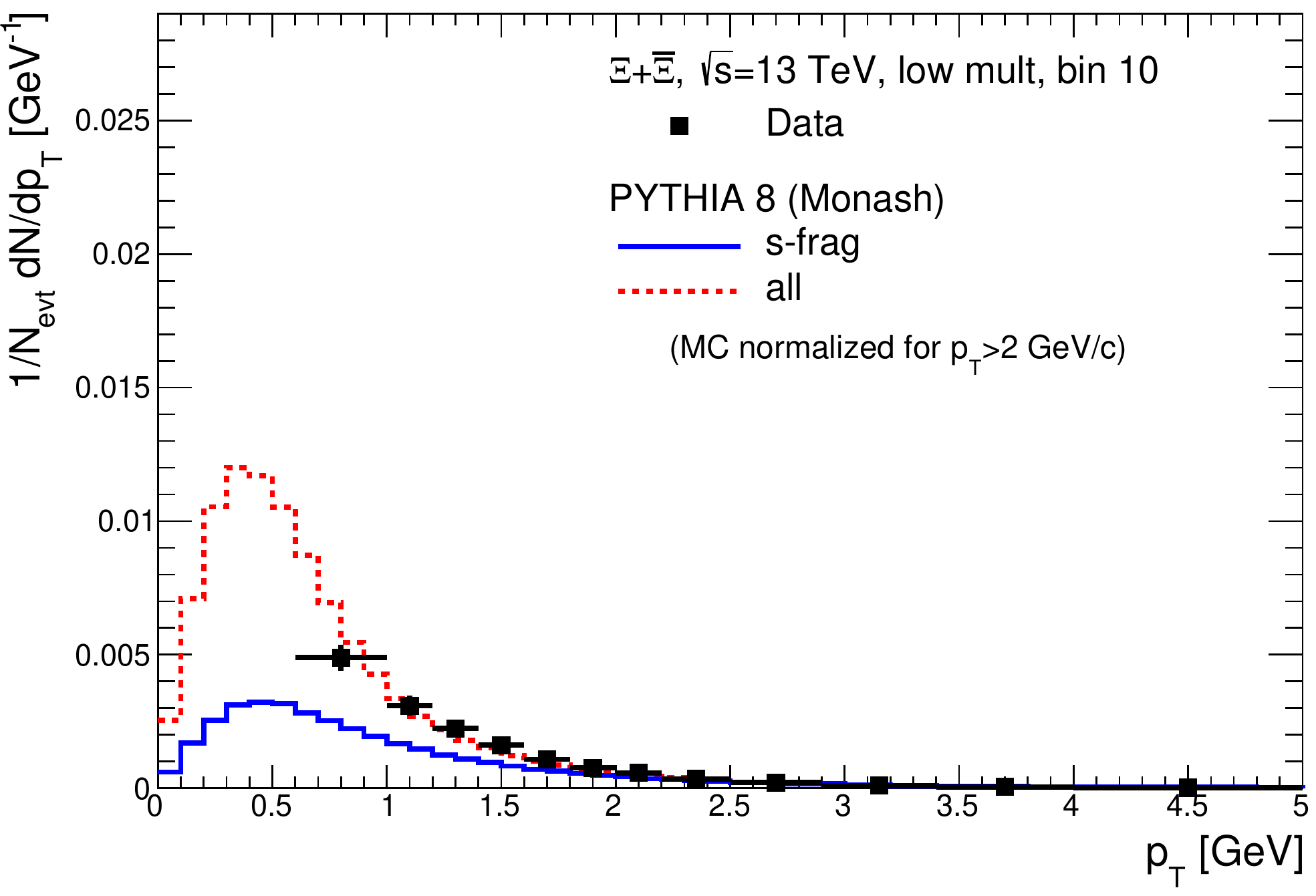}
\includegraphics[width=0.32\linewidth]{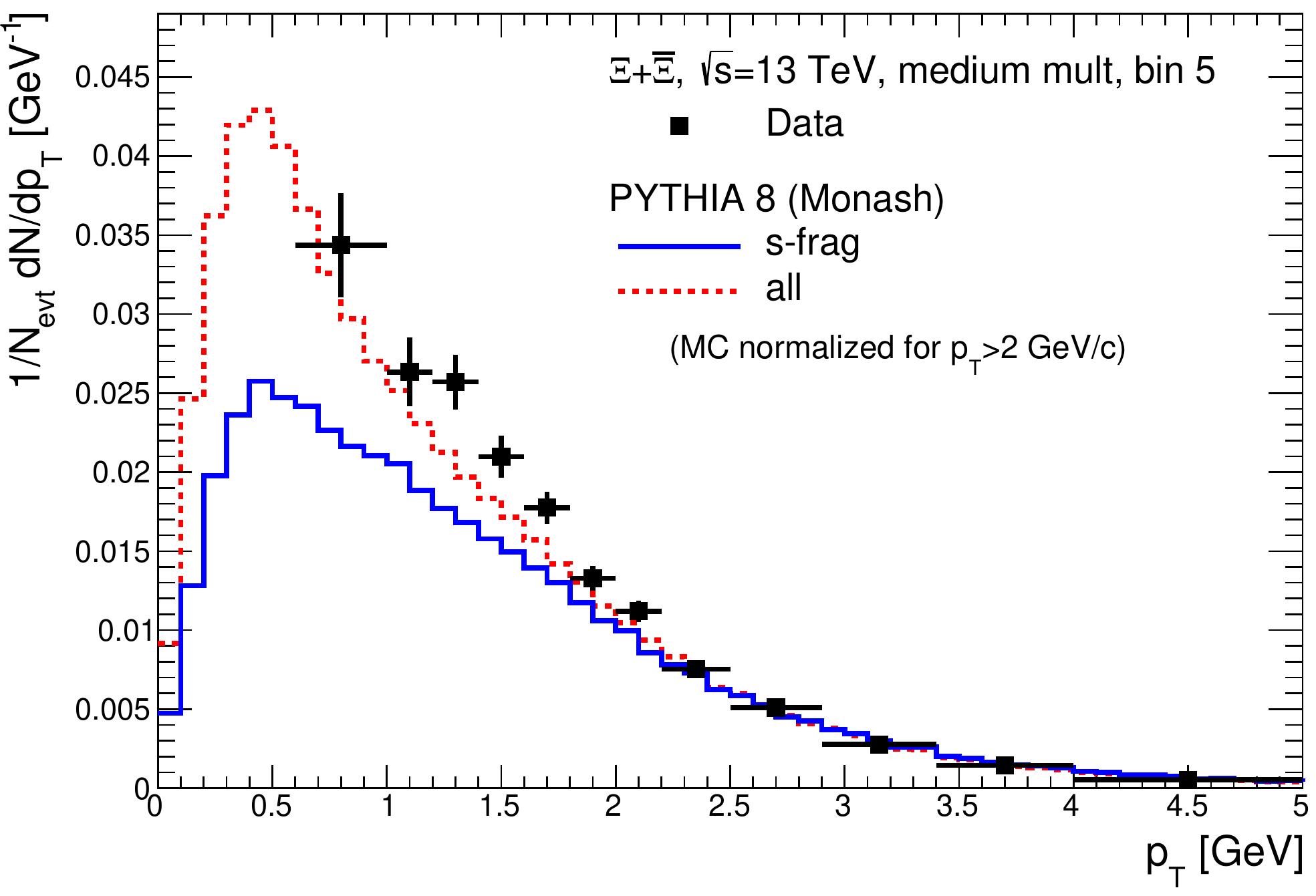}   
\includegraphics[width=0.32\linewidth]{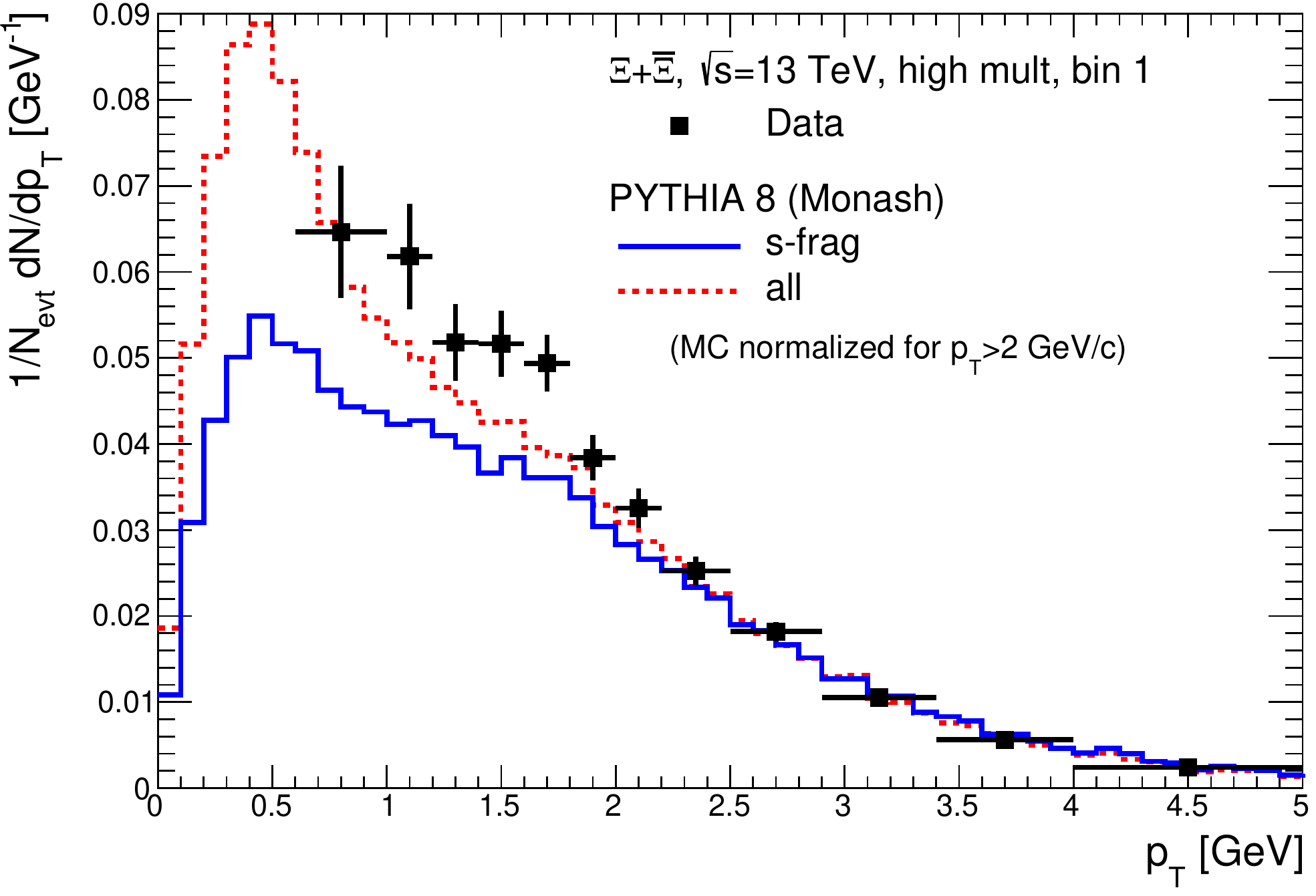}       
\includegraphics[width=0.32\linewidth]{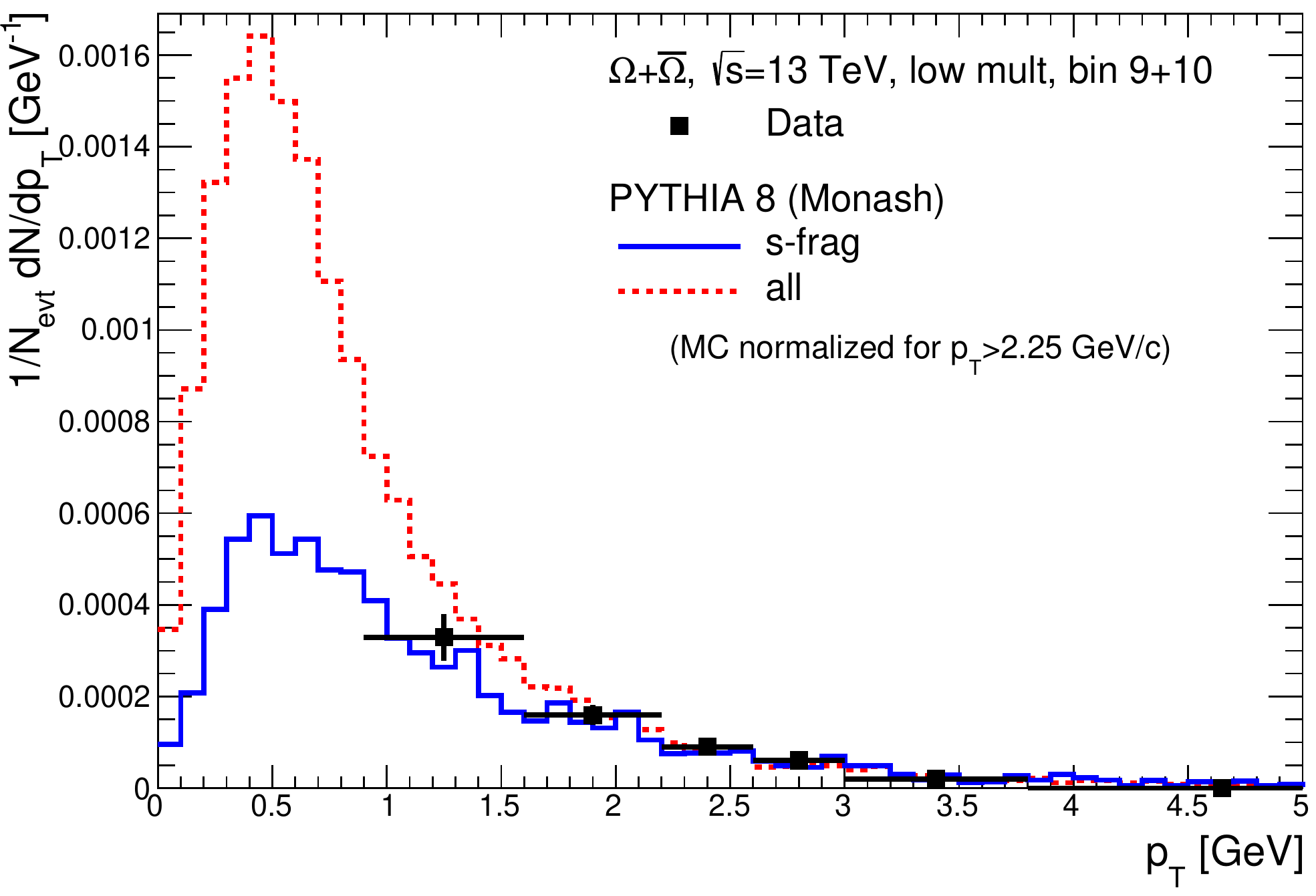}
\includegraphics[width=0.32\linewidth]{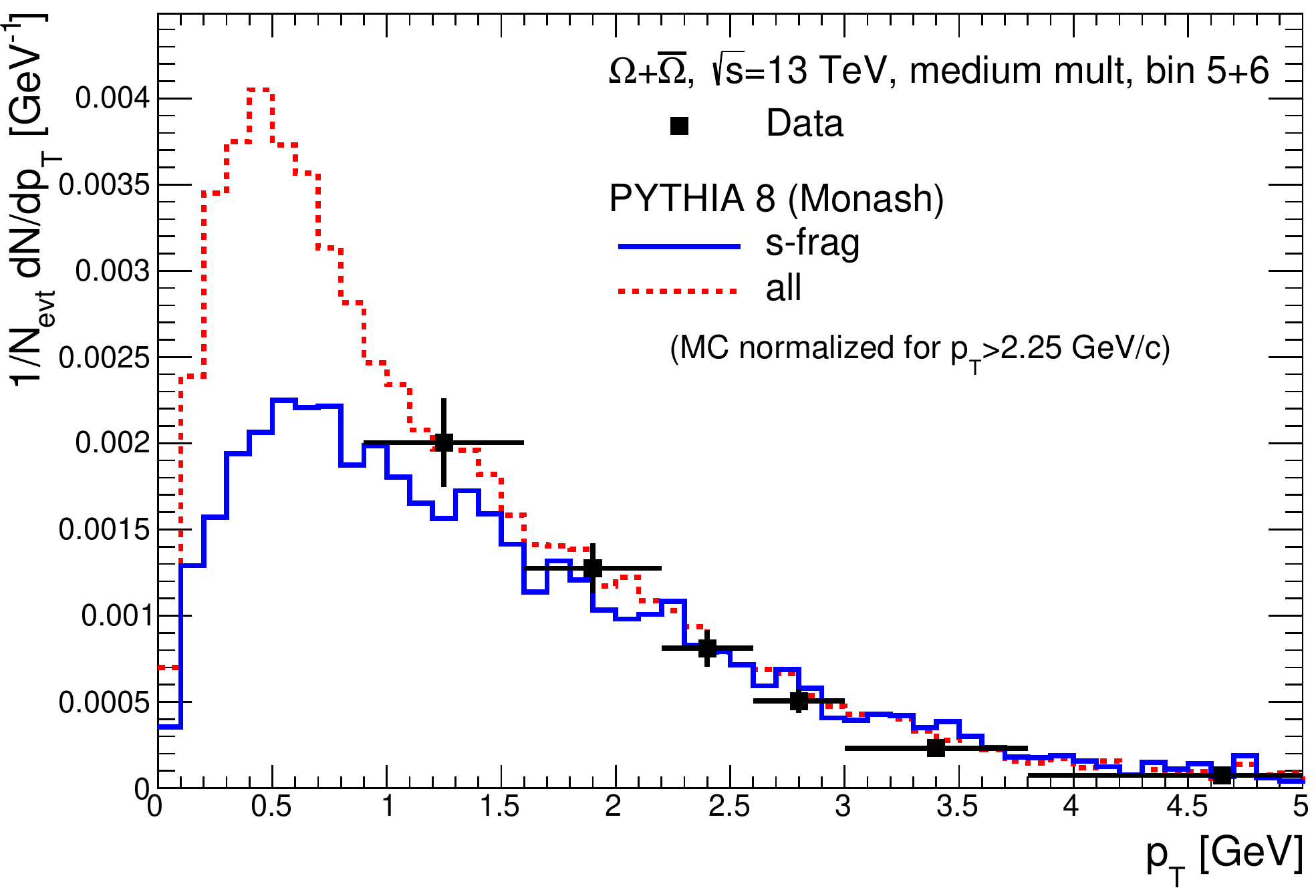}   
\includegraphics[width=0.32\linewidth]{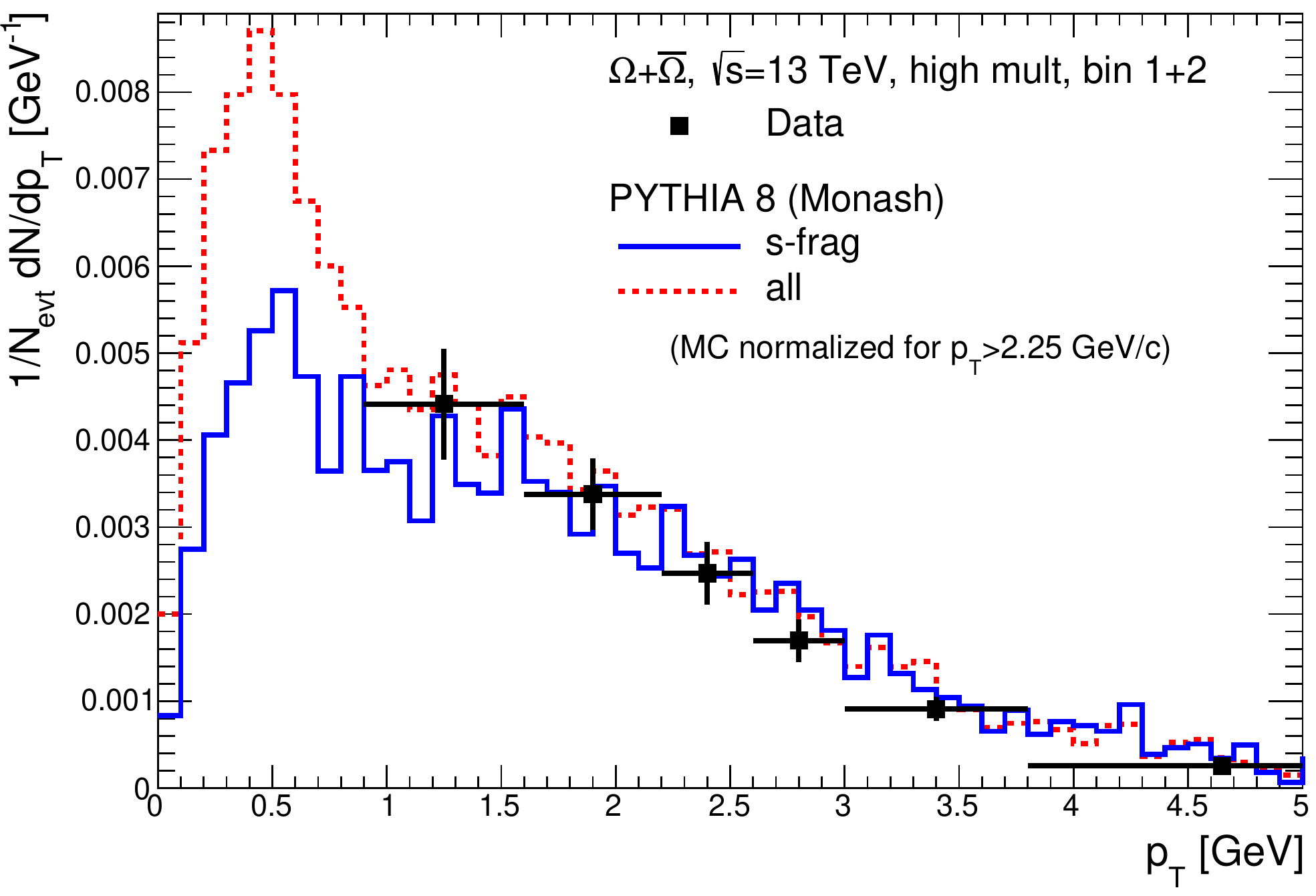}
\else
\includegraphics[width=0.32\linewidth]{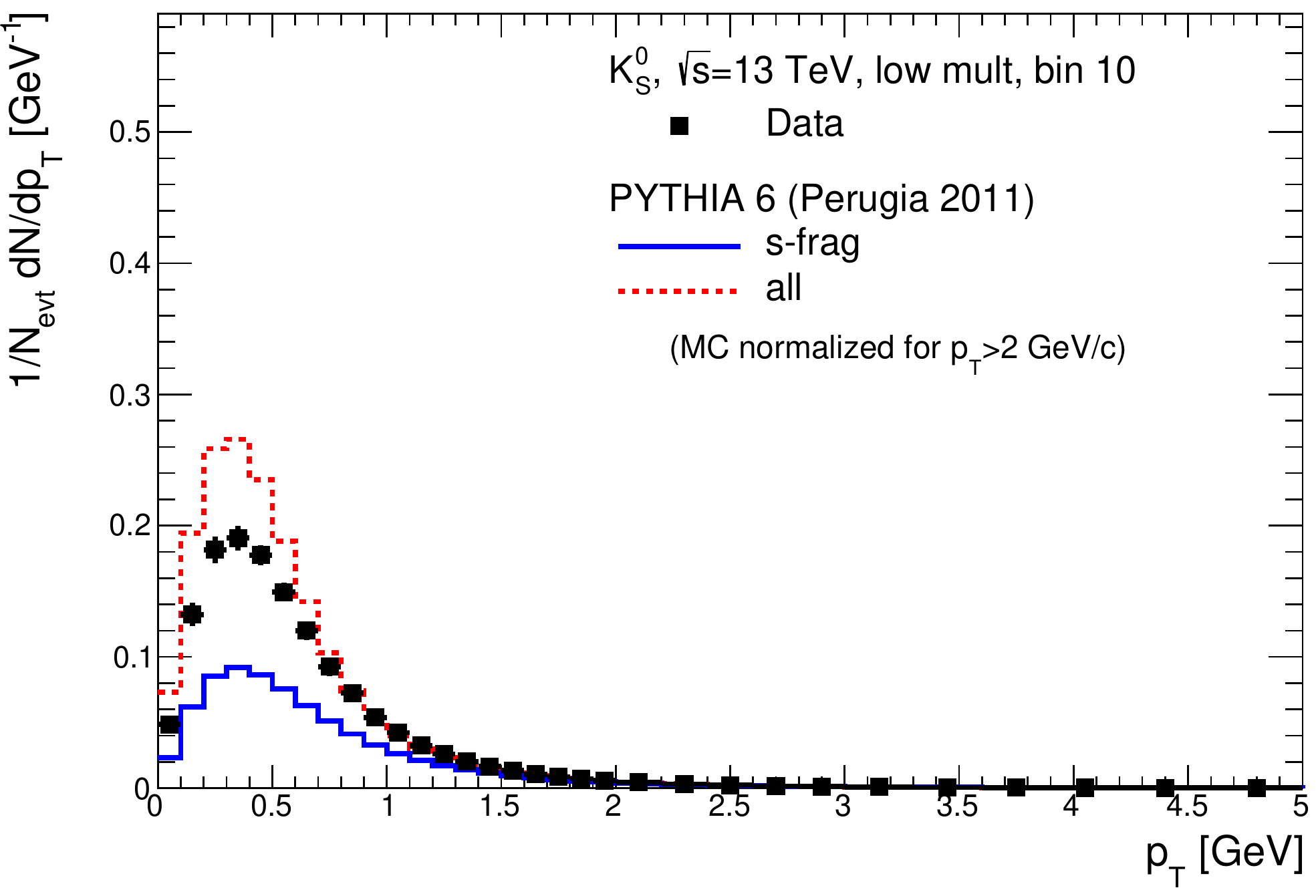}
\includegraphics[width=0.32\linewidth]{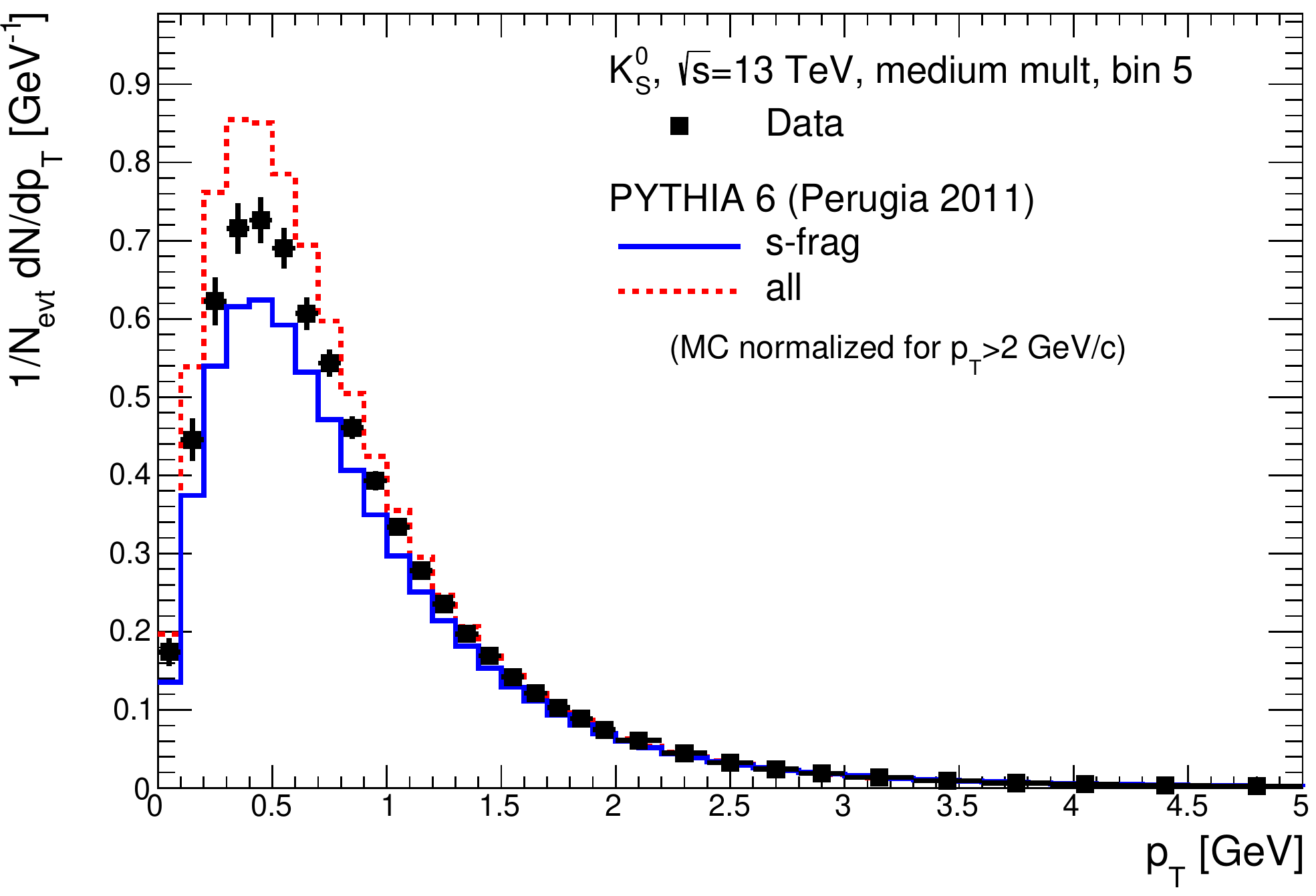}   
\includegraphics[width=0.32\linewidth]{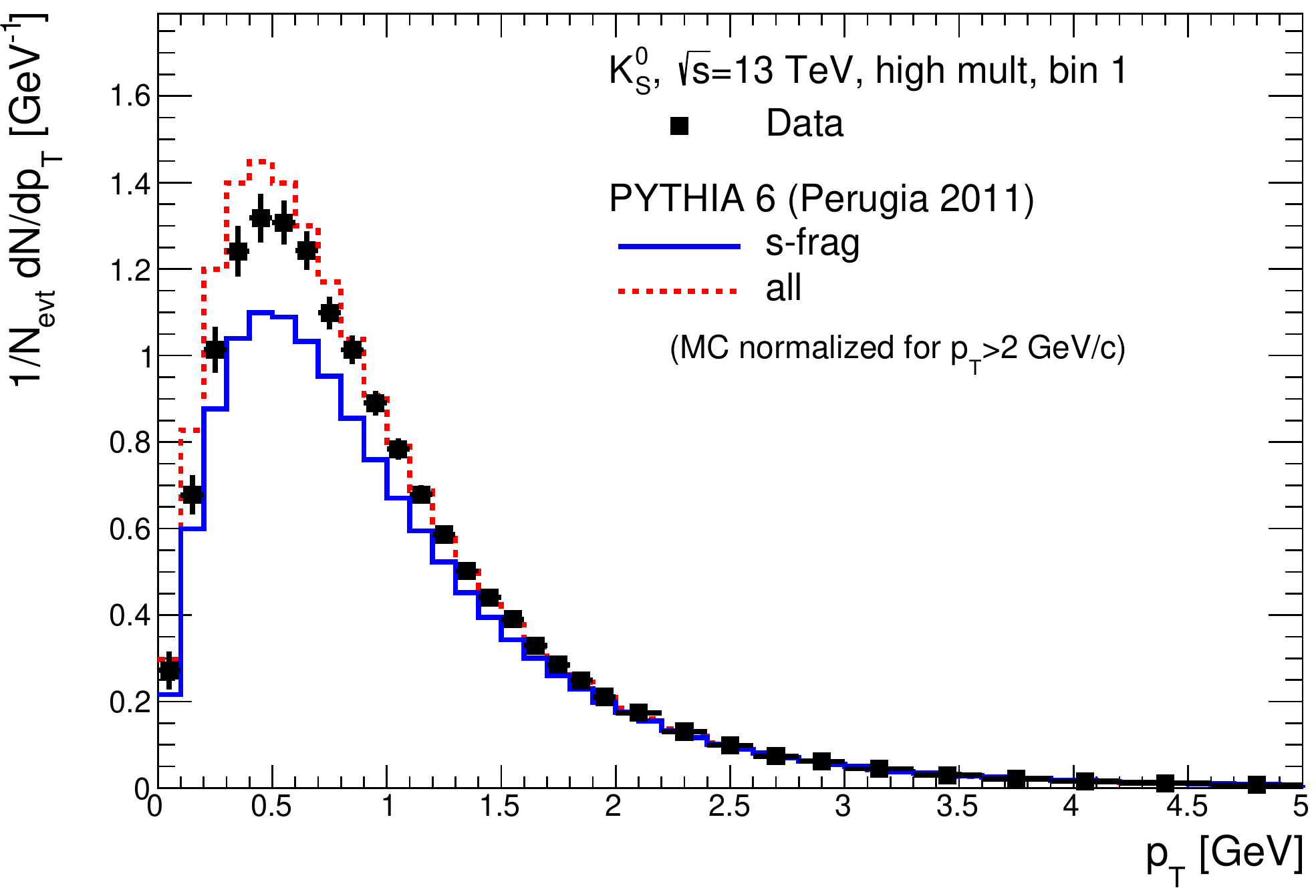}       
\includegraphics[width=0.32\linewidth]{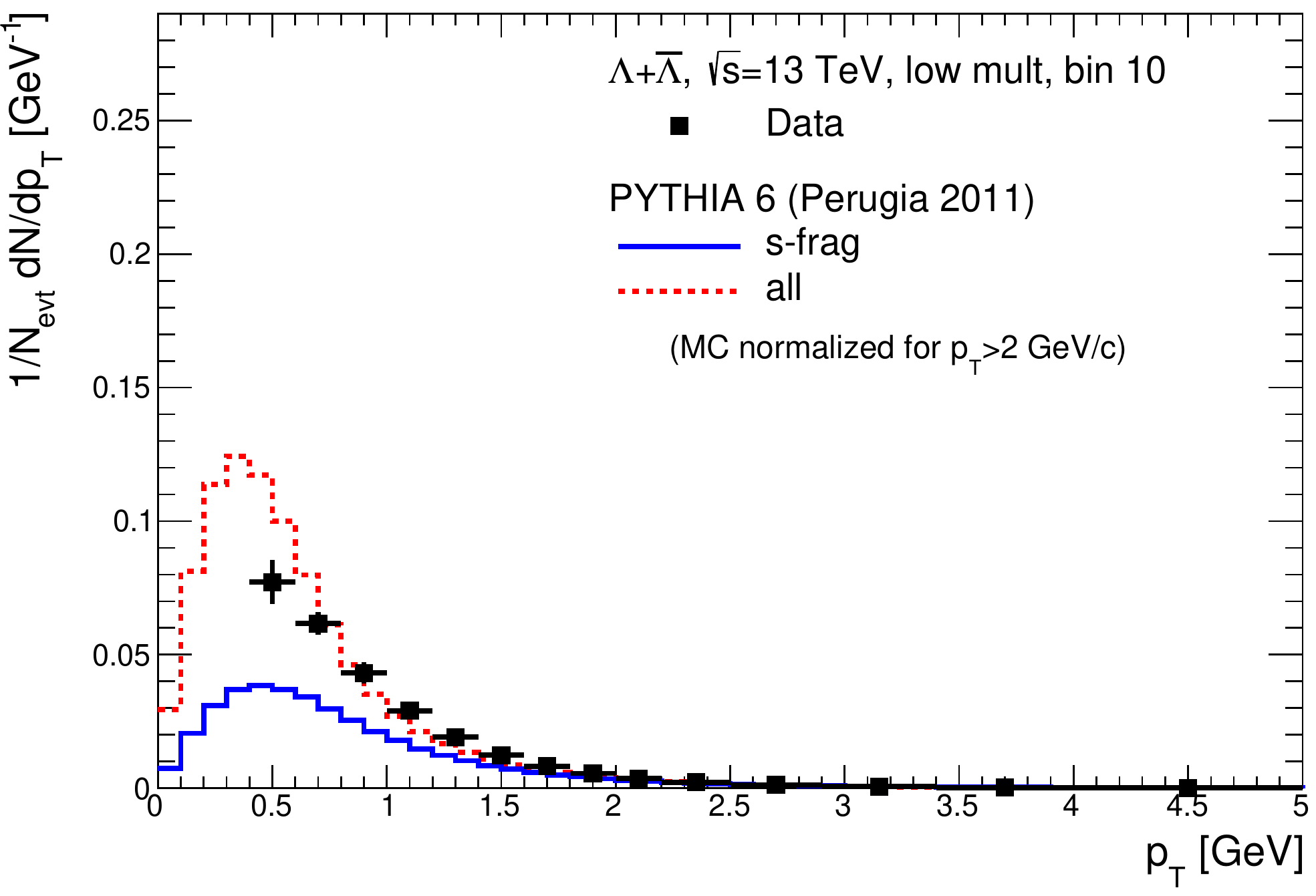}
\includegraphics[width=0.32\linewidth]{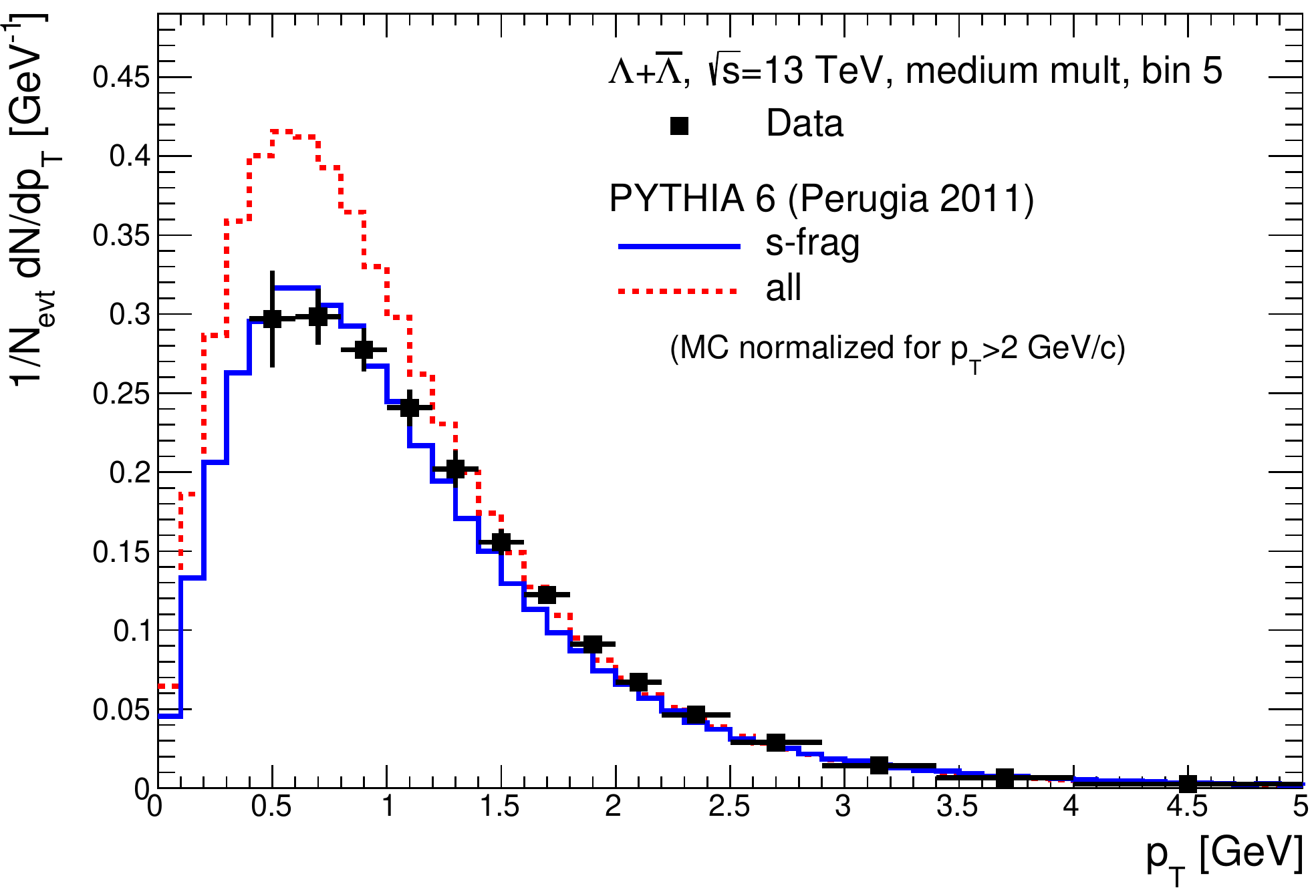}   
\includegraphics[width=0.32\linewidth]{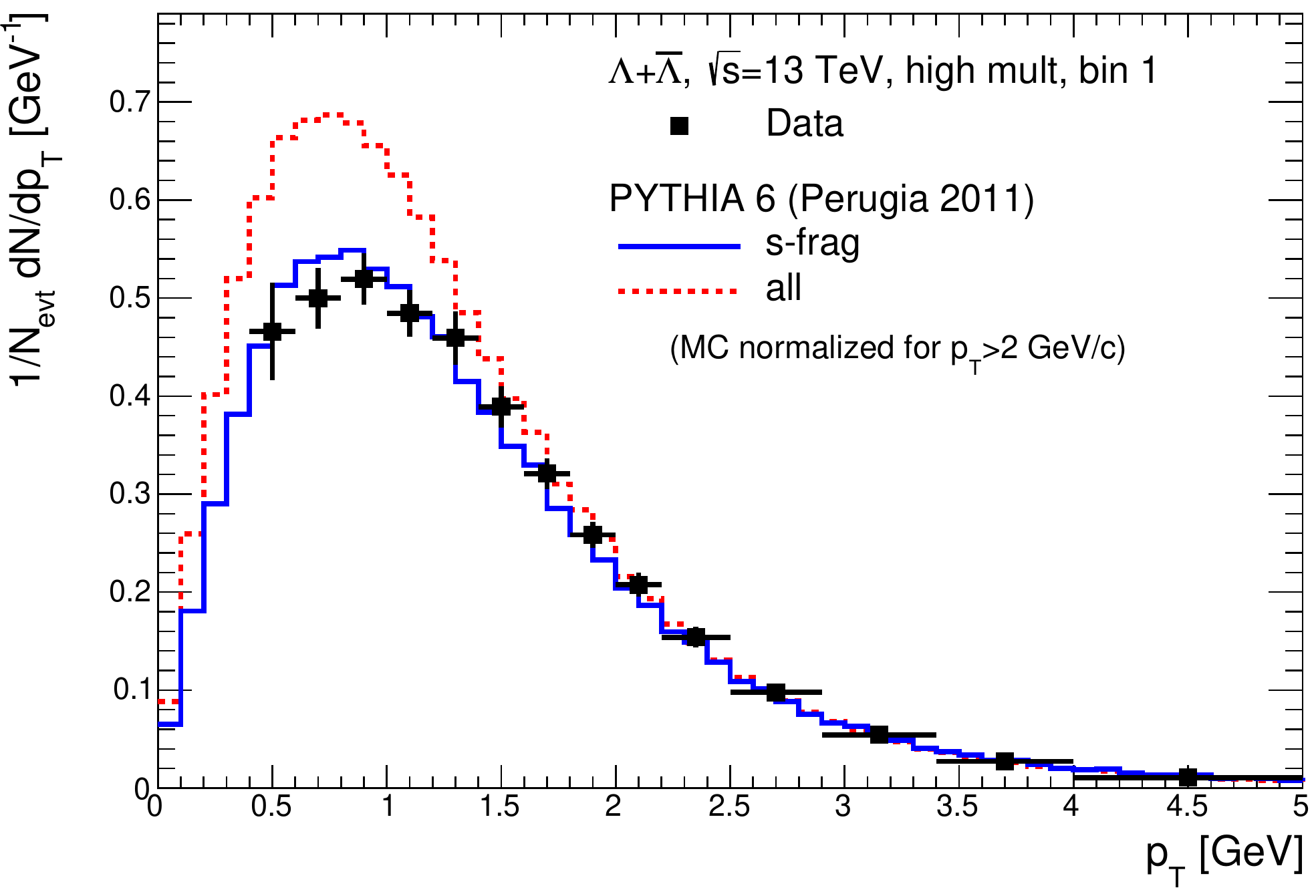}
\includegraphics[width=0.32\linewidth]{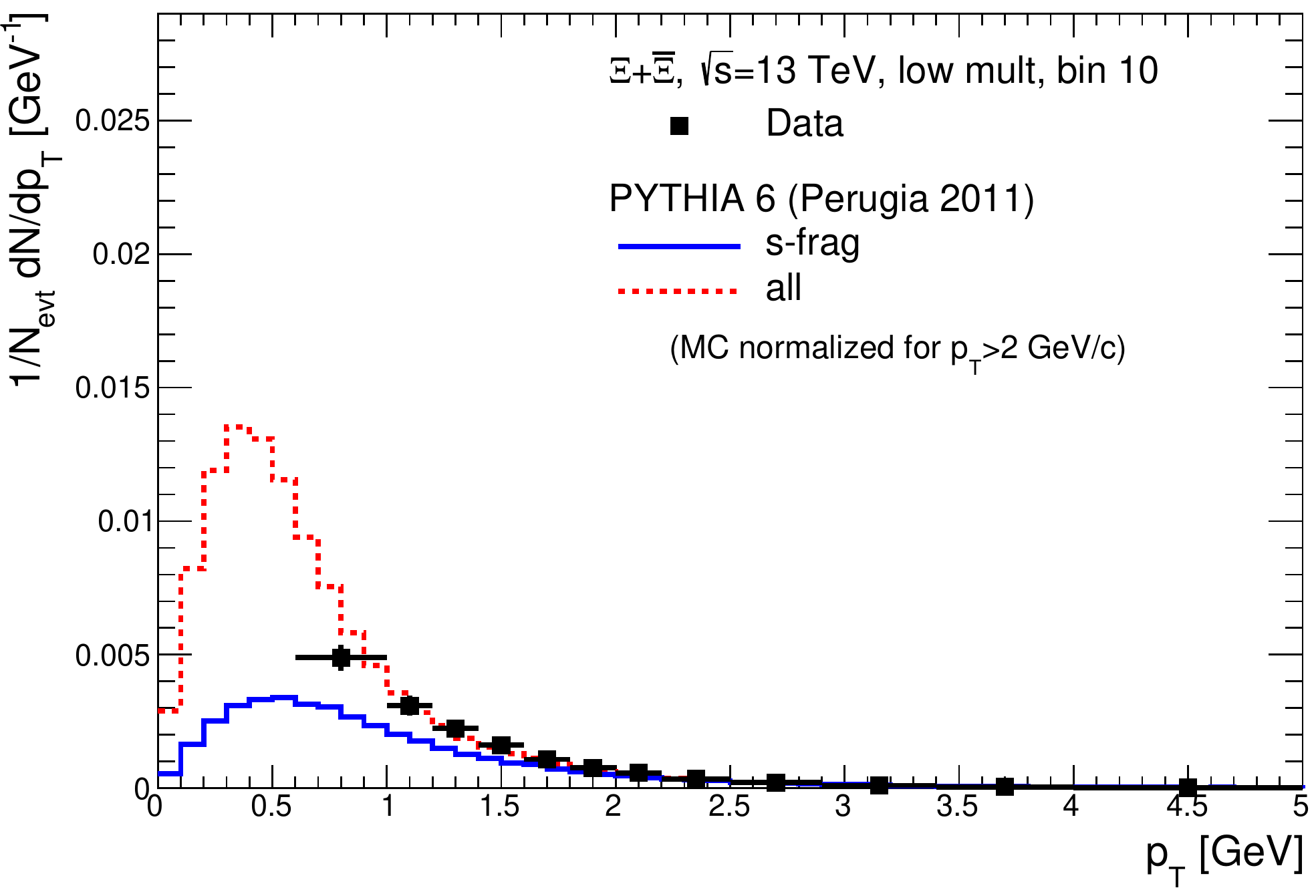}
\includegraphics[width=0.32\linewidth]{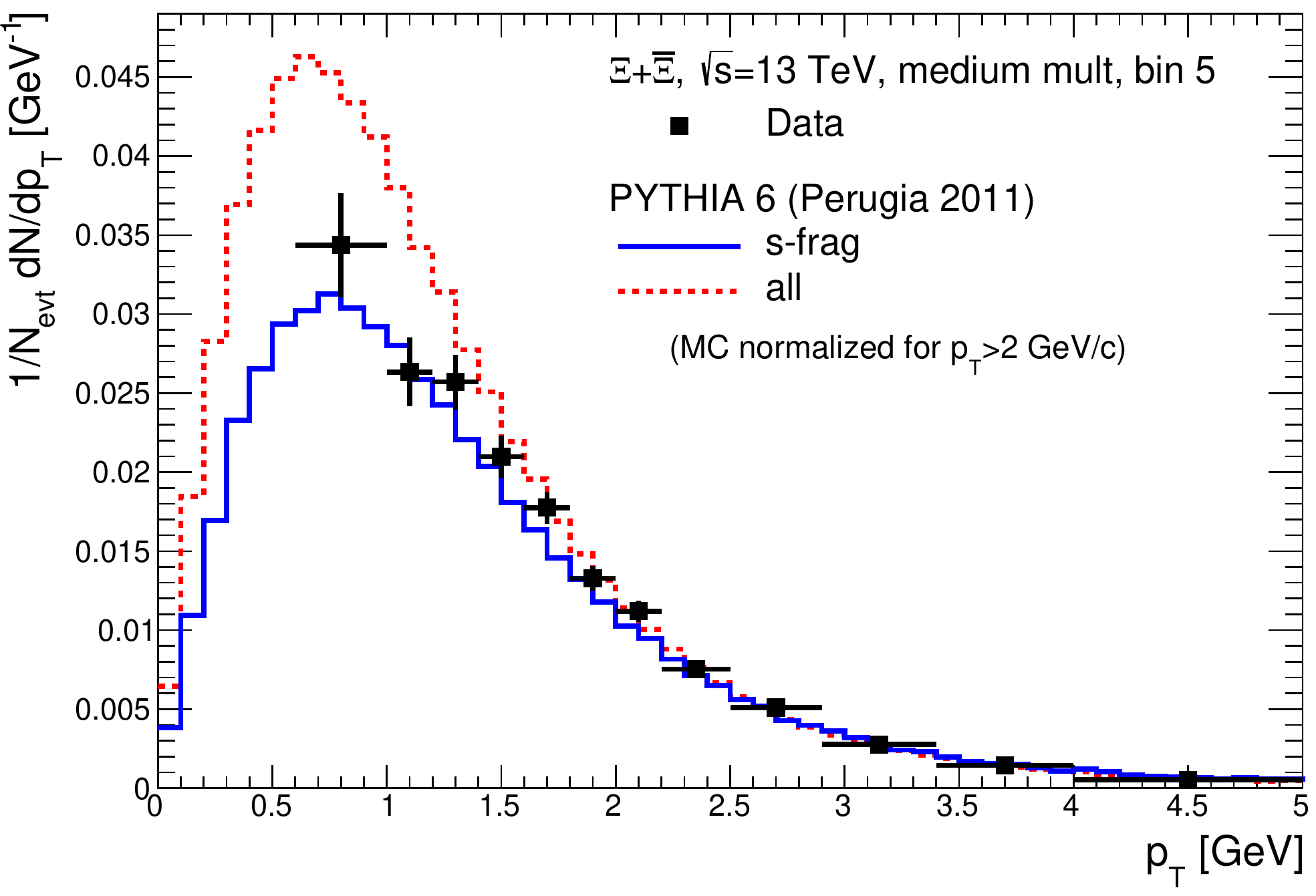}   
\includegraphics[width=0.32\linewidth]{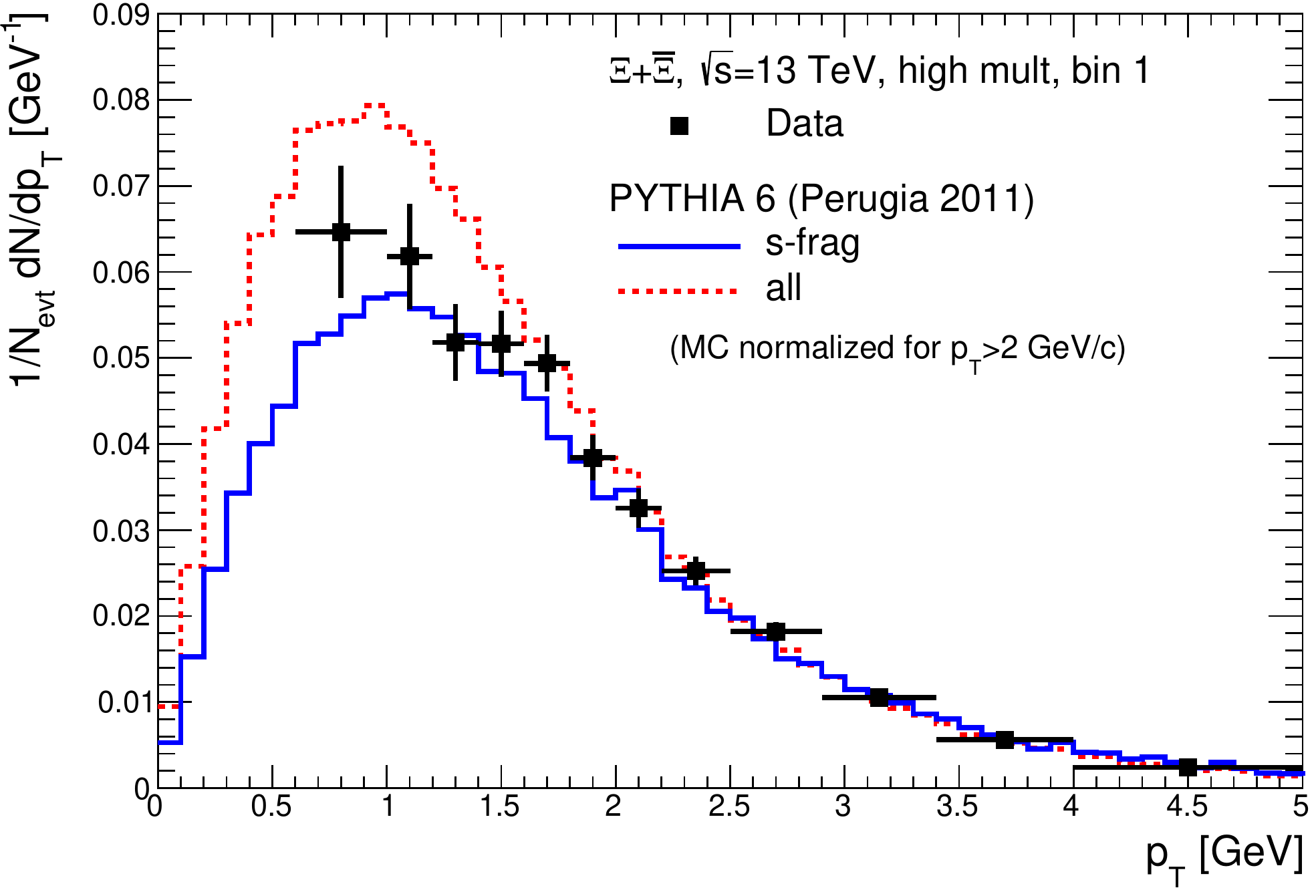}       
\includegraphics[width=0.32\linewidth]{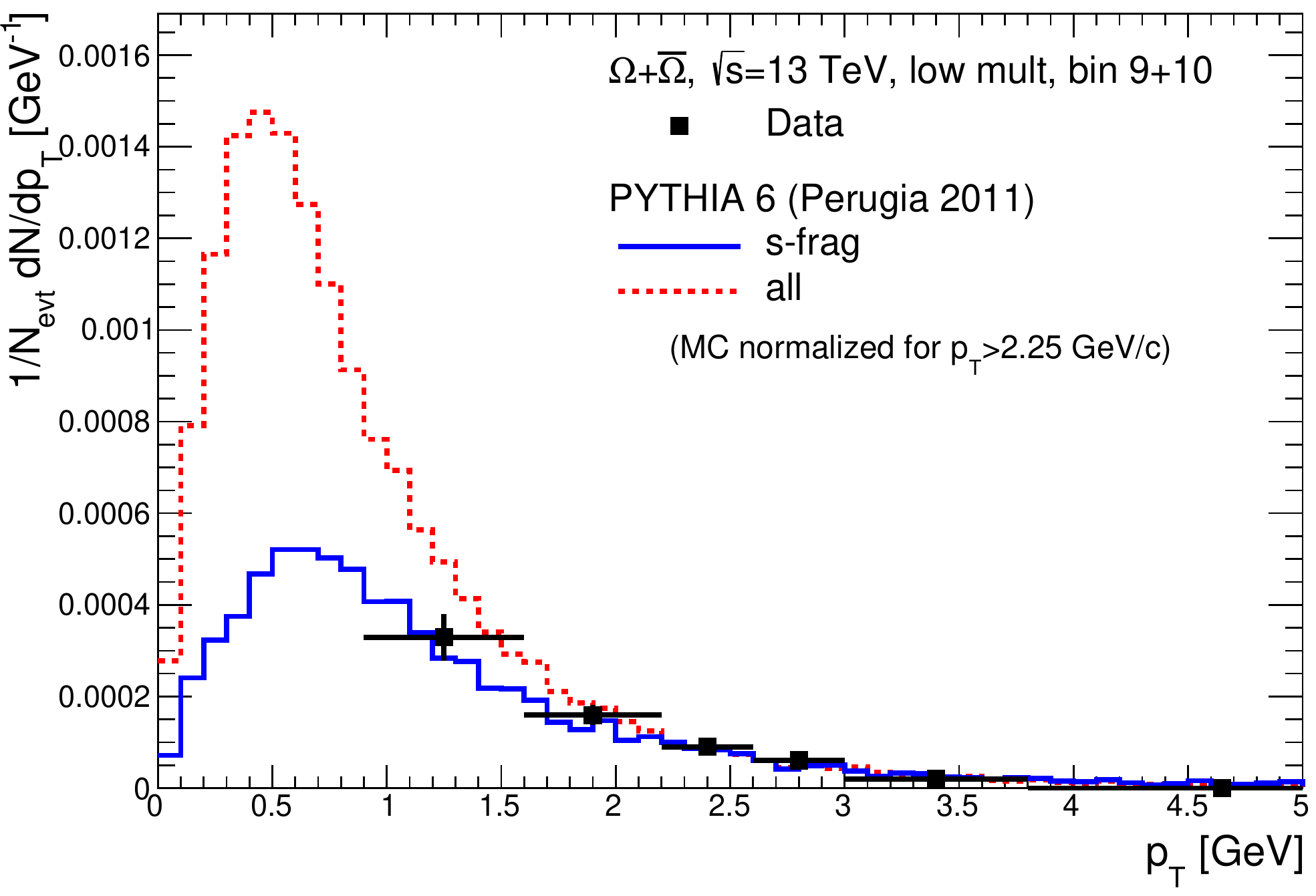}
\includegraphics[width=0.32\linewidth]{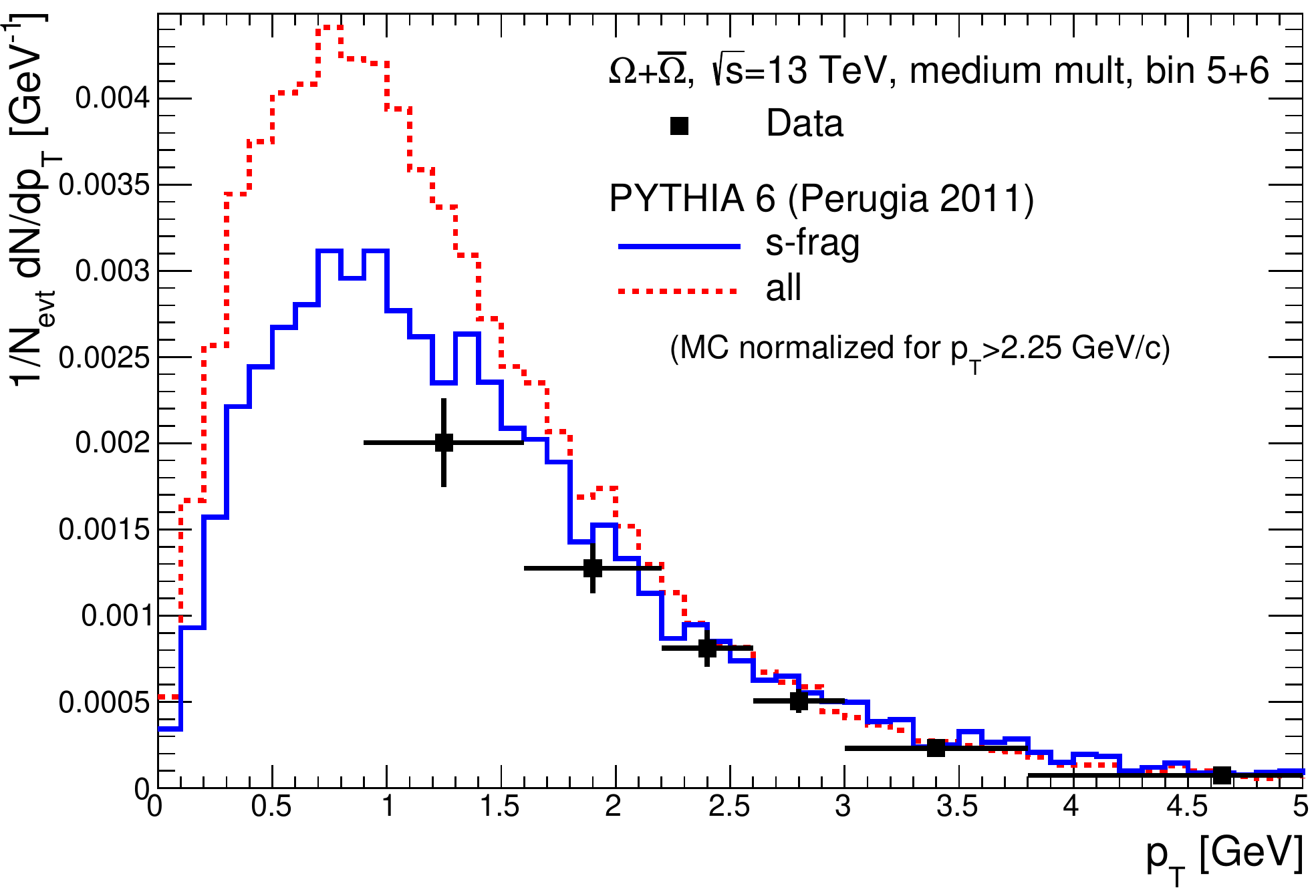}   
\includegraphics[width=0.32\linewidth]{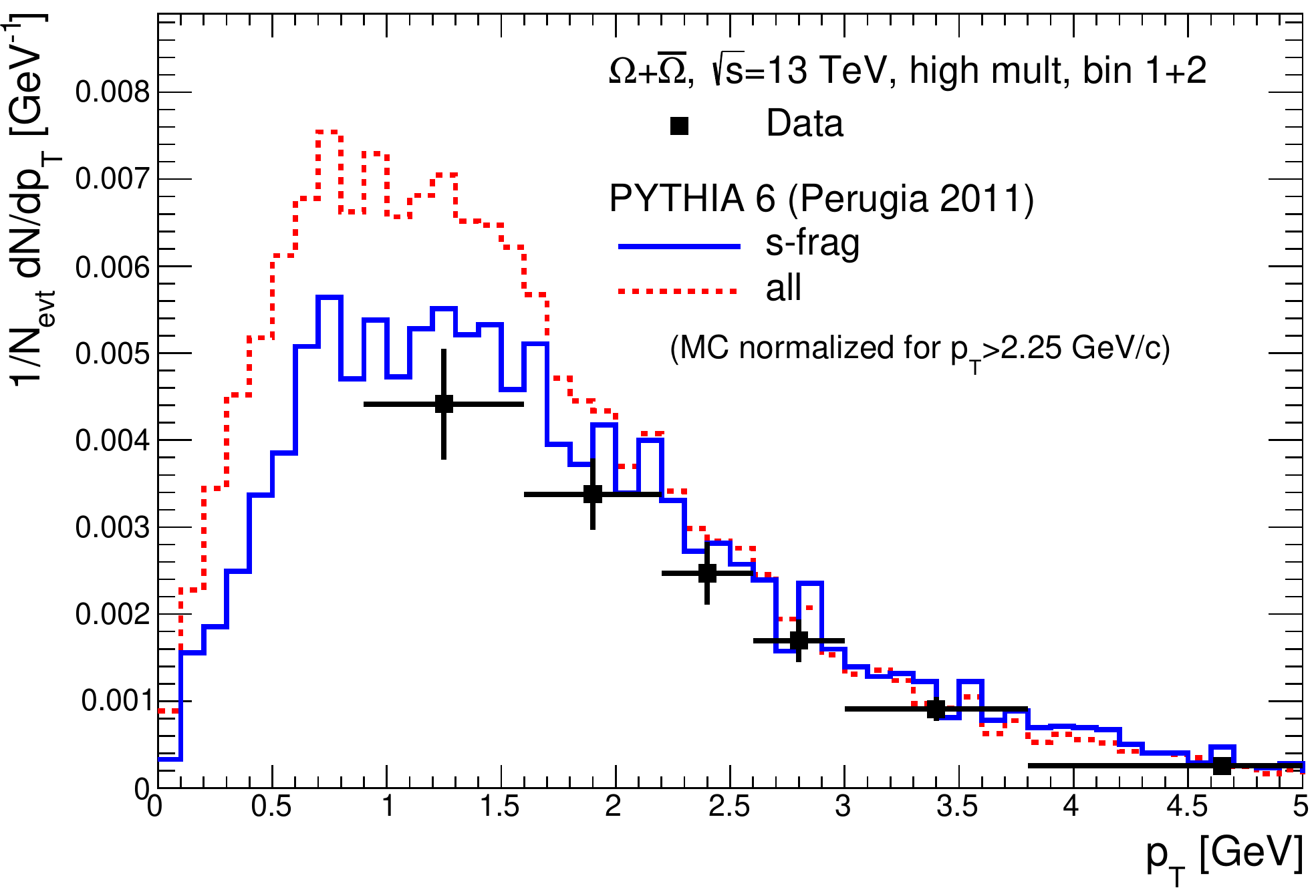}
\fi
\caption{Strange particle $\pt$-spectra~($K_{\rm S}^0$, $\Lambda$, $\Xi$, $\Omega$) measured in low~(left panels), medium~(middle panels) and high~(right panels) multiplicity \pp\ collisions at $\s=13$ TeV~\cite{Acharya:2019kyh} compared with \pytstrb\ calculations. The calculated distributions, which show all produced particles of a given type, as well as those produced by $s$-quark fragmentation alone, are normalized to the data in the region $\pt\gtrsim2$~GeV/$c$. See \Fig{fig:multdep} for the corresponding \pytstr\ calculations.
} 
\label{fig:multdep_app}
\end{center}
\end{figure}

\begin{figure}[htb!]
\begin{center}
\ifplotpold
\includegraphics[width=0.75\linewidth]{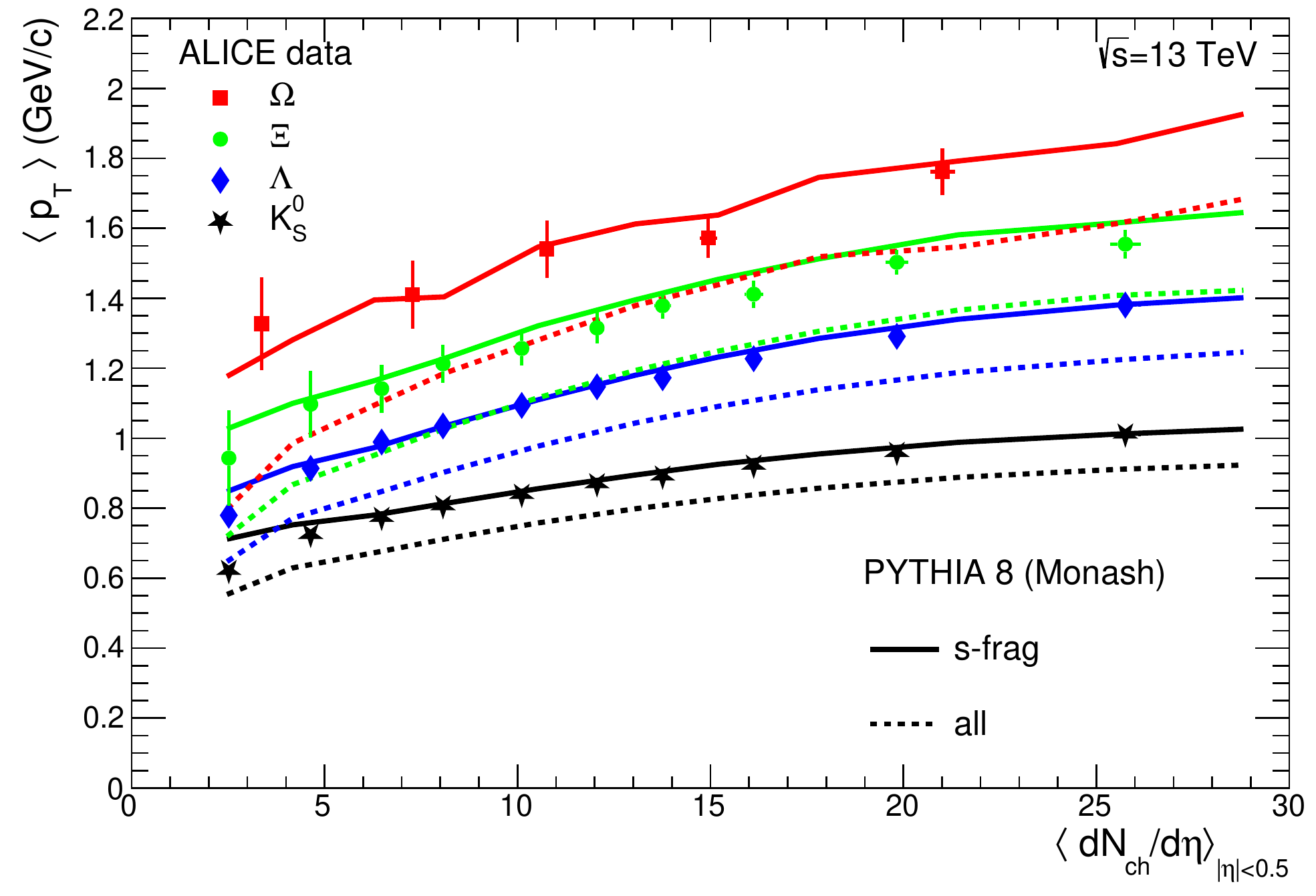}
\else
\includegraphics[width=0.75\linewidth]{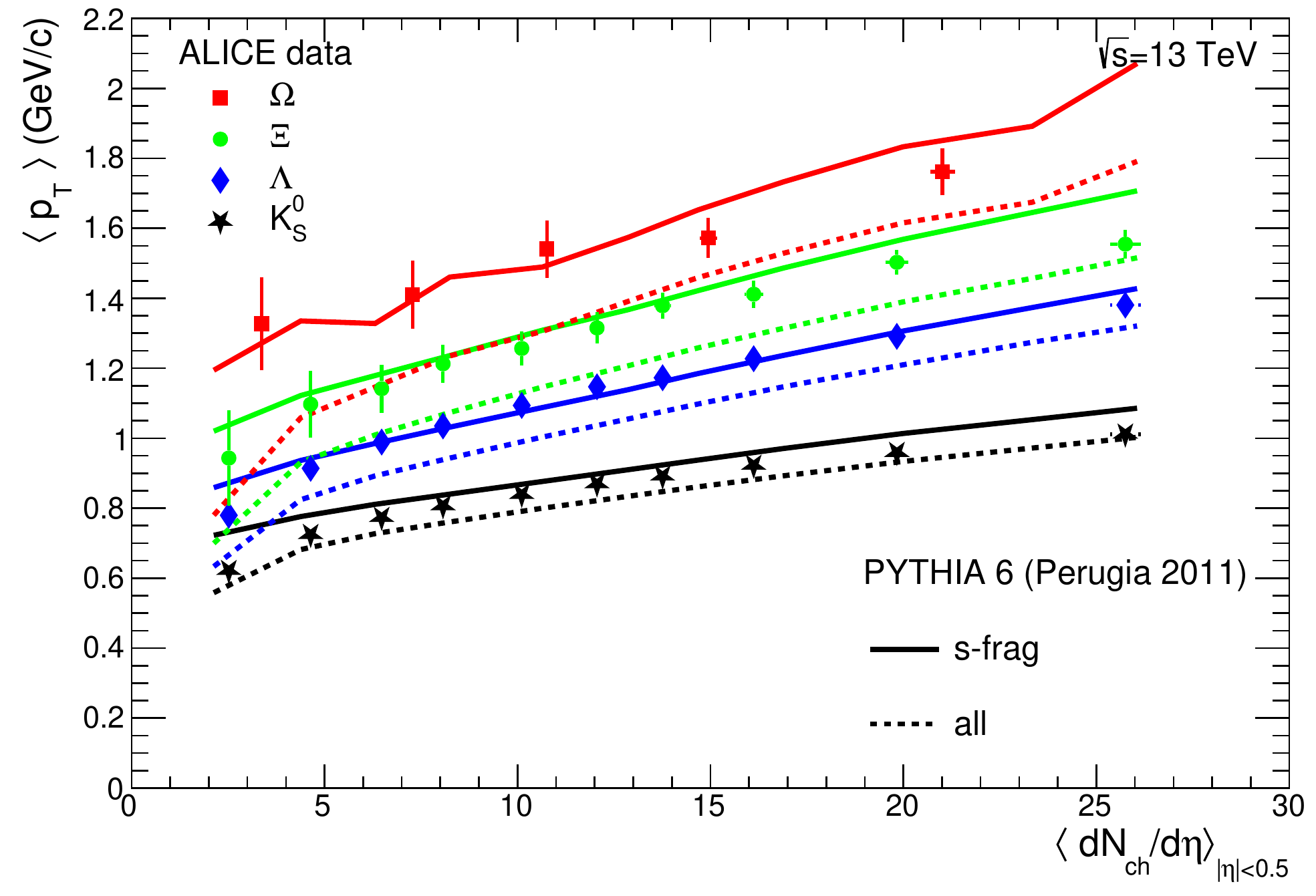}
\fi
\caption{Measured average transverse momenta of strange particles~($K_{\rm S}^0$, $\Lambda$, $\Xi$, $\Omega$) versus charged-particle multiplicity at midrapidity in \pp\ collisions at $\s=13$ TeV~\cite{Acharya:2019kyh} compared with \pytstrb\ calculations. 
%The calculation shows the $\avg{\pt}$ of all produced particles of a given type, as well as the $\avg{\pt}$ of those produced by $s$-quark fragmentation alone. 
See \Fig{fig:meanpt} for the corresponding \pytstr\ calculations.
} 
\label{fig:meanpt_app}
\end{center}
\end{figure}

\begin{figure}[htb!]
\begin{center}
\ifplotpold
\includegraphics[width=0.75\linewidth]{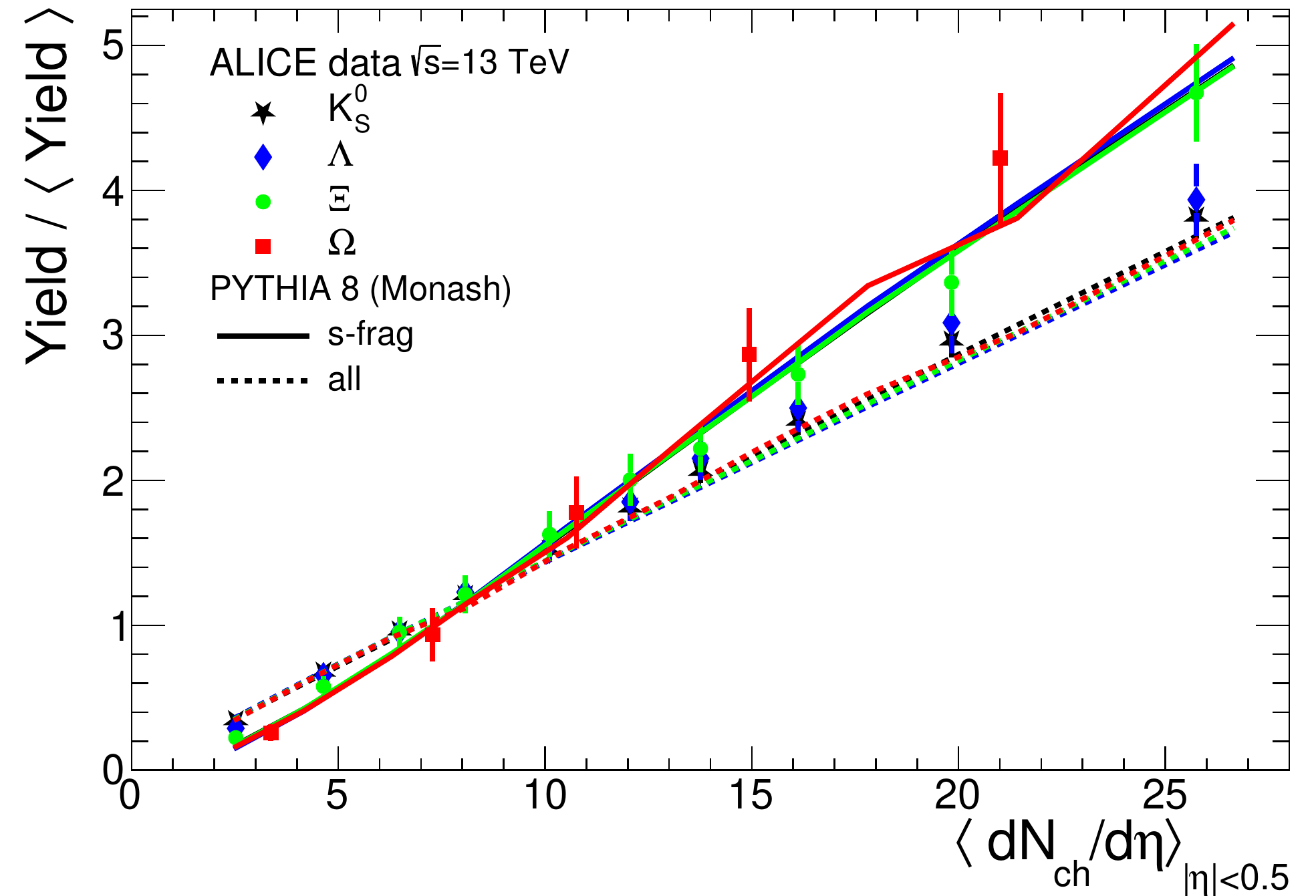}
\else
\includegraphics[width=0.75\linewidth]{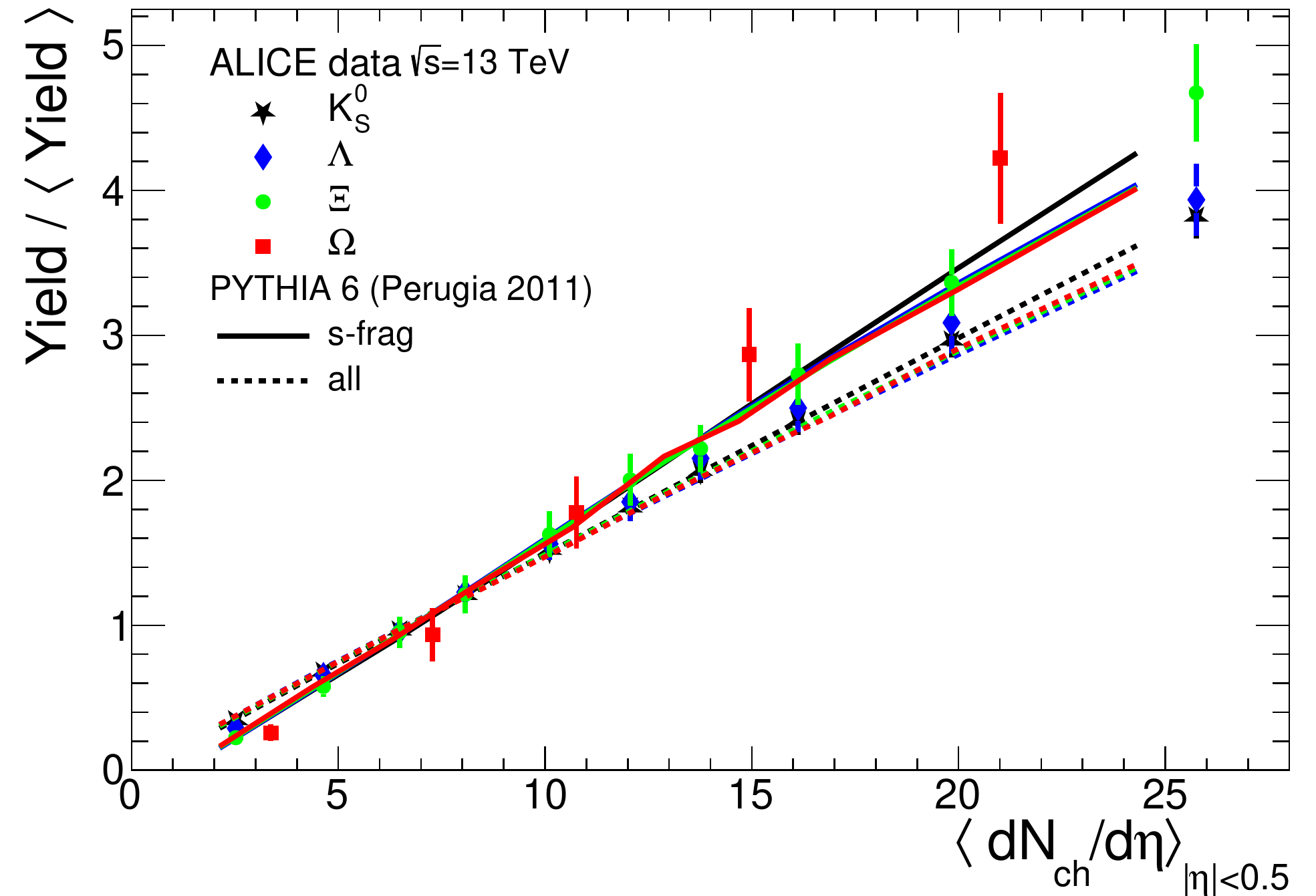}
\fi
\caption{Self-normalized yields of strange particles~($K_{\rm S}^0$, $\Lambda$, $\Xi$, $\Omega$) versus charged-particle multiplicity at midrapidity in \pp\ collisions at $\s=13$ TeV~\cite{Acharya:2019kyh} compared with \pytstrb\ calculations. 
%The calculation shows the yield of all produced particles of a given type, as well as the yield of those produced by $s$-quark fragmentation alone.
See \Fig{fig:selfnormyields} for the corresponding \pytstr\ calculations.
} 
\label{fig:selfnormyields_app}
\end{center}
\end{figure}

%%%%%%%%%%%%%%%%%%%%%%%%%%%%%%%%%%%%%%%%%%%%%%%%%%%%%%%%%%%%%%%%%%%%%%%%%%%%%%%%%%%%%%%%%%%
\ifarxiv
We thank J.\ Schukraft for many fruitful discussions on this topic over the past years and C.\ Bierlich for providing information on the rope hadronisation implementation in PYTHIA 8. We also thank S.\ Dash and R.\ Nayak for making available the PYTHIA 8 settings used in \cite{Nayak:2018xip} and N.\ Zardoshti for the proofreading of the manuscript.
\else
We thank J.\ Schukraft for many fruitful discussions on this topic over the past years and and N.\ Zardoshti for the proofreading of the manuscript.
\fi
C.L.~acknowledges financial support by the U.S.\ Department of Energy, Office of Science, Office of Nuclear Physics, under contract number DE-AC05-00OR22725. 
%%%%%%%%%%%%%%%%%%%%%%%%%%%%%%%%%%%%%%%%%%%%%%%%%%%%%%%%%%%%%%%%%%%%%%%%%%%%%%%%%%%%%%%%%%%
\bibliographystyle{utphys}
\bibliography{biblio}
%%%%%%%%%%%%%%%%%%%%%%%%%%%%%%%%%%%%%%%%%%%%%%%%%%%%%%%%%%%%%%%%%%%%%%%%%%%%%%%%%%%%%%%%%%%
\ifarxiv
\newpage
\appendix
\section*{Appendix}

\section{Additional figures}
\label{app:addfigs}
We show here the figures \Fig{fig:cmscomparison_app}, \Fig{fig:multdep_app}, \Fig{fig:meanpt_app} and \Fig{fig:selfnormyields_app} which were obtained with \pytstrb\ instead of \pytstr.
For \Pythia~6 we used version 6.4.25, while for \Pythia~8, version 8.243. 
We checked that we get the same results for \Pythia~8 with the latest version 8.306.
\begin{figure}[htb!]
\begin{center}
\ifplotpold    
\includegraphics[width=0.6\linewidth]{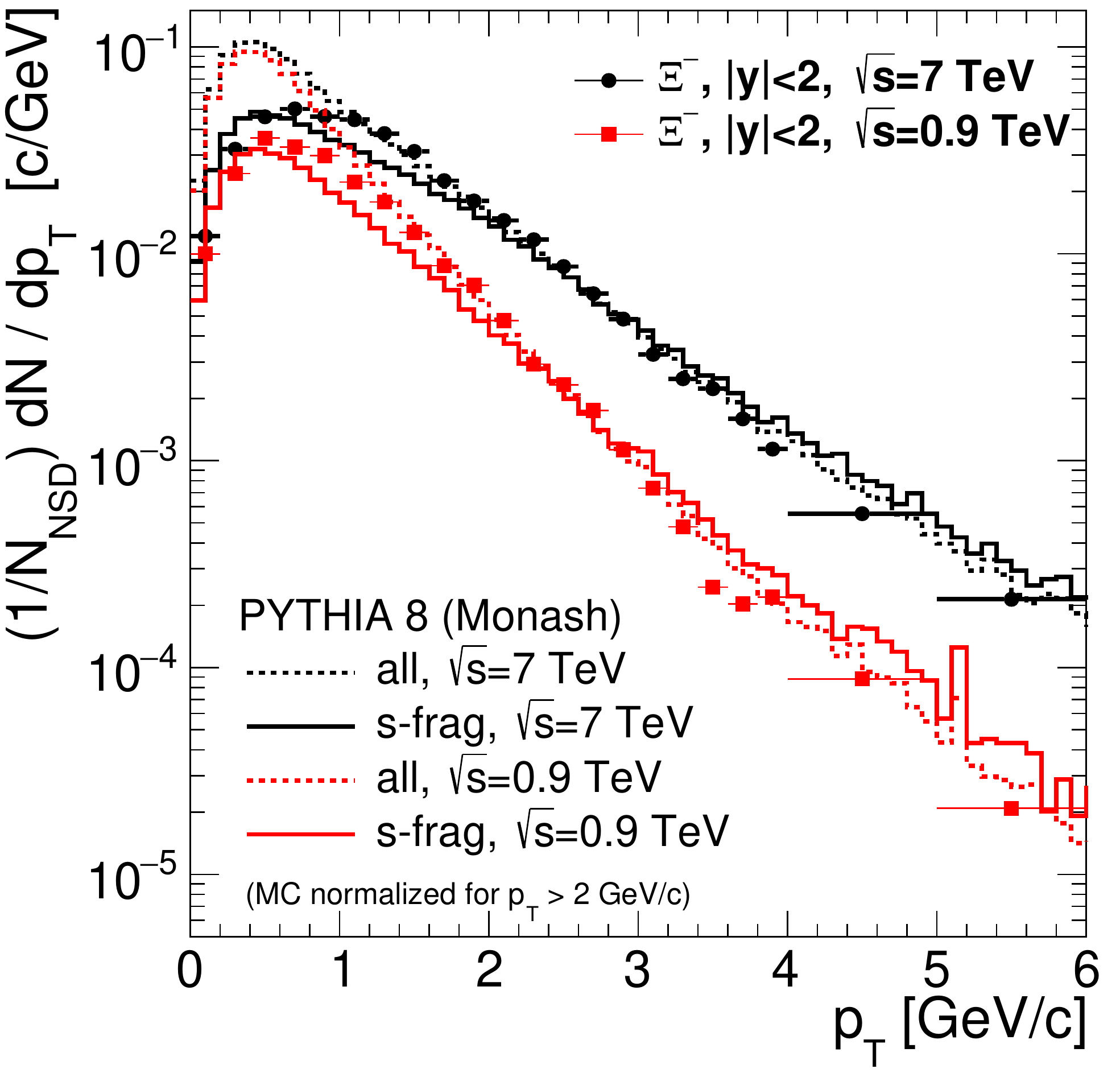}
\else
\includegraphics[width=0.6\linewidth]{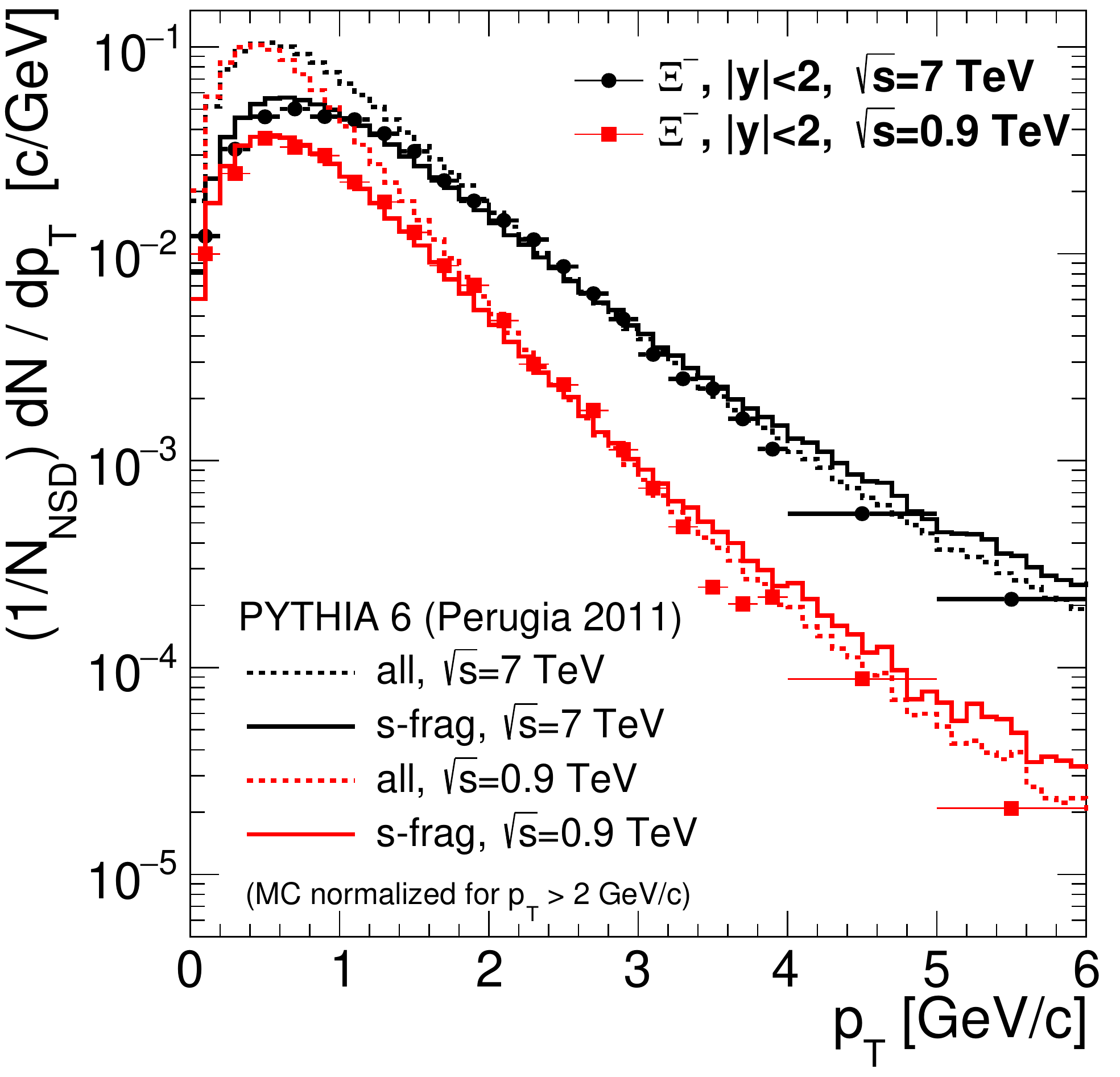}
\fi
\caption{$\Xi^{-}$ spectra in \pp\ collisions at 0.9 and 7~TeV measured by CMS~\cite{Khachatryan:2011tm} down to nearly zero $\pt$ compared with  \pytstrb\ calculations. The calculated spectra, which are shown for all produced $\Xi^{-}$ as well as for those produced by s-quark fragmentation alone, are normalized to the data in the region $\pt>2$~GeV/$c$. See \Fig{fig:cmscomparison} for the corresponding \pytstr\ calculations.
}
\label{fig:cmscomparison_app}
\end{center}
\end{figure}
\fi

\begin{figure}[htb!]
\begin{center}
\includegraphics[width=0.75\linewidth]{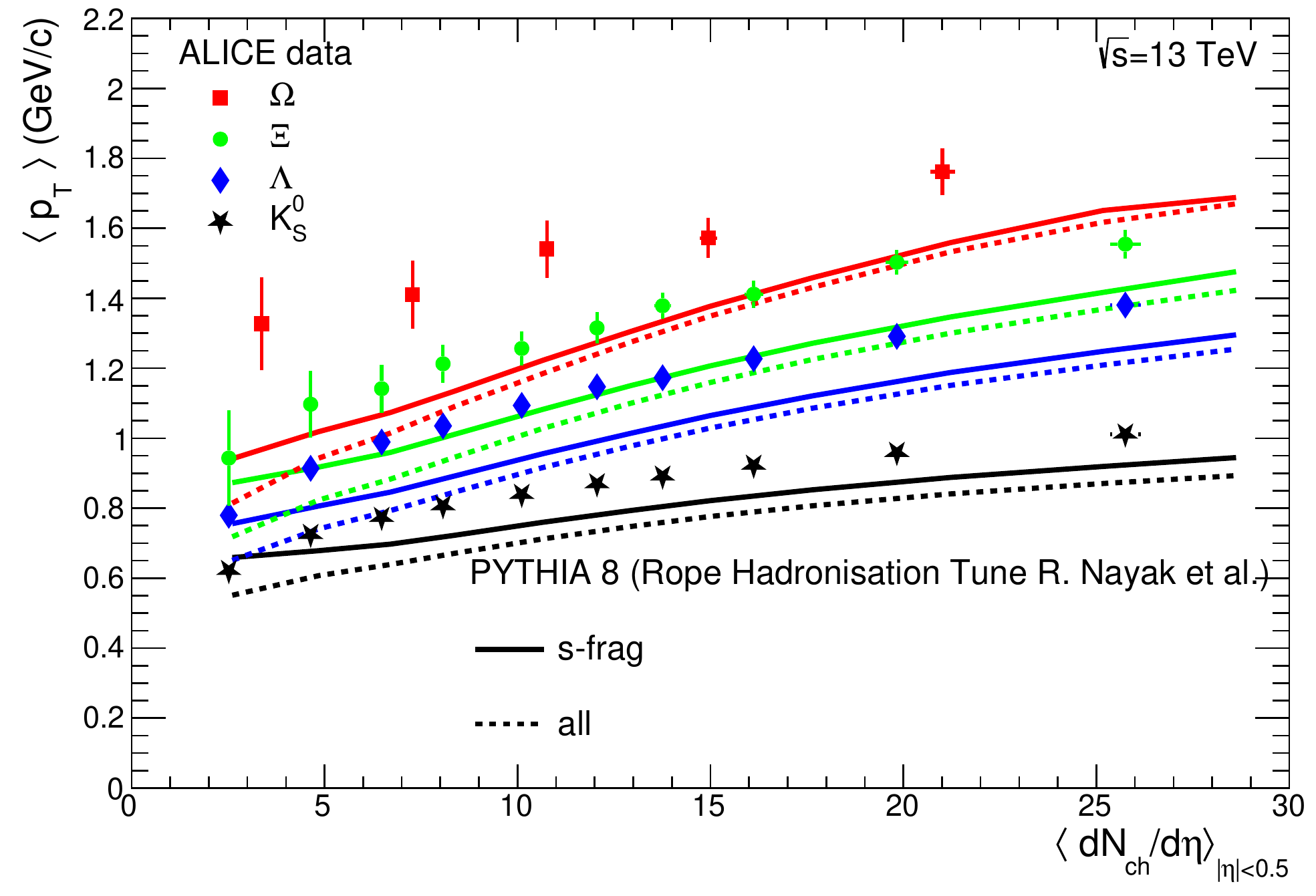}
\caption{Measured average transverse momenta of strange particles~($K_{\rm S}^0$, $\Lambda$, $\Xi$, $\Omega$) versus charged-particle multiplicity at midrapidity in \pp\ collisions at $\s=13$ TeV~\cite{Acharya:2019kyh} compared with calculations of \Pythia~8.306 with rope hadronization, with parameters by Nayak et al.\ as described in the text.
} 
\label{fig:meanpt_app_nayak}
\end{center}
\end{figure}
\begin{figure}[h!]
\begin{center}
\includegraphics[width=0.75\linewidth]{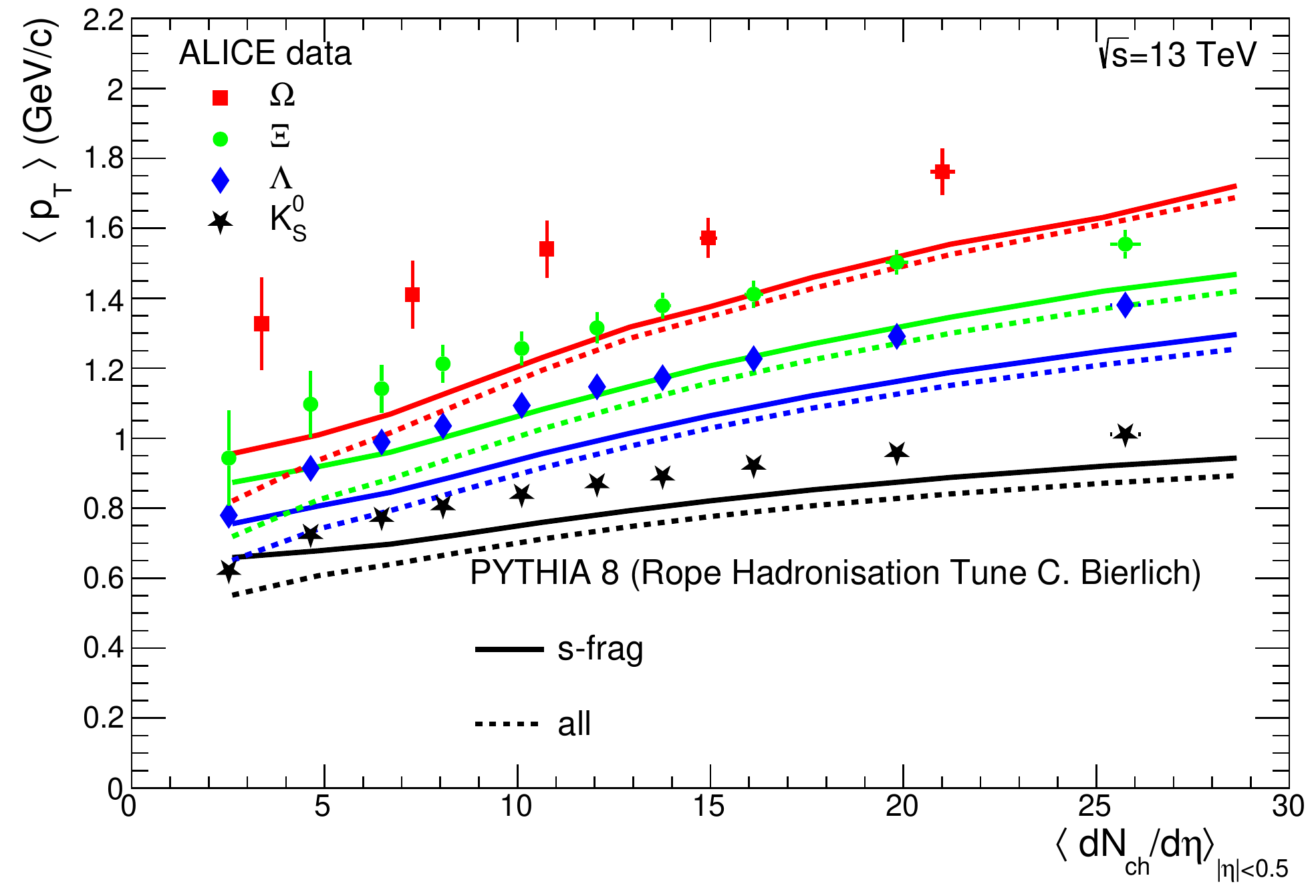}
\caption{Same as \Fig{fig:meanpt_app_nayak} but with parameters from Bierlich.
} 
\label{fig:meanpt_app_ropes1}
\end{center}
\end{figure}

\clearpage
\section{\Pythia\ with rope hadronization}
\label{app:rope}
In \Refe{Nayak:2018xip} it is claimed that \Pythia\ including the rope hadronization mechanism, which was introduced in \cite{Bierlich:2014xba}, is able to describe yields and mean $\pt$ of strange and multi-strange particles versus multiplicity in \pp\ collisions at 7 and 13~TeV.
We therefore repeated our calculations with the latest \Pythia~8.306 and the following explicit settings for color reconnection and ropt hadronization:
\begin{verbatim}
// QCD based CR
pythia->ReadString("MultiPartonInteractions:pT0Ref = 2.15");                                                                                             
pythia->ReadString("BeamRemnants:remnantMode = 1");                                                                                                      
pythia->ReadString("BeamRemnants:saturation = 5");                                                                                                       
pythia->ReadString("ColourReconnection:mode = 1");                                                                                                       
pythia->ReadString("ColourReconnection:allowDoubleJunRem = off");                                                                                        
pythia->ReadString("ColourReconnection:m0 = 0.3");                                                                                                       
pythia->ReadString("ColourReconnection:allowJunctions = on");
pythia->ReadString("ColourReconnection:junctionCorrection = 1.2");                                                                                       
pythia->ReadString("ColourReconnection:timeDilationMode = 2");                                                                                           
pythia->ReadString("ColourReconnection:timeDilationPar = 0.18");                                                                                         
// Rope Hadronization 
pythia->ReadString("Ropewalk:RopeHadronization = on");
pythia->ReadString("Ropewalk:doShoving = on");
pythia->ReadString("Ropewalk:doFlavour = on");
pythia->ReadString("Ropewalk:r0 = 0.5");
pythia->ReadString("Ropewalk:m0 = 0.2");
pythia->ReadString("Ropewalk:beta = 1.0");      
// Set shoving strength to 0 explicitly                                                                 
pythia->ReadString("Ropewalk:gAmplitude = 0."); 
// Parton Vertex
pythia->ReadString("PartonVertex:setVertex = on");                                                                                                       
pythia->ReadString("PartonVertex:protonRadius = 0.7");        
pythia->ReadString("PartonVertex:emissionWidth = 0.1");
\end{verbatim}
For the Nayak et al. parameters~\cite{Nayak:2018xip} we used
\begin{verbatim}
// Nayak et al. parameters
pythia->ReadString("Ropewalk:tInit = 1.0");     
pythia->ReadString("Ropewalk:tShove = 10.");   
pythia->ReadString("Ropewalk:deltat = 0.05");
\end{verbatim}
For the Bierlich parameters~(which he communicated to us by email) we used
\begin{verbatim}
// Bierlich parameters
pythia->ReadString("Ropewalk:beta = 0.1");
pythia->ReadString("Ropewalk:tInit = 1.5");
pythia->ReadString("Ropewalk:tShove = 0.1");
\end{verbatim}
The respective results for the mean $\pt$ versus multiplicity are shown in \Fig{fig:meanpt_app_nayak} for the Nayak et.\ al.\ and \Fig{fig:meanpt_app_ropes1} for the Bierlich settings.
To our surprise the results that are very similar for the two settings do not reveal the reported agreement with the data.
\end{document}